\input harvmac
\input epsf.tex
\overfullrule=0mm

\def\gl{\lambda}

\def\ge{\epsilon}
\def\gs{\sigma}
\def\gd{\delta}
\def\gr{\rho}
\def\ga{\alpha}
\def\gb{\beta}
\def\gc{\gamma}

\def\encadre#1{\vbox{\hrule\hbox{\vrule\kern8pt\vbox{\kern8pt#1\kern8pt}
\kern8pt\vrule}\hrule}}
\def\encadremath#1{\vbox{\hrule\hbox{\vrule\kern8pt\vbox{\kern8pt
\hbox{$\displaystyle #1$}\kern8pt}
\kern8pt\vrule}\hrule}}

\newcount\figno
\figno=0
\def\fig#1#2#3{
\par\begingroup\parindent=0pt\leftskip=1cm\rightskip=1cm\parindent=0pt
\baselineskip=11pt
\global\advance\figno by 1
\midinsert
\epsfxsize=#3
\centerline{\epsfbox{#2}}
\vskip 12pt
{\bf Fig. \the\figno:} #1\par
\endinsert\endgroup\par
}
\def\figlabel#1{\xdef#1{\the\figno}}
\def\encadremath#1{\vbox{\hrule\hbox{\vrule\kern8pt\vbox{\kern8pt
\hbox{$\displaystyle #1$}\kern8pt}
\kern8pt\vrule}\hrule}}

\def\tvi{\vrule height 12pt depth 6pt width 0pt}
\def\tv{\tvi\vrule}

\def\frac#1#2{\scriptstyle{#1 \over #2}}

%
%

	\def\CT{{\cal T}}

\def\({ \left( }
\def\){ \right) }
%


\def\IR{\relax{\rm I\kern-.18em R}}
\font\cmss=cmss10 \font\cmsss=cmss10 at 7pt
\def\IZ{\relax\ifmmode\mathchoice
{\hbox{\cmss Z\kern-.4em Z}}{\hbox{\cmss Z\kern-.4em Z}}
{\lower.9pt\hbox{\cmsss Z\kern-.4em Z}}
{\lower1.2pt\hbox{\cmsss Z\kern-.4em Z}}\else{\cmss Z\kern-.4em Z}\fi}
\def\inbar{\,\vrule height1.5ex width.4pt depth0pt}
\def\IB{\relax{\rm I\kern-.18em B}}
\def\IC{\relax\hbox{$\inbar\kern-.3em{\rm C}$}}
\def\ID{\relax{\rm I\kern-.18em D}}
\def\IE{\relax{\rm I\kern-.18em E}}
\def\IF{\relax{\rm I\kern-.18em F}}
\def\IG{\relax\hbox{$\inbar\kern-.3em{\rm G}$}}
\def\IH{\relax{\rm I\kern-.18em H}}
\def\II{\relax{\rm I\kern-.18em I}}
\def\IK{\relax{\rm I\kern-.18em K}}
\def\IL{\relax{\rm I\kern-.18em L}}
\def\IM{\relax{\rm I\kern-.18em M}}
\def\IN{\relax{\rm I\kern-.18em N}}
\def\IO{\relax\hbox{$\inbar\kern-.3em{\rm O}$}}
\def\IP{\relax{\rm I\kern-.18em P}}
\def\IQ{\relax\hbox{$\inbar\kern-.3em{\rm Q}$}}
\def\IGa{\relax\hbox{${\rm I}\kern-.18em\Gamma$}}
\def\IPi{\relax\hbox{${\rm I}\kern-.18em\Pi$}}
\def\ITh{\relax\hbox{$\inbar\kern-.3em\Theta$}}
\def\IOm{\relax\hbox{$\inbar\kern-3.00pt\Omega$}}






\def\dim{{\rm dim\,}}

\def\KK{Kontsevich}

%
\Title{SPhT 94/111}
{{\vbox {
\bigskip
\centerline{Quantum intersection rings}
\bigskip
}}}

\bigskip

\centerline{P. Di Francesco,}
\bigskip
\centerline{and}
\bigskip
\centerline{C. Itzykson}
\bigskip
\centerline{ \it Service de Physique Th\'eorique de Saclay
\footnote*{Laboratoire de la Direction des Sciences 
de la Mati\`ere du Commissariat \`a l'Energie Atomique.},}
\centerline{ \it F-91191 Gif sur Yvette Cedex, France}

\vskip .2in

\noindent

\Date{09/94}
%

\lref\KON{M. \KK,
\it Intersection theory on the moduli space of curves\rm,
Funk. Anal.\& Prilozh., {\bf 25} (1991) 50-57\semi 
\it Intersection theory on the moduli space of curves and the matrix Airy
function\rm, 
lecture at the Arbeitstagung, Bonn, June 1991 and 
Comm. Math. Phys. {\bf 147} (1992) 1.}
\lref\Wun{E. Witten, 
\it Two dimensional gravity and intersection theory on moduli space\rm,  
Surv. in Diff. Geom. {\bf 1} (1991) 243-310.}
\lref\WIT{E. Witten,
\it On the \KK\ model and other models of two dimensional gravity\rm,
preprint IASSNS-HEP-91/24}
\lref\Wdep{E. Witten, 
\it The $N$ matrix model and gauged WZW models\rm,
preprint IASSNS-HEP-91/26, to appear in Nucl. Phys. B. }
\lref\Wtr{E. Witten,
\it Algebraic geometry associated with matrix models of two dimensional 
gravity, \rm
 preprint IASSNS-HEP-91/74.}
\lref\CJB{ C. Itzykson and J.-B. Zuber,
\it Combinatorics of the Modular Group II: The \KK\ integrals. \rm, 
to appear in Int. J. Mod. Phys. }
\lref\INT{Harish-Chandra, 
{\it Differential operators on a semisimple Lie algebra}, 
Amer.J.Math. {\bf 79} (1957) 87-120 \semi
C. Itzykson and J.-B. Zuber, 
\it The planar approximation II,\rm\  J. Math. Phys. {\bf 21} (1980) 411-421.}
\lref\DVV{R. Dijkgraaf, H. Verlinde and E. Verlinde, \it  
Topological strings in $d<1$, \rm Nucl. Phys. {\bf B352} (1991) 59-86.}
\lref\DUB{B. Dubrovin, \it Integrable systems in topological field
theory, \rm\ Nucl. Phys. {\bf B379} (1992) 627; \it Geometry of 2D topological
field theories, \rm preprint SISSA--89/94/FM.}
\lref\KM{M. \KK\ and Y. Manin, \it Gromov--Witten classes,
quantum cohomology and enumerative geometry, \rm\ preprint MPI--94--21.}
\lref\CITZ{C. Itzykson, \it Counting rational curves on rational surfaces, \rm\ Int. J. of Mod. Phys. {\bf A} to appear.}
\lref\SR{J. Semple and L. Roth, \it Introduction to algebraic geometry, \rm\
Clarendon Press, Oxford (1987).}
\lref\SCHUB{H. Schubert, \it Kalk\"ul der abz\"ahlenden Geometrie, \rm\ reprint,
Springer, New York (1979).}
\lref\YM{Y. Manin, \it Cubic forms: algebra, geometry, arithmetics, \rm\
North Holland, Amsterdam (1974).}
\lref\MAC{I. Macdonald, \it Notes on Schubert polynomials, \rm\ LACIM,
Universit\'e du Qu\'ebec \`a Montr\'eal (1991).}
\lref\WG{E. Witten, \it The Verlinde algebra and the cohomology of 
the Grassmannian, \rm IASSNS preprint.}
\lref\HAR{J. Harris, \it Algebraic geometry--a first course, \rm Springer,
New York (1992).}
\lref\VAIN{I. Vainsencher, \it Enumeration of n-fold tangent hyperplanes
to a surface, \rm Journal of Algebraic Geometry, to appear.}
\lref\KLEI{S. Kleiman and R. Piene, private communication.}
\lref\FOK{A. Fokas and M. Ablowitz, \it On a unified approach
to transformations and elementary solutions of Painlev\'e equations, \rm
J. Math. Phys. {\bf 23} (1982) 2033; A. Fokas and Y. Yortsos, \it
The transformation properties of the sixth Painlev\'e
equation and one--parameter families of solutions, \rm Lett.
Nuovo Cimento {\bf 30} (1981) 539-544; 
A. Fokas, R. Leo, L. Martina and G. Soliani,
\it The scaling reduction of the three--wave resonant system and the Painlev\'e
VI equation, \rm Phys. Lett. {\bf A115} (1986) 329.}
\lref\RP{R. Piene, \it On the enumeration of algebraic curves- from circles to 
instantons, \rm to appear in the Proceedings of the first European Congress of Mathematics, Paris, 1992 (Birkh\"auser).}
\lref\HH{ J. Harris, \it Galois groups of enumerative problems, \rm Duke Math.
Journ. {\bf 46} (1979) 685-724.}
\lref\KQG{M. Kontsevich \it Intersection theory on the moduli space of curves
and matrix Airy function, \rm Comm. Math. Phys. {\bf 147} (1992) 1.}
\lref\IZQG{C. Itzykson and J.-B. Zuber, \it Combinatorics of the
modular group II, \rm Int. Jour. Mod. Phys. {\bf A7} (1992) 5661.}


\listtoc

\newsec{Introduction}

Within the broadly defined subject of topological field theory E. Witten
suggested in \Wun\ to study generalized ``intersection numbers"
on a compactified moduli space ${\bar {\cal M}}_{g,n}$ of 
Riemann surfaces. These are computed by integrating
pullbacks of appropriate forms on a target K\"ahler manifold obtained through
holomorphic maps of marked surfaces. The corresponding axioms, 
discussed by R. Dijkgraaf and E. and H. Verlinde \DVV, were investigated 
by M. Kontsevich and Y. Manin \KM\ and lead for a subclass
of targets to surprising results on
the enumeration of rational curves.
Our purpose here is to study a few illustrative examples
and to check some of them using the most primitive tools of 
geometry.

While modern algebraic and differential geometry has reached such a 
high degree of abstraction and sophistication it is refreshing to 
return with methods inspired from physics to problems of elementary
enumerative geometry such as those dealing with curves in projective 
spaces, initiated by Chasles, 
Schubert and Zeuthen. It is not clear to the authors when these lines 
are written that complete proofs are in print to support the conjectures 
that the numbers computed below are indeed what they are meant to be.
But it is likely that such proofs, if not available yet, should appear 
pretty soon in the literature.

Dealing with concrete examples of projective algebraic 
varieties covered by rational curves, we shall describe -- to the best of our
knowledge -- the classical cohomology ring $H^*(M)$. The latter corresponds
to some of
the quantum observables of the topological field theory. Their
correlation functions are however affected by quantum corrections
due to non trivial maps $\IP_1 \to M$ with the necessary
markings to rigidify them.
This leads to a deformation of the ring structure on
$H^*(M)$ explaining the name ``quantum cohomology ring".
This deformation, parametrized by $H^*(M)$, 
should satisfy conditions expressing commutativity,
associativity and the existence of a unit (the ``puncture operator").
The structure constants of this deformed ring 
(the analog of a fusion ring in conformal theory) are derived from a
generating function: the (perturbed) free energy. Once these elements
are  put together, one derives differential equations for the free energy.
These generate recursion relations for the enumerative numbers, therefore
entirely specified by boundary conditions.

The original example of the projective plane $\IP_2$ due to Kontsevich 
will be discussed first as it provides the simplest case and can be 
studied in some detail. We will comment on various aspects of
the sequence $\{ N_d \}$ of numbers of
rational plane curves of degree $d$ passing through $(3d-1)$ points
in general position. In particular we investigate
the asymptotic behaviour of $N_d$ for large $d$. 
Following Dubrovin \DUB, we also consider an associated flat
connection on a trivial bundle over $H^*(M) \times \IC$, and record for 
completeness in appendix A his geometric interpretation of the
differential equation leading to an equivalence with a Painlev\'e VI
equation.
The generalization of a geometric argument, 
in support of the basic differential
equation, produces a formula constraining the ``characteristic numbers"  
of rational plane curves which seems to be new.

To illustrate the generality of the method,
the subsequent sections are devoted to higher 
projective spaces, quadrics and cubics in $\IP_3$, the Pl\"ucker
quadric in $\IP_5$ describing line geometry in $\IP_3$, 
finally the flag variety of $\IP_2$.
In a final section we turn to open 
questions and relations.

As physicists we are more interested in
results that in the (painful and obviously necessary) process of 
justification of the interpretation. This might serve as an excuse for not
quoting adequately the extensive mathematical literature, especially
on intersection theory.
Finally a word of caution. 
As one of the authors was warned by the referee of a previous version
\CITZ, it would be often more appropriate to use the word
``conjecture" than ``proposition" in some statements. 
The reader will correct for himself.

C.I. is indebted to M. Kontsevich, S. Kleiman, R. Piene and I. Vainsencher
who
offered generously their help -- and time -- to introduce a novice to
the subject. Most of what follows is to be considered as an application
of their ideas, but the authors share the sole responsability for
the presentation. 
P.D.F. thanks C. Procesi for illuminating discussions.

\newsec{Rational curves in $\IP_2$}

Let $N_d$ be the number of rational irreducible plane curves
of degree $d$ through $3d-1$ points in general position.
\bigskip

\noindent{\bf Proposition 1.} (Kontsevich)
$$\eqalign{(i) \ N_1&=1 \cr
(ii)\ N_d&=\sum_{d_1+d_2=d \atop d_1,d_2 \geq 1} N_{d_1} N_{d_2}
\bigg[ d_1^2 d_2^2 { 3d-4\choose 3 d_1 -2 }-d_1^3 d_2 {3d-4 \choose 3d_1 -1}
\bigg] \ , \  d>1. \cr}
$$

Supplied by the initial condition $(i)$ (i.e. there exists a
single line through two distinct points in the plane, or dually two 
distinct lines intersect in a single point), the recursion relation $(ii)$
yields the set of $N_d$'s, $d>1$, as positive integers. It gives 
$N_2=1$ (a single conic through $5$ points), $N_3=12$ (uninodal cubics 
through $8$ points), $N_4=620$ (trinodal quartics through $11$ points), 
a result due to Zeuthen and recorded on page $186$ of 
Schubert's famous treatise \SCHUB. 
The next number $N_5=87304$ ($6$--nodal quintics
through $14$ points) and has been recently confirmed by Vainsencher \VAIN.

\subsec{Preliminaries} 

An irreducible algebraic plane curve of degree $d$ is described 
by the vanishing of an irreducible homogeneous polynomial of degree $d$ in
$3$ variables $f(x_0,x_1,x_2)=0$. The set of such curves is embedded in
a projective space $\IP_D$ describing homogeneous polynomials, of dimension
\eqn\dimo{ D= {d(d+3) \over 2} .}
Smooth curves form an open dense set in $\IP_D$, the complement
of a ``discriminant locus" of singular curves for
which either the polynomial is not irreducible and/or the tangent (line)
is not defined at certain points, i.e. the three partial derivatives
${\partial f \over \partial x_i}(x_0,x_1,x_2)=0$ admit simultaneous 
non trivial solutions.
Assuming that the only singularities of an irreducible curve are 
$\delta$ ordinary double points (i.e. with distinct tangents) 
its class $c$, the degree of the dual curve, is
\eqn\class{ c=d(d-1)- 2 \delta}
from which the Riemann--Hurwitz theorem yields the genus $g$ of the curve
\eqn\genus{ g={(d-1)(d-2) \over 2} - \delta}
Since each double point reduces the freedom of curves by one unit we expect the set of irreducible curves with $\delta$ double points to depend on 
\eqn\homog{ {d(d+3) \over 2}- \delta= 3d-1+g }
parameters, and therefore by imposing $3d-1+g$ independent conditions
on such curves to find a finite number of those.
We define $N_d^{(g)}$, and for short $N_d^{(0)}\equiv N_d$, 
to be the number of 
irreducible curves of degree $d$, genus $g$, with only simple nodes,
through $3d-1+g$ points ``in general position",
since requiring that the curve go through
a point implies a single linear condition. The statement ``in general
position" should be made precise in each concrete situation: it implies that
the conditions as applied to irreducible curves are independent.
Note that curves with simple nodes form the smallest codimension
set among curves of a fixed degree having a given genus $g \leq (d-1)(d-2)/2$.
In this sense they are the generic ones.

When $\delta=(d-1)(d-2)/2$, $g=0$, we require that curves go through
$3d-1$ points. Alternatively such a parametrized
rational curve is prescribed by $3$ 
homogeneous polynomials of degree $d$ in $2$ variables, hence depends on $3(d+1)$ 
parameters. Quotienting by linear transformations on the
homogeneous coordinates of $\IP_1$ leaves again $3d-1$ parameters.
The values $N_1=N_2=1$ are classical. To go slightly beyond, let 
$n_{d,\delta}\equiv N_d^{(g)}$, $g+\gd=(d-1)(d-2)/2$,
denote the number of irreducible degree $d$ plane curves with $\delta$ nodes
through $d(d+3)/2 -\delta$ points in general position.
For $\delta$ up to $6$ the following general results hold.
\bigskip

\noindent{\bf Proposition 2.}(Kleiman and Piene \KLEI, Vainsencher \VAIN\ )
$$\eqalign{ (i) \ n_{d,1}&= 3(d-1)^2,\ d \geq 3 \cr
(ii) \ n_{d,2}&= {3 \over 2} (d-1)(d-2)(3d^2-3d-11), \ d\geq 4 \cr
(iii) \ n_{d,3}&={9 \over 2} d^6 -27 d^5+{9 \over 2} d^4+{423 \over 2}d^3
-229 d^2-{829 \over 2}d+525 -\delta_{d,4}{11 \choose 2}, \ d\geq 4 \cr
(iv) \ n_{d,4}&={27 \over 8} d^8-27 d^7+{1809 \over 4}d^5-642 d^4 
-2529d^3+ \cr
&+{37881\over 8}d^2+{18057\over 4}d-8865 
-\gd_{d,5} {16 \choose 2}, d\geq 5 \cr
(v) \ n_{d,5}&={81 \over 40}d^{10}-{81 \over 4}d^9-{27 \over 8}d^8+
{2349 \over 4}d^7-1044 d^6-{127071 \over 20}d^5+{128859 \over 8}d^4 + \cr
&+{59097 \over 2}d^3-{3528381 \over 40}d^2-{946929 \over 20}d+153513
-\gd_{d,5}27 {15 \choose 2}, d \geq 5 \cr
(vi) \ n_{d,6}&={81 \over 80} d^{12}-{243 \over 20}d^{11}-{81 \over 20}d^{10}
+{8667 \over 16}d^9-{9297 \over 8}d^8-{47727 \over 5}d^7+\cr
&+{2458629 \over 80}d^6+{3243249 \over 40}d^5-{6577679 \over 20}d^4
-{25387481 \over 80}d^3+\cr
&+{6352577 \over 4}d^2+{8290623 \over 20}d-2699706-\gd_{d,5}
\bigg[{14 \choose 5}+225{14 \choose 2}\bigg], d\geq 5
\cr}$$

For $d=3$, $(i)$ yields $N_3=12$, from $(iii)$ for $d=4$
one recovers $N_4=675-55=620$, while for $d=6$, $(vi)$ yields
$N_5=87304$.

\noindent{\bf Remarks.} 

a) The polynomial parts of these formulas include reducible curves.
For instance,
for $d=3$, formula $(ii)$ yields the ${7 \choose 2}=21$ reducible
cubics formed by a conic through $5$ points and a line through 
the remaining two points. Similarly the subtractive term of $(iii)$
for $d=4$ represents the ${11 \choose 2}=55$ reducible quartics formed
by a cubic through $9$ points and a line through the remaining two,
with analogous interpretations for the remaining relations.

b) Including some reducible cases for small $d$ these results would suggest
that $n_{d,\delta}$, $1 \leq \delta\leq (d-1)(d-2)/2$, is a 
polynomial in $\IQ[d]$ of degree $2 \delta$ with leading term 
of the form $(3d^2)^\delta / \delta! 
\simeq n_{d,1}^\delta/\delta!$, corresponding to the fact
that one has essentially to pick $\delta$ points on a Jacobian variety
(see below) of degree $3(d-1)^2$. We conjecture the following general 
structure for the first few coefficients of the polynomial part
of $n_{d,\gd}$, denoted $z_{d,\delta}$
\eqn\conjnd{
\eqalign{z_{d,\gd}&={3^\gd \over \gd !} 
\bigg[ d^{2 \gd}-2 \gd d^{2 \gd -1}+{\gd (4-\gd) \over 3}d^{2\gd -2}+
{\gd(\gd-1)(20 \gd -13) \over 6}d^{2 \gd -3}+\cr
&-{\gd(\gd-1)(69 \gd^2-85 \gd +92) \over 54}d^{2 \gd -4}-
{\gd (\gd-1)(\gd-2)(702 \gd^2-629 \gd -286) \over 270}d^{2 \gd-5}+\cr
&+{\gd(\gd -1)(\gd -2)(6028 \gd^3-15476 \gd^2+11701 \gd+4425) \over 3240}
d^{2\gd -6} +\cdots \bigg] \ ,\cr}
}
in agreement with the data of proposition 2.

An elementary derivation of $(i)$ and $(ii)$ 
(see for instance the book by Semple and Roth \SR\ )
will illustrate how rapidly these problems become intricate.

Degree $d$ curves through $d(d+3)/2-1$ points form a 
linear pencil
\eqn\penc{ \gl_0 f_0 + \gl_1 f_1=0}
For nodal curves we look for simultaneous solutions of the system
\eqn\jaco{ \gl_0 {\partial f_0 \over \partial x_i}+ \gl_1 
{\partial f_1 \over \partial x_i}=0, \ 0 \leq i \leq 2 ,}
which entails \penc\ by virtue of Euler's relation. 
Each solution yields a value of the ratio $\gl_0$:$\gl_1$
provided
\eqn\deter{\left\vert \matrix{{\partial f_0 \over \partial x_0}&
{\partial f_1 \over \partial x_0}\cr
{\partial f_0 \over \partial x_2} & {\partial f_1 \over \partial x_2}\cr }
\right\vert = \left\vert \matrix{ {\partial f_0 \over \partial x_1}&
{\partial f_1 \over \partial x_1}\cr
{\partial f_0 \over \partial x_2} & {\partial f_1 \over \partial x_2}\cr }
\right\vert =0 \ ,}
omitting the $(d-1)^2$ points where ${\partial f_0 \over \partial x_0}=
{\partial f_1 \over \partial x_2}=0$. {}From Bezout's theorem, this gives
$[2(d-1)]^2-(d-1)^2=3(d-1)^2$ points, and proves $(i)$.

To derive $(ii)$ for binodal curves of degree $d \geq 4$, consider the 
two dimensional system of degree $d$ curves ${\cal C}_{\gl}$
through
\eqn\pts{ \nu_d={d(d+3) \over 2} -2={(d-1)(d+4) \over 2} }
points, represented as
\eqn\faisc{ {\cal C}_{\gl}: \qquad \sum_{0\leq i \leq 2} \gl_i f_i(x)=0. }
with $f_i(x)$ three homogeneous polynomials of degree $d$, generically
such that $f_i(x)=0$ is irreducible and smooth. Consider a map
\eqn\maptoit{
\eqalign{ T \ : \ \IP_2(x) &\to \IP_2(f) \cr
x_0,x_1,x_2 &\to f_0(x),f_1(x),f_2(x) \cr } 
}
sending curves ${\cal C}_{\gl}$ to lines in $\IP_2(f)$, or points of
the dual 
${}^*\IP_2(f)=\IP_2(\gl)$, the space of curves ${\cal C}_{\gl}$. Different
choices of bases $f_0,f_1,f_2$ for ${\cal C}_{\gl}$ amount to a
$PGL(3)$ automorphism of $\IP_2(f)$. The map $T$  is locally one to one 
except on the Jacobian curve $J \subset \IP_2(x)$
\eqn\jacob{ J \ : \ {D(f) \over D(x) } \equiv 
\det({\partial f_i \over \partial x_j})=0 \ \ {\rm deg}(J)=3(d-1).}
Let
$\Gamma\equiv T(J) \subset \IP_2(f)$ and $\tilde \Gamma$ its dual in $\IP_2(\gl)$.
The assigned points \pts\ are generically simple nodes of $J$. Indeed
if $x^{(0)}$ is an assigned point, from Euler's identity
\eqn\euler{ {1 \over d} \sum_{0 \leq j \leq 2} x_j^{(0)} 
{\partial f_i \over \partial x_j^{(0)}}=f_i(x^{(0)})=0 \ ,}
it belongs to $J$. On the other hand $J$ is the locus of nodes of curves 
of the family ${\cal C}_{\gl}$. Hence there exists a curve in ${\cal C}_{\gl}$
with a node at $x^{(0)}$ and we can choose it to be of the form $f_0=0$ (this
induces the harmless presence of a singular curve in the basis). Pick a
coordinate system in $\IP_2(x)$ such that $x^{(0)}=(0,0,1)$ and such
that $x_0=0$ is tangent to $f_1=0$ (assumed smooth at $x^{(0)}$). Thus
\eqn\coor{ 
\eqalign{
f_0(x)&= x_2^{d-2} (a x_0^2+2 b x_0 x_1+c x_1^2)+ \cdots \cr
f_1(x)&= x_2^{d-1} (\alpha x_0) + \cdots \cr
f_2(x)&= x_2^{d-1} (\beta x_0+ \gamma x_1) + \cdots \cr}
}
Calculation yields $D(f)/D(x)=x_2^{3d-5} p_2(x_0,x_1)+ \cdots$ with 
$p_2(x_0,x_1)$ an homogeneous second degree polynomial generically irreducible,
thus confirming that $x^{(0)}$ is a simple node of $J$. 
Consequently, the genus of $J$ reads
\eqn\genJ{
\eqalign{ g(J)&= {(3d-4)(3d-5) \over 2}-{(d-1)(d+4) \over 2}=4d^2
-15d+12 \cr
2g(J)-2&= 2(d-1)(4d-11) \ .\cr}
}
The degree of $\Gamma=T(J)$ is the number of variable points in the intersection
of a curve ${\cal C}_{\gl}$ with $J$
\eqn\degam{ m={\rm deg}(\Gamma)=3(d-1)\times d -2 \times 
{(d-1)(d+4) \over 2}=2(d-1)(d-2) }
where in the subtraction the assigned points, as nodes of $J$,
are counted twice.
On the other hand the degree of $\tilde \Gamma$ is the number of singular 
(hence generically uninodal) curves in a linear pencil of curves ${\cal C}_{\gl}$
computed in $(i)$
\eqn\duadeg{ n= {\rm deg}({\tilde \Gamma}) = 3(d-1)^2 }
while
\eqn\eqgenera{ g({\tilde \Gamma})\equiv g(\Gamma)=g(J) .}
Clearly the number $n_{d,2}$ of binodal curves is the number of nodes of 
$\tilde \Gamma$ which are distinct from the $\nu_d$ ones, images of those of $J$.
With the following notations
\eqn\notplu{
\eqalign{\delta&= n_{d,2}+\nu_d = \ \hbox{number of nodes of } 
{\tilde \Gamma} \cr
\kappa&= \ \hbox{number of cusps of } {\tilde \Gamma}
=\ \hbox{number of cuspidal curves in } {\cal C}_\gl \cr
\tau&= \ \hbox{number of bitangents of } {\tilde \Gamma}
=\ \hbox{number of nodes of } \Gamma \cr
i&= \ \hbox{number of flexes of } {\tilde \Gamma}
= \ \hbox{number of cusps of } \Gamma \cr
g&=g({\tilde \Gamma})={(n-1)(n-2) \over 2} - \delta - \kappa\cr
&=g(\Gamma)={(m-1)(m-2) \over 2} -\tau -i \cr}
}
the Pl\"ucker formulas \SR\ read
\eqn\pluc{
\eqalign{
m&= 2 n + 2 g -2 - \kappa \cr
n&= 2 m + 2g-2 -i  \ .\cr}
}
Since we know $m$, $n$ and $g$, this gives the number of cuspidal curves 
of degree $d$ through $(d-1)(d+4)/2$ points
\eqn\kapcal{
\eqalign{ \kappa&= 2n-m+2g-2=6(d-1)^2-2(d-1)(d-2)+2(d-1)(4d-11) \cr
&=12(d-1)(d-2) \ ,\cr}
}
from which the formula for the genus $g$ yields 
\eqn\delcal{ \delta=n_{d,2}+ \nu_d = {(n-1)(n-2) \over 2} -g - \kappa}
Finally the number of binodal curves through $(d-1)(d+4)/2$ points reads
\eqn\ndcal{
\eqalign{ n_{d,2}&={(3d^2-6d+2)(3d^2-6d+1) \over 2} 
-(4d^2 -15d+12)
-12(d-1)(d-2)-{(d-1)(d+4) \over 2} \cr
&={3 \over 2} (d-1)(d-2)[3d(d-1)-11] \ . \cr}
}
Although more rigor must be provided to justify a number of implicit assumptions, 
this gives at least a heuristic proof of part $(ii)$ of 
Proposition 2 and gives a
flavour of how intricate the proof of parts $(iii)$--$(vi)$ can be.

\subsec{Quantum ring}

A ``topological $\sigma$--model coupled to two dimensional gravity" is a field
theory defined on some covering of the moduli space of marked 
Riemann surfaces  equipped with metrics, with 
target space a K\"ahler manifold-- here $\IP_2$.
This means that the basic fields are maps to the target $\IP_2$ 
together with the metric.
A suitable compactification of the space of maps is necessary. 
Critical points of the action correspond to conformal metrics on the
surface and maps to stable irreducible curves, satisfying certain 
conditions to be specified below. 
Finally a set of observables is selected with the
property that their correlations are pure numbers, generally rational. 
This is where the topological nature of the model manifests itself. 
The concept of short distance expansion for products of fields, familiar from the field theory point of view, translates into a consistent 
``fusion ring". A subset of  
observables are in correspondence with cohomology classes of the target, 
and for genus $0$ source, the corresponding
fusion ring is a deformation -- due to 
quantum corrections -- of the classical cohomology (or intersection) ring of
the target. This ``jargon" can eventually be translated into well defined axioms,
the role of ``physical intuition" being reduced to the interpretation,
which remains to be put on a more respectable mathematical footing.
Following Kontsevich and Manin, the deformation is parametrized by the (dual of)
the cohomology space as a complex vector space (as we will only deal here with 
even cohomology, this requires no  $\IZ_2$ grading of the space).

For $\IP_2$ the cohomology ring is $\IC[u]/u^3$ with basis
$t_i=u^i$, $i=0,1,2$. The multiplication table is therefore $t_i t_j=t_{i+j}$
for $i+j \leq 2$, and $0$ otherwise. The only non--trivial
relation $t_1^2=t_2$ is dually equivalent to the fact that two lines intersect
in a point in $\IP_2$.
A general element in $H^*$ is therefore of the form 
$\sum_{0 \leq i \leq 2} y_i t_i$\foot{For typographical reasons we use 
subscripts to index the coordinates $y_i$ instead of the more 
appropriate $y^i$, which might induce a confusion when raised 
to some powers. We return to upper indices in the appendix.}.

The free energy of the would--be topological field theory is split into 
contributions of genus $0$, $1$, ... and we call 
$F$ the genus $0$ contribution.
The formal power series $F(y_0,y_1,y_2)$ defined
up to a second degree polynomial, is a sum of a ``classical" and 
a ``quantum" part
\eqn\split{ F=f_{\rm cl} + f }
where $f_{\rm cl}$ is a cubic polynomial encoding the 
multiplication rules of the classical intersection ring, namely
\eqn\enco{ {\partial f_{\rm cl} \over \partial y_i \partial y_j \partial y_k}=
\ \hbox{coeff. of the top class $t_2$ in the product } t_i t_j t_k \ ,}
in accordance with the fact that it requires at least $3$ points to
stabilize $\IP_1$. In particular the intersection form is
\eqn\interform{ \eta_{ij} \equiv {\partial f_{\rm cl} \over 
\partial y_0 \partial y_i \partial y_j} }
For $\IP_2$ as a target the only non--vanishing 
values are $\eta_{02}=\eta_{11}=1$,
hence the inverse $\eta^{ij} =(\eta^{-1})_{ij}=\eta_{ij}$, and
\eqn\clatwo{ f_{\rm cl}(y_0,y_1,y_2)= { 1 \over 2} (y_0^2 y_2+ y_0 y_1^2) \ .}
The fact that all relations derived below only involve third derivatives 
of $F$ 
explains why it is defined only up to a second degree polynomial. 
The splitting
of the free energy into $f_{\rm cl}$ and $f$ is according to maps for 
which the image of $\IP_1$ is respectively a point or an irreducible curve.
An obvious invariant of the latter is its degree,
proportional to the integral over the pre--image of the K\"ahler class of 
$\IP_2$ represented by $t_1$.
In a path integral each ``instanton configuration" (non--trivial critical
point of the action or alternatively rational curve) would appear 
weighted by its ``area" (K\"ahler class). To rigidify the curve one 
requires that it intersects $(3d-1)$
indistinguishable points, dual to the class $t_2$.
This serves as motivation to postulate that 
\eqn\energ{ f= \sum_{d=1}^{\infty} N_d {y_2^{3d-1} \over (3d-1)!} e^{d y_1} \ .}
The factorial in the denominator accounts for the 
indiscernability of the points and a physicist would say that in \energ\
$N_d$ counts the number of degree $d$ rational curves through $(3d-1)$ points.
In particular $N_1=1$. Fusion implies the following:
\bigskip

\noindent{\bf Axiom.}
There exists a commutative, associative ring with a unit (rather an algebra over
$\IQ[[y_0,y_1,y_2]]$) with basis $T_0$ (the unit), $T_1$, $T_2$, such that
\eqn\qring{ T_i T_j = F_{ijk}(y_0,y_1,y_2) \eta^{kl} T_l \ .}

For short, subindinces on $F$ stand for derivatives w.r.t. 
the corresponding $y$ 
variables. Requiring $T_0$ to be the unit is equivalent to $F_{0ij}=\eta_{ij}=$
constant, thus agrees with $\partial f/\partial y_0=0$.
Commutativity is explicit. 
The most important constraint is the associativity of the deformed ring.
For $\IP_2$, this reduces to  a single equation
\eqn\assop{ 
\encadremath{f_{222}= f_{112}^2 - f_{111} f_{122}}
}
which from the expansion \clatwo\ is equivalent to the statement of proposition 1.

\bigskip
This is obviously {\it not} a proof but, from the suggestion of an underlying 
path integral, a strong hint -- supported as we saw by checks  
for $N_d$, $d \leq 5$.
We give below a sketch of a direct enumerative proof, 
along lines suggested by Kontsevich.

There is an important homogeneity relation crucial
in later generalizations. It is possible to assign 
weights denoted as $[.]$ to $y_0$, $e^{y_1}$, $y_2$ so as to make $F$ 
homogeneous up to a second degree polynomial,
which does not affect third derivatives. These are
\eqn\deghom{ [y_0]=1 \quad [e^{y_1}]=3 \quad [y_2]=-1 \quad [F]=1 \ .}
This means that $y_1$ is a parabolic
element of weight $0$, i.e. under $y_0 \to \gl y_0$, $y_1 \to y_1+ 3 {\rm ln}
\gl$, $y_2 \to \gl^{-1} y_2$, then $F \to \gl F+$ second degree polynomial.
This type of property will appear 
again and again in further examples.

\fig{Duality relations in a topological field theory. 
Each trivalent vertex stands for a $3$ point correlation of observables 
in the theory. The intermediate link symbolizes the sum over intermediate
states of the form $(ab)=a \eta^{ab} b$, where  
$\eta^{ab}$ plays the role of propagator between the two states.}{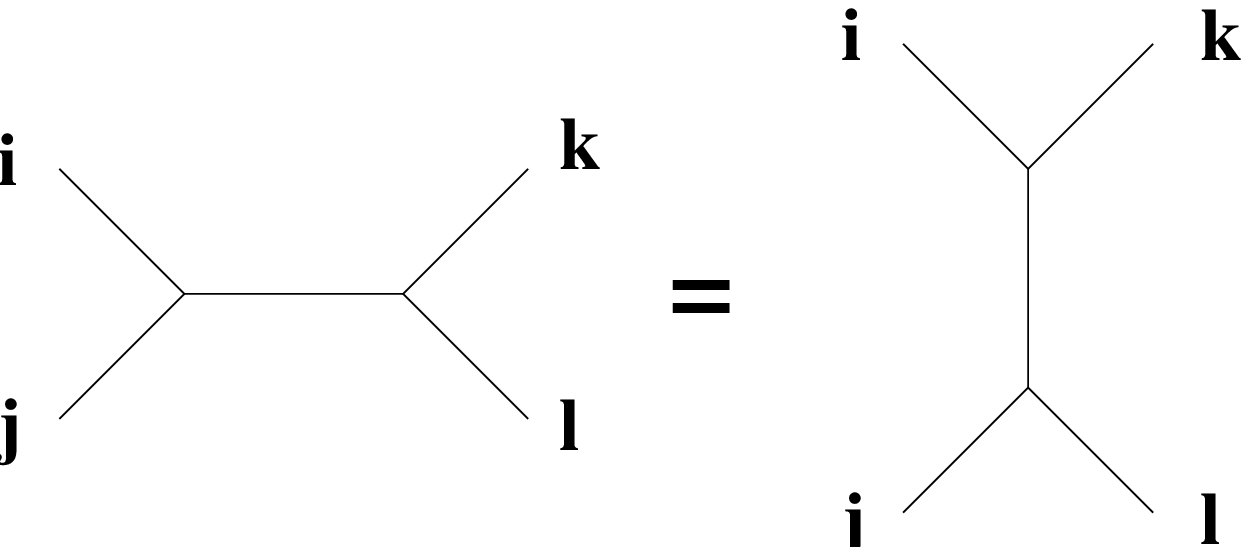}{5cm}
\figlabel\crossing

In topological field theories, 
the associativity conditions above are the ``duality" relations 
familiar to physicists. They express the equality between the two different
ways of writing a $4$ point correlation 
by decomposing it onto a complete set of
``intermediate states", leading to products of $3$ point correlations  
$F_{ijk}$ interpreted as correlations of the observables dual 
to $y_i$, $y_j$, $y_k$, namely the deformed classes 
$\langle T_i T_j T_k \rangle$ (Fig.\crossing\ ). 
Here the only non--trivial relation is obtained for
$i=j=1$, $k=l=2$, and the intermediate states to be summed over are $(02)$, $(11)$
and $(20)$ respectively, so that 
\eqn\croptwo{ \sum_{0 \leq j \leq 2} F_{11j} F_{2-j22}= 
\sum_{0 \leq j \leq 2} F_{12j} F_{2-j12} \ ,}
equivalent to \assop\ by writing $F=f_{\rm  cl}+f$, with $f_{\rm cl}$ as in 
\enco.

\subsec{Sketch of an enumerative proof (according to Kontsevich)}

An enumerative proof of proposition 1 is suggested by the very form of the 
recursion relation for $N_d$.
It is obtained by studying degenerate cases i.e. boundary cycles
on the moduli space.
Consider an algebraic family of rational irreducible curves through $(3d-2)$ points
split into a subset of $(3d-4)$ points $\{ q_* \}$ and two 
distinguished points $p_1$ and $p_2$. Let two fixed lines $l_3$
and $l_4$ intersect in $p_0$. For each curve $C$ of the family pick a point 
$p_i \in C\cap l_i$. Since $\{ p_1,p_2,p_3,p_4 \}$ all lie
on $C \simeq \IP_1$, it makes sense to consider their cross ratio
\eqn\crora{ x = {p_1-p_3 \over p_1-p_4}{p_2-p_4 \over p_2-p_3}}
which defines a map from the family to $\IP_1 -\{0,1,\infty \}$.
One computes its degree in two different ways by letting $x$ go to $0$ or $1$.
For $x=0$, $C$ degenerates in all possible ways
into a union of two necessarily rational curves 
$C_1$, $C_2$ of degrees $d_1$, $d_2$, with $d_1+d_2=d$. $C_1$ contains 
$p_1$, $p_3$ and $(3d_1-2)$ points among the $q_*$, which together with
$p_1$ make up the required number $(3d_1-1)$ to fix $N_{d_1}$ rational
curves. Similarly there are $N_{d_2}$ possibilities for $C_2$, 
containing $p_2$, $p_4$ and the other $(3d_2-2)$ $q_*$ points.
These two curves $C_1$ and $C_2$ correspond to pinching a generic curve $C$
at one of their $d_1d_2$ intersection points, while $p_3$ (resp. $p_4$)
is among the $d_1$ points in $C_1\cap l_3$ (resp. $C_2\cap l_4$).
This results in the following degree of the map
\eqn\degmap{ \sum_{d_1+d_2=d \atop d_1,d_2 \geq 1} N_{d_1} N_{d_2}
{ 3d-4 \choose 3d_1-2 }\times d_1 \times d_2 \times d_1d_2}

As $x \to 1$, there are two
possibilities.
Either $p_3=p_4=p_0=l_3\cap l_4$, giving a contribution $N_d$
to the degree, or the
curve degenerates with $p_3$, $p_4$ on $C_1$ of degree $d_1$, which should
therefore also contain $(3d_1-1)$ of the $q_*$, and $p_1$, $p_2$ on $C_2$
of degree $d_2$, which contains the $(3d_2-3)$ other $q_*$. The pinching point
is again one of the $d_1d_2$ points of $C_1\cap C_2$ 
while this time $p_3$ and $p_4$
are to be chosen among the intersections of $C_1$ with 
$l_3$ and $l_4$, $d_1^2$
in number. Altogether the degree is
\eqn\mapdeg{ N_d + \sum_{d_1+d_2=d \atop d_1,d_2 \geq 1} N_{d_1} N_{d_2}
{3d-4 \choose 3d_1-1 } d_1 \times d_2 \times d_1^2}
Equating \degmap\ and \mapdeg, we find the recursion 
relation of proposition 1.

The very spirit of this reasoning is not so different from the 
one which led to the consideration of the quantum cohomology 
ring in the first place, and the
degenerations of $C$ leading to the expressions \degmap\ and \mapdeg\
are reminiscent of the two diagrams of Fig.\crossing, which correspond
to two ways of degenerating a $4$ point correlation function.

\subsec{Asymptotics}

The solutions of the recursion formula for $N_d$ exhibits a factorial 
growth shown up to degree $12$ in table I.

%
$$\vbox{\offinterlineskip
\halign{\tv\quad # & \quad\tv \quad 
# & \quad \tv \quad  # & \quad \tv #\cr 
\noalign{\hrule}
\tvi $d$ & $N_d$& $(3d-1)$ &\cr
\noalign{\hrule}
\tvi  $1$ &  $1$  &         $2$ &\cr
\tvi  $2$ &  $1$ &         $5$ &\cr
\tvi  $3$ &  $12$ &        $8$ &\cr
\tvi  $4$ &  $620$ &       $11$ &\cr
\tvi  $5$ &  $87304$ &      $14$ &\cr
\tvi  $6$ &  $26312976$ &      $17$ &\cr
\tvi  $7$ &  $14616808192$ &     $20$ &\cr
\tvi  $8$ &  $13525751027392$ &    $23$ &\cr
\tvi  $9$ &  $19385778269260800$ &   $26$ &\cr
\tvi  $10$ & $40739017561997799680$ & $29$ &\cr
\tvi  $11$ & $120278021410937387514880$ & $32$ &\cr
\tvi  $12$ & $482113680618029292368686080$ & $35$ &\cr
\noalign{\hrule} }} $$
\noindent{\bf Table I:} The numbers $N_d$ of irreducible, degree $d$,
rational, plane curves through $(3d-1)$ points, up to $d=12$.
\vskip 1cm
%

\noindent{\bf Proposition 3.}
There exists two real positive numbers $a$ and $b$ such that 
\eqn\asymp{ {N_d \over (3d-1)!} = a^d d^{-7/2} b \big(1+O(d^{-1})\big)}

\bigskip
Numerically the asymptotic behavior is in agreement with the exponent $7/2$ and
yields
\eqn\numerics{ a=0.138 \qquad b=6.1 }

The proposition implies that, as a function of the variable
$y_2^3 e^{y_1}$, $y_2f(y_1,y_2)$ admits a convergent power series in a disk
of radius $a^{-1}$, with a singularity on the real axis 
(a square root branch point). 
One might think of ${\rm ln}a$ as an analog of entropy of 
rational curves of large degree. 

The proof is divided into two parts. The first uses the ``explicit" 
solution of the 
recursion relation for $N_d$ to prove the convergence of 
the power series for
$f$. The second is based on an analysis of the differential
equation to extract the exponent $7/2$.

Let us rewrite the recursion relation in proposition 1 by 
symmetrizing the sum
on the r.h.s. as
\eqn\symnd{
\eqalign{{N_d \over (3d-1)!} &= 
\sum_{d_1+d_2=d \atop d_1,d_2 \geq 1} {N_{d_1}\over (3d_1-1)!} 
{N_{d_2}\over (3d_2-1)!} q(d_1,d_2) \cr
q(d_1,d_2)&= {d_1d_2 (3d_1d_2(d+2)-2d^2)\over 2(3d-1)(3d-2)(3d-3)}\cr
&={d_1d_2[(3d_1-2)(3d_2-2)(d+2)+8(d-1)]\over 6(3d-1)(3d-2)(3d-3)}\cr}
}
where $d=d_1+d_2$. The second form exhibits the positivity of $q(d_1,d_2)$.
In general, consider a recursion relation of the type
\eqn\recufor{ 
\eqalign{ u_1&>0 \cr
u_d&=\sum_{d_1+d_2=d \atop d_1,d_2 \geq 1} u_{d_1} u_{d_2}  X(d_1,d_2) 
\ d\geq 2\cr}
}
where $X(d_1,d_2)$ need not be symmetric.

\bigskip
\noindent{\bf Lemma 1.} 

(i) For $d>1$, $u_d$ is $(u_1)^d$ times a sum of contributions, 
each one assigned to a rooted trivalent tree graph with oriented and labeled edges.
At each vertex there is one incoming edge labeled $d_a$, and there are two outgoing
edges labeled $d_b$, $d_c$, with $d_a=d_b+d_c$. Vertices, $(d-1)$ in number, are
assigned a weight $X(d_b,d_c)$. The contribution of a tree is the product of
its vertex weights.

(ii) The number of trees contributing to $u_d$ is the Catalan number
$(2d-2)!/[d! (d-1)!]$.

(iii) Let $u_d^{(1)}$ and $u_d^{(2)}$ be two solutions of a recursion relation of the
type \recufor\ with kernels $X^{(1)}$ and $X^{(2)}$, such that 
$u_1^{(1)}\geq u_1^{(2)}$, and $X^{(1)}(d_1,d_2)\geq X^{(2)}(d_1,d_2)$, then
for all $d\geq 1$, $u_d^{(1)}\geq u_d^{(2)}$.

Part (i) is obtained by iterating the recursion relation until one 
gets rid of all $u_d$, $d>1$.
As for (ii), it is sufficient to replace $X(d_1,d_2)$ by a constant $X$ and 
to consider the generating function
\eqn\catagen{ \Phi(t)= \sum_{d=1}^{\infty} u_d t^d }
satisfying the quadratic relation 
\eqn\quacat{ \Phi(t)-tu_1=X \Phi(t)^2 }
hence
\eqn\catphi{ \Phi(t)={1-\sqrt{1-4Xtu_1} \over 2X}=\sum_{d=1}^{\infty}
{(2d-2)! \over d! (d-1)!} u_1^d X^{d-1} t^d }
proving (ii). Property (iii) follows immediately from (i).

\fig{Rooted trees, the flow of degrees and the corresponding ``parenthesing"
for $d=2$, $3$ and $4$.}{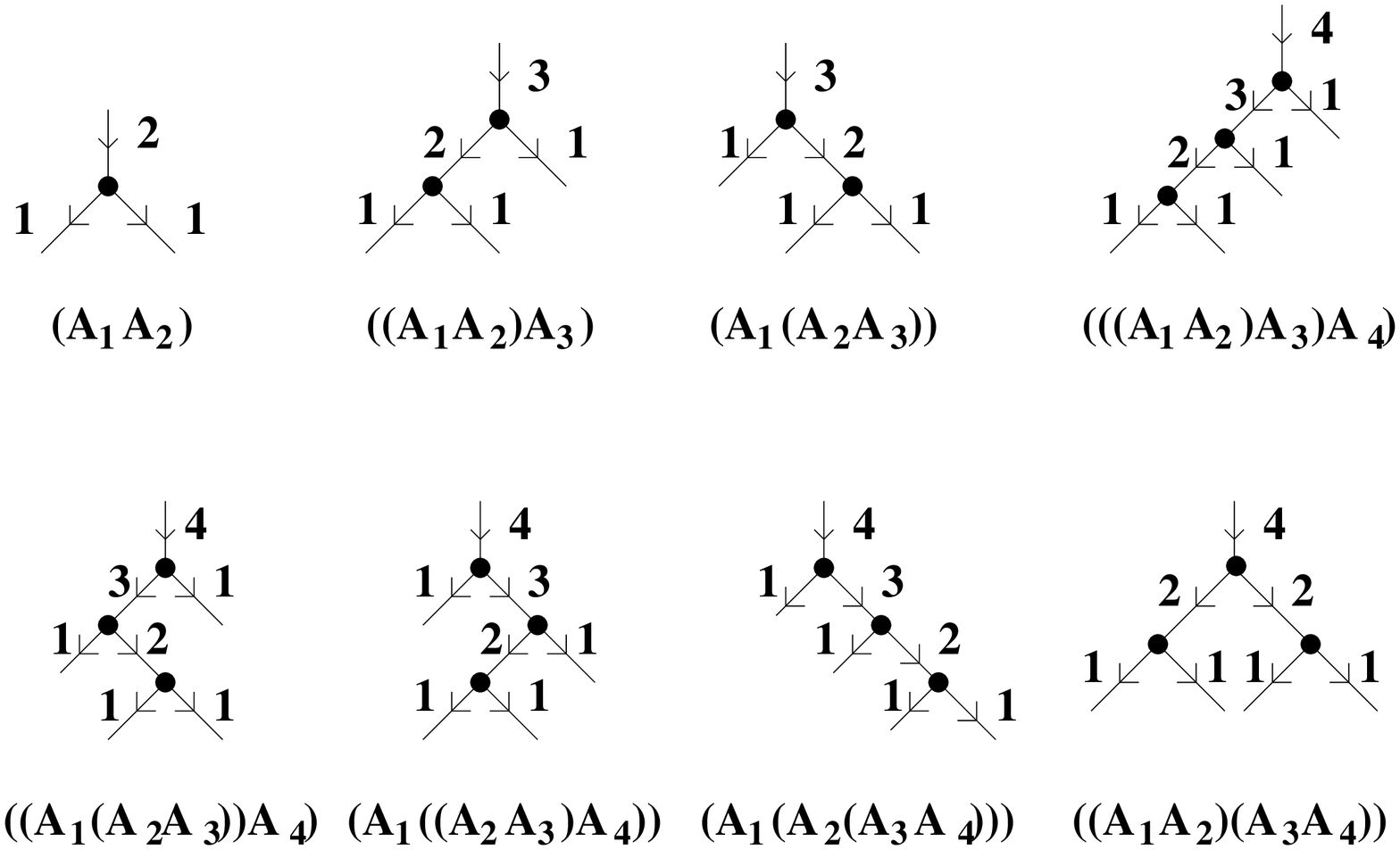}{10cm}
\figlabel\parent

The above trees are in one to one correspondence with ``parenthesings". Given
a sequence of symbols $A_1$, $A_2$, ...,$A_d$, we insert pairs of open/closed
parentheses in all possible ways and compute the corresponding weight
accordingly, as exemplified in Fig.\parent, for $d=2$, $3$, 
and $4$. Thus, with 
$Q(d_1,d_2)=Q(d_2,d_1)\equiv q(d_1,d_2) (3d-1)!/[(3d_1-1)!(3d_2-1)!]$, and
$Q(1,1)=1$, $Q(2,1)=6$, $Q(2,1)=224$, $Q(3,1)=33/2$, $N_1=1$, we find
\eqn\calcatnd{
\eqalign{N_2&=Q(1,1)=1\cr
N_3&=2 Q(1,1)Q(2,1)=12\cr
N_4&=4Q(1,1)Q(2,1)Q(3,1)+Q(1,1)^2Q(2,2)=4.6.33/2+224=620\cr}
}
in agreement with table I.

There is a natural equivalence relation on the kernels $X(d_1,d_2)$, namely
\eqn\eqrel{ X(d_1,d_2) \sim X(d_1,d_2) {X(d_1) X(d_2) \over X(d_1+d_2)}}
for arbitrary non vahishing $X(d)$. It amounts to replace $u_d \to X(d) v_d$.
The equivalence class of $X(d_1,d_2)=X=$ constant yields the Catalan numbers.

Since $N_d/(3d-1)!$ is positive, to show that the series for 
$y_2f(y_1,y_2)$ converges as claimed in proposition 3, 
it is sufficient to prove, 
for large $d$, inequalities of the type
\eqn\fram{ (a_<)^d \leq {N_d \over (3d-1)!} \leq (a_>)^d \ .}
To obtain the lower bound we start from
\eqn\kernel{ q(d_1,d_2)={d_1 d_2 \over 6}
{[(3d_1-2)(3d_2-2)(d+2)+8(d-1)] \over (3d-1)(3d-2)(3d-3)} \ .}
In the numerator we omit the positive term $8(d-1)$ and replace $(d+2)$ by $(d-1)$
while in the denominator we replace $(3d-1)$ by $3d$ to obtain
\eqn\minker{q(d_1,d_2)> {d_1 d_2 \over 54 d} {(3d_1-2)(3d_2-2) \over (3d-2)}
\equiv q_{<}(d_1,d_2) }
with $q_<(d_1,d_2)$ in the equivalence class of a constant.
According to lemma 1, taking into account $N_1/(3.1-1)!=1/2$, we get
\eqn\minnd{ {N_d \over (3d-1)!} > {27 (2d)! \over (108)^d d(2d-1)(3d-2)[d!]^2} }
and using Stirling's formula $(2d)! /[d!]^2\sim 2^{2d}/\sqrt{\pi d}$,
for $d$ large enough
\eqn\minimnd{{N_d \over (3d-1)!} > {9 \over 2 \sqrt{\pi} (27)^d d^{7/2}}(1+
O(1/d))}
proving that the series for $F$ has a non--vanishing radius
of convergence.
Similarly to obtain a rough upper bound starting from $q(d_1,d_2)$, we increase
the numerator by replacing $8(d-1)$ by $8(d-1)(3d_1-2)(3d_2-2)$, $(d+2)$ by
$4(d-1)$ and $d_1 d_2$ by $(3d_1-1)(3d_2-1)/4$.
Hence
\eqn\upker{ q(d_1,d_2)< {(3d_1-1)(3d_1-2)(3d_2-1)(3d_2-2)\over 6(3d-1)(3d-2)}
\equiv q_>(d_1,d_2)}
so that
\eqn\maxnd{ {N_d \over (3d-1)!} < {3 (2d)! \over 6^d (2d-1)(3d-1)(3d-2) [d!]^2}}
and asymptotically
\eqn\ndsup{ {N_d \over (3d-1)!} < {1 \over 6 \sqrt{\pi} d^{7/2}}
\left({2 \over 3}\right)^d (1+O(1/d)) }
{}From this we conclude that the series for $y_2 f$ has a finite radius of convergence $a={\rm lim}_{d \to \infty} [N_d/(3d-1)!]^{1/d}$, satisfying
\eqn\boundrad{ {1 \over 108} < a < {2 \over 3} \ .}
Our numerical value $.138$ satisfies these bounds. 

To find the power law
prefactor $d^{-7/2}$, we set
\eqn\defG{ G(x)={1 \over 3} \sum_{d=1}^{\infty} {N_d \over (3d-1)!} e^{dx} }
with a finite abscissa of convergence $Re(x)<x_0={\rm ln}1/a$ and a 
behavior $e^x/6$ for $x \to -\infty$. It satisfies the translation 
invariant differential equation
\eqn\diffG{ (9+2G'-3G'')G'''=2G+11G'+18G''+(G'')^2 \ .}
The functions $G$, $G'$, $G''$, $G'''$ are all positive with $G<G'<G''<G'''$
for real $x<x_0$. Therefore  $G$, $G'$ and $G''$ remain
finite at $x_0$ with $G'''$ blowing up. With $0<\alpha<1$, we have in the 
vicinity of $x_0$
\eqn\vici{ G(x)=g_0+g_1(x_0-x)+g_2{(x_0-x)^2 \over 2} +\gl (x_0-x)^{2+\alpha}
+\cdots}
hence
\eqn\calcvici{
\eqalign{ -[(9-2g_1-3g_2)-3 \gl(1+\alpha)(2+\alpha)&(x_0-x)^{\alpha}+\cdots]
\gl \alpha (1+\alpha)(2+\alpha)(x_0-x)^{\alpha-1} \cr
&=2g_0-11g_1+18g_2+g_2^2+O(x)}
}
leading to
\eqn\calccivi{
\eqalign{ 
9-2g_1-3g_2=0 &\qquad 2 \ga -1 =0 \cr
3 \ga (1+\ga )^2(2+\ga )^2 \gl^2&=2g_0-11g_1+18g_2+g_2^2 \cr}
}
thus $\alpha=1/2$ which corresponds to a behavior 
\eqn\behav{ \lim_{d \to \infty} {d^{7/2} N_d \over a^d (3d-1)!} ={\rm const.} }
This concludes the proof of proposition 3. 
It is remarkable that our very poor upper and lower bounds both exhibit 
the same power law exponent $7/2$. If, as physicists, we write it as 
$3-\gamma_{\rm string}$ 
(the expected area dependence of the free energy of a two dimensional 
quantum gravity theory), we find $\gamma_{\rm string}=-1/2$, 
which corresponds to
the ``pure gravity" case\foot{Using this analogy with quantum gravity,
we are led to the following
conjecture for the asymptotics of higher genus $g>0$ numbers
of curves through $(3d-1+g)$ points 
$$ {N_d^{(g)} \over (3d-1+g)!} \propto {a^d \over d^{1+{5 \over 2}(1-g)}}$$.}.
As shown in appendix A, following Dubrovin \DUB,
the differential equation \diffG\ is related to a particular case of 
a Painlev\'e VI equation.

\subsec{Flat connection}

For generic
values of $y_i$ within the domain of convergence of $f$, the commutative
algebra generated by the $T_i$'s is semi--simple. Its regular representation
as linear operators acting on the basis $(T_2,T_1,T_0)^T$ reads
\eqn\repsq{ T_0=\pmatrix{1&0&0\cr 0&1&0\cr 0&0&1\cr} \ \ 
T_1=\pmatrix{0&f_{112} &f_{122}\cr
1&f_{111}&f_{112}\cr
0&1&0\cr} \ \ 
T_2=\pmatrix{0&f_{122}& f_{222}\cr
0&f_{112}&f_{122}\cr
1&0&0\cr}}
Associativity is granted for multiplication of linear operators, the 
actual constraint is the commutativity $[T_i,T_j]=0$, and reduces to
$[T_1,T_2]=0$, equivalent to \assop.
Moreover from \repsq\ we get 
\eqn\crossder{ 
{\partial T_i \over \partial y_j}={\partial T_j \over \partial y_i} \ .}
Conversely, let
\eqn\aprio{ T_0=\pmatrix{1&0&0\cr 0&1&0\cr 0&0&1\cr} \ \ 
T_1=\pmatrix{0&a_{11} &a_{12}\cr
1&a_{21}&a_{22}\cr
0&1&0\cr} \ \ 
T_2=\pmatrix{0&b_{11}& b_{12}\cr
0&b_{21}&b_{22}\cr
1&0&0\cr}}
where the matrix elements $a_{ij}$, $b_{ij}$ are a priori functions of $y_0$,
$y_1$, $y_2$. For $a_{ij}=b_{ij}=0$, we find the classical cohomology ring 
of $\IP_2$. Let us require that
\eqn\requ{ [T_i,T_j]=0 \ \ \ {\rm and} \ \ \ 
{\partial T_i \over \partial y_j}=
{\partial T_j \over \partial y_i} \ \ \ i\neq j \ .}
Hence 
$0=\partial T_0/\partial y_j -\partial T_j / 
\partial y_0=-\partial T_j/ \partial y_0$ so $a_{ij}$ and $b_{ij}$ 
are functions of $y_1$, $y_2$ only while
\eqn\crossab{ {\partial a_{ij} \over \partial y_2}=
{\partial b_{ij} \over \partial y_1} \ . }
Commutativity requires
\eqn\commcons{ 
\eqalign{ a_{12}=b_{11}=b_{22}  &\qquad a_{11}=a_{22}=b_{21} \cr
b_{12} &=(a_{11})^2 -a_{12}a_{21} \ ,\cr}
}
which combines with \crossab\ to yield 
\eqn\derconstr{
\eqalign{
{\partial a_{11} \over \partial y_2}&={\partial a_{22} \over \partial y_2}
={\partial a_{12} \over \partial y_1} \cr
{\partial a_{11} \over \partial y_1}&={\partial a_{21} \over \partial y_2}\cr
{\partial a_{12} \over \partial y_2}
&={\partial b_{12} \over \partial y_1}\ .\cr}
}
This implies the existence of a function $\Psi(y_1,y_2)$ such that
\eqn\solv{ 
a_{11}=a_{22}=b_{21}={\partial^2 \Psi \over \partial y_1 \partial y_2}
\ ; \ a_{21}={\partial^2 \Psi \over \partial y_1^2} \ ; \ 
a_{12}=b_{11}=b_{22}={\partial \Psi \over \partial y_2^2} }
while $\partial b_{12} / \partial y_2=\partial^3 \Psi/\partial y_2^3$,
which is solved if we write $\Psi=\partial f /\partial y_1$.
Finally
\eqn\expref{ a_{11}=a_{22}=b_{21}=f_{112} \ ; \ a_{21}=f_{111} \ ; \
a_{12}=b_{11}=b_{22}=f_{122} \ ; \ b_{12}=f_{222} \ . }
The last condition from commutativity amounts to the familiar relation \assop.
\bigskip

\noindent{\bf Proposition 4.} (Dubrovin)

(i) The necessary and sufficient condition that the covariant derivatives 
\eqn\covdef{ D_i \equiv {\partial \over \partial y_i}- z T_i \ ,}
with $T_i$ of the form \aprio, commute for arbitrary $z$ is that there exists a 
function $f(y_1,y_2)$ solution of \assop\ such that the $T$'s have the form
\repsq.

(ii) Set 
\eqn\deftone{ T_{-1}=\pmatrix{-1&0&0\cr 0&0&0\cr 0&0&1\cr} \ .}
With $f=\sum_{d \geq 1} N_d y_2^{3d-1} e^{d y_1}/(3d-1)!$
(in its region of convergence), the $D_i$ commute with 
\eqn\opdiff{D_z\equiv z {\partial \over \partial z} -\big[ T_{-1}+z(y_0 T_0+
3 T_1- y_2 T_2) \big] \ .}

The commutation relations $[D_i,D_j]=0$ for arbitrary $z$ amount to 
$\partial_i T_j=\partial_j T_i$ and $[T_i,T_j]=0$ proving (i).
As for (ii), it follows from the homogeneity of $f$, namely
$f(y_1+3 {\rm ln}\gl,\gl^{-1} y_2)=\gl f(y_1,y_2)$. We have$[D_0,D_z]=0$, 
since $\partial_0 T_1=\partial_0 T_2=0$. Moreover from $[T_1,T_2]=0$ and 
$\partial_2 T_1=\partial_1 T_2$ we get
\eqn\verif{
\eqalign{[D_1,D_z]&=-z\partial_1(3T_1-y_2T_2)+zT_1+z[T_1,T_{-1}]\cr
&=z\big[ -3 \partial_1 T_1+y_2 \partial_2 T_2+T_1+[T_1,T_{-1}] \big] \cr
&=z \pmatrix{0&(2-3\partial_1+y_2\partial_2)f_{112}
&(3-3\partial_1+y_2\partial_2)f_{122}\cr
0&(1-3\partial_1+y_2\partial_2)f_{111}&
(2-3\partial_1+y_2\partial_2)f_{112}\cr
0&0&0\cr} }
}
Since the degrees of homogeneity are $[f_{111}]=1$, $[f_{112}]=2$ and
$[f_{122}]=3$, all the terms vanish, for instance
\eqn\inst{\big( 1-3 \partial_1+y_2\partial_2)f_{111}
=\gl {\partial \over \partial \gl} \big[ \gl f_{111}(y_1,y_2)-
f_{111}(y_1+3 {\rm ln} \gl, \gl^{-1} y_2) \big]_{\gl =1}=0 .}
Similarly one checks that $[D_2,D_z]=0$.

The system 
\eqn\system{ D_i L=D_z L=0 \ ,}
where $L$ is a $3\times 3$ matrix, admits therefore a consistent
solution and we have a trivial bundle over a covering of that part of the dual
$H^*(\IP_2) \times (\IC-\{0\})$ (for the variable $z$), where $f$ is defined.
Thus, at least locally, $T_i$ is a pure gauge
$T_i=z^{-1} L^{-1} \partial_i L$.

On the plane $y_2=0$ the ring generated by $\{ T_i \}$ reduces to
a simpler one since the only non--vanishing $f_{ijk}$ is
$f_{122}=e^{y_1}\equiv q^3$.
Its multiplication laws read
\eqn\reduc{ T_1^2=T_2 \ ;\ T_1 T_2=q^3 T_0 \ ; \ T_2^2=q^3 T_1 }
i.e. it is the ring $\IC[x]/(x^3-q^3)$ with the identification $T_i \to x^i$.
In the vicinity of this plane the 
covariant derivatives $D_i$ generate ``monodromy preserving" flows of 
the solutions of $D_z L=0$. 
To identify this monodromy, it is therefore 
sufficient to look at this equation
at a specific point, say $y_0=y_2=0$, where it takes the form ($z \neq 0$)
\eqn\redform{ \left[ {\partial \over \partial z} -\pmatrix{-1/z&0 &3 q^3\cr
3&0&0\cr
0&3&1/z\cr} \right] L =0 \ .}
This equation is easily solved as follows.
For $z \neq 0$, set 
\eqn\multiset{
\eqalign{A(z)&=z \sum_{n=0}^{\infty} {(zq)^{3n} \over [n!]^3} \cr
B(z)&= -2z \sum_{n=1}^{\infty}{(zq)^{3n} \over [n!]^3}\sum_{p=1}^n {1 \over p}\cr
C(z)&=z\sum_{n=1}^{\infty}{(zq)^{3n} \over [n!]^3}\bigg[
\big(\sum_{p=1}^n {1 \over p}\big)^2+{1 \over 3} \sum_{p=1}^n {1 \over p^2} \bigg] \cr}
}
and define the matrix $L_0$ as
\eqn\deflo{
\pmatrix{{1 \over 9} \partial(\partial-1/z)A
&{4 \over 9z}(\partial-1/z)A+{1 \over 9} \partial(\partial-1/z)B
&{2A \over 9 z^2}+{2 \over 9z}(\partial-1/z)B+{1 \over 9}
\partial(\partial-1/z)C\cr
{1 \over 3} (\partial -1/z)A&{2A \over 3z}+{1 \over 3} (\partial -1/z)B&
{B \over 3z}+{1 \over 3} (\partial -1/z)C\cr
A&B&C\cr} 
}
where we use the shorthand notation $\partial$ for $\partial / \partial z$.
Finally let us introduce the matrix
\eqn\matX{ X\equiv\pmatrix{0&2&0\cr 0& 0& 1\cr 0&0&0\cr} \quad , 
\quad  X^3=0 \ .}
\bigskip
\noindent{\bf Lemma 2.}

(i) Up to multiplication to the right by a $z$--independent matrix, 
the solution $L$ of \redform\ is of the form
\eqn\formsol{ L= L_0 z^X \ .}

(ii) Therefore, up to conjugacy the generator of the monodromy of the solution 
$L$ around the origin is
\eqn\monoL{ L \ \to  \  L \ e^{2i \pi X} \ .}

The matrix $\exp(2i \pi X)$ is unipotent
\eqn\unipo{ e^{2i \pi X} = 1+2i\pi X -2 \pi^2 X^2=\pmatrix{1&4i\pi&-4\pi^2\cr
0&1&2i\pi\cr 0&0&1\cr} }
It is perhaps not unexpected that $X$ is a generator of the nilpotent ring
$\IC[x]/x^3$, i.e. of the classical cohomology ring.

The monodromy around the origin is independent 
of $q^3=e^{y_1}$. This is part of a more general property (Dubrovin).
Namely, returning to the case of general $(y_0,y_1,y_2)$ we can solve locally
{\it simultaneously} $D_z L=D_iL=0$, since all the differential operators commute
when they are defined. Thus to lowest order in 
$\ge \equiv (\ge_0,\ge_1,\ge_2)$
\eqn\loorder{ L(y+\ge,z)=L(y,z)+\sum_{0 \leq i \leq 2} z \ge_i T_i(y) L(y,z)+
O(\ge^2) }
implying that $L(y+\ge,z)$ has the same monodromy around the origin as $L(y,z)$.
Up to conjugation this is the same as on the line $y_0=y_2=0$, computed above.

In the completed plane the only other singularity of the general equation 
$D_z L=0$ is an essential singularity at infinity. 
Kontsevich and Manin suggest to perform  a formal Fourier transform
\eqn\four{ L(y,z)=\int dp e^{pz} {\tilde L}(y,p) \ .}
We should then solve for 
\eqn\transfou{ \left[ {\partial \over \partial p} - 
\big( p -(y_0 T_0+3 T_1-y_2 T_2) \big)^{-1}(1+T_{-1}) \right] {\tilde L}(y,p)=0 \ .}
This differential equation has now $4$ singular points, namely the roots of
\eqn\rootp{ \det\bigg[ p -(y_0T_0+3T_1-y_2T_2) \bigg]=0}
and the point at infinity.
Again we see that as $y$ varies the monodromy is preserved. 
In particular when $y_0=y_2=0$, \transfou\ reduces to
\eqn\foutrans{ \left[ {\partial \over \partial p} -
{1 \over p^3 -27 q^3} \pmatrix{ 0&9q^3 &6q^3 p\cr 0& p^2&18q^3\cr 0&3p&2p^2\cr}
\right] {\tilde L}(y,p)=0 \ , }
exhibiting the three simple poles at $p=3q$ and $p=3q e^{\pm 2 i \pi /3}$.

\bigskip
\noindent{\bf Remark.}
For generic values of $y$, the commutative algebra generated by the $T$'s over $\IC$
is semi--simple and assumes the form $\IC[x]/P(x)$, where $P$ is a 
monic cubic polynomial obtained as follows.
Set 
\eqn\potring{
\eqalign{ T_0&=1 \  ,\cr
T_1&=x \ ,  \cr T_2&=x^2-f_{111}x-f_{112} \ ,\cr}
}
{}From 
$T_1 T_2=f_{112}T_1 +f_{122} T_0$, we deduce that 
\eqn\polpot{P(x)\equiv x^3-f_{111} x^2-2 f_{112} x -f_{122}=0.}
When $y_0=y_2=0$, $P$ reduces to
$x^3 -e^{y_1}=x^3-q^3$. 
Such a property will extend to projective spaces in arbitrary dimension.
Let $P(x)$ be a monic polynomial with distinct roots of arbitrary degree $n$, then the 
commutative algebra $\IC[x]/P(x)$ admits a set of idempotent generators $\CT_1$, $\CT_2$,...
$\CT_n$, satisfying
\eqn\idempo{ \CT_{\alpha}\  \CT_{\beta}\ = \ \CT_{\beta}\ \CT_{\alpha} 
\ = \ \delta_{\alpha,\beta} \ \CT_{\alpha} }
with the unit being $1=\sum_{1 \leq \alpha \leq n} \CT_{\alpha}$. Indeed let $a_{\alpha}$ be the distinct 
roots of $P(x)=\prod_{1 \leq \alpha \leq n} (x-a_{\alpha})$, one has
\eqn\idem{ \CT_{\alpha}\equiv \prod_{\beta \neq \alpha} {x-a_{\beta} \over a_{\alpha}-a_{\beta}} 
\qquad {\rm mod} \ P(x) }
hence for any polynomial 
$R(x)=\sum_{1 \leq \alpha \leq n} R(a_{\alpha}) \CT_{\alpha}$ mod $P(x)$.
In particular, $x^k=\sum_{1 \leq \alpha \leq n} \CT_{\alpha} a_{\alpha}^k$. Presented in the form \idempo,
the algebra conveys no information except $n$ (the degree of the polynomial $P(x)$),
and its semi--simplicity. The crucial information is coded in the $T$'s as linear
combinations of the $\CT$'s or equivalently in the metric $\eta$. 
We refer to appendix A for a discussion of the matter.

\subsec{Higher genera}

A generalization of the preceding
enumerative theory in higher genus is lacking at present. However we can
collect some data. Set
\eqn\genf{ f^{(g)}= \sum_{d:(d-1)(d-2)\geq 2g} N_d^{(g)} {y_2^{3d-1+g} \over (3d-1+g)!}
e^{d y_1} }
where $f^{(0)}\equiv f$, $N_d^{(0)}=N_d$, and $N_d^{(g)}$ counts the number of 
degree $d$, genus $g$, stable irreducible curves through $(3d-1+g)$ points in  
$\IP_2$, i.e. with $\delta=(d-1)(d-2)/2-g$ ordinary double points.
The sum in \genf\ therefore starts from $d=1$ for $g=0$, and from integer 
$d \geq [3+\sqrt{1+8g}]/2$ for $g \geq 1$. 
{}From propositions 1 and 2, the available data are, up to degree $5$
\eqn\datag{
\eqalign{
{y_2^2 \over 2}e^{y_1}+ {y_2^5 \over 5!}e^{2 y_1}+ 12 {y_2^8 \over 8!} e^{3 y_1} 
+620 {y_2^{11} \over 11!}e^{4 y_1} + 87304 {y_2^{14} \over 14!} e^{5 y_1}+ ...&=f^{(0)}\cr
{y_2^9 \over 9!} e^{3 y_1}+225 {y_2^{12} \over 12!}e^{4 y_1}
+87192 {y_2^{15} \over 15!}e^{5 y_1} + ...&=f^{(1)}\cr
27 {y_2^{13} \over 13!} e^{4 y_1} + 36855{y_2^{16} \over 16!} 
e^{5 y_1}+ ...&=
f^{(2)}\cr
{y_2^{14} \over 14!} e^{4 y_1} + 7915 {y_2^{17}\over 17!}
e^{5 y_1}+ ...&=f^{(3)}\cr
882 {y_2^{18} \over 18!} e^{5 y_1}+ ...&=f^{(4)}\cr
48 {y_2^{19} \over 19!} e^{5 y_1}+...&=f^{(5)}\cr
{y_2^{20} \over 20!} e^{5 y_1}+...&=f^{(6)} \ .\cr}
}
The challenge is to find the topological recursion relation, or the non--linear
extension of the Dubrovin flows, or else a manageable path integral which 
would generate these higher genus contributions (on the so--called ``little phase space").
Note that all the enumerative coefficients $N_d^{(g)}$ are presumed to be 
non--negative integers. This is to be contrasted with
general intersection numbers on the orbifold compactification
of the moduli space of punctured Riemann surfaces of fixed genus, 
which tend to be rational.

\subsec{Characteristic numbers}

Let us return to rational curves of degree $d$.
When discussing the deformed ring we only dealt with the number 
$N_d$ of such curves through $3d-1$ points.
Classically one also considered mixed conditions involving points
and lines. Let $N_{\ga,\gb}$, $\ga+\gb=3d-1$, denote the number
of rational plane curves through $\ga$ fixed points and tangent to $\gb$
fixed lines, so that $N_{3d-1,0}=N_d$. The generating function
\eqn\genezf{ \phi(y_1,y_2,z)=\sum_{d \geq 1 \ \ga,\gb \geq 0 \atop
\ga+\gb=3d-1} N_{\ga,\gb} {y_2^\ga \over \ga !} {z^\gb \over \gb !}
e^{d y_1} }
reduces to $f(y_1,y_2)$ defined in \energ\ for $z=0$.
The integers $N_{\ga,\gb}$ are refered to as {\it characteristic numbers}
(an analogous definition holds for higher genera).

It is possible to obtain for 
the derivatives of $\phi$ a quadratic relation which generalizes equation 
\assop\ by extending the argument of section 2.3 to mixed conditions.
We have only to modify the ``spectator" set of conditions
(which included $3d-4$ fixed points $q_*$), to involve $\ga -3$ points
and $\gb$ lines of tangency.
The reasoning is completely analogous provided we take into account
degenerating curves intersecting on a line of tangency or on one of their 
intersections and their multiplicities.
If this is done carefully, the
result
reads
\eqn\tangpt{\encadremath{
\phi_{222}=(\phi_{112}^2 -\phi_{111}\phi_{122})+2z(\phi_{112}\phi_{122}-\phi_{111}\phi_{222})+
2z^2(\phi_{112}^2 -\phi_{112}\phi_{222}) } }
Readily available initial data are
\eqn\inidatab{\eqalign{ 
d=1 \ &: \ \quad N_{2,0}=1 \quad N_{1,1}=N_{0,2}=0 \cr
d=2 \ &: \ \quad N_{\ga,\gb}=N_{\gb,\ga} \cr}}
where the second line expresses duality for the conics.
\bigskip

\noindent{\bf Proposition 5.}
Equation \tangpt\ determines all characteristic numbers $N_{\ga,\gb}$ for 
$\gb \leq 3d_0-1$, in terms of the $N_{\ga,\gb}$ in degree $d \leq d_0$.
Beyond, i.e. for $\gb \geq 3d_0$, one only gets quadratic relations.
In particular the initial conditions \inidatab\ determine 
all the $N_{\ga,\gb}$ for $\gb \leq 5$.
\bigskip

For low degree we present below
the characteristic numbers for $\gb \leq 5$ using only eq.\inidatab. 
In degree $d \leq 4$ they agree with known results \SCHUB\ \RP.
Complementing these data with the missing values in degrees 
$3$ and $4$ from the above references, we obtain 
$N_{14-\gb,\gb}$ in degree $5$ up to $\gb=11$. In the process we
also check the relations arising from powers of $z^6$ to $z^8$ in $\phi$ in 
degree $4$.
\eqn\respropx{\eqalign{
\hbox{\bf d=1 : }\ N_{2,0}&=1 \quad
N_{1,1}=0  \quad
N_{0,2}=0  \cr
\hbox{\bf d=2 : }\ N_{5,0}&=1 \quad
N_{4,1}=2 \quad
N_{3,2}=4 \cr
N_{2,3}&=4  \quad
N_{1,4}=2  \quad
N_{0,5}=1 \cr
\hbox{\bf d=3 : }\ N_{8,0}&=12 \quad
N_{7,1}=36 \quad
N_{6,2}=100  \cr
N_{5,3}&=240 \quad
N_{4,4}=480 \quad
N_{3,5}=712  \cr
N_{2,6}&=756 \quad
N_{1,7}=600 \quad
N_{0,8}=400  \cr 
\hbox{\bf d=4 : }\ N_{11,0}&=620 \quad
N_{10,1}=2184 \quad
N_{9,2}=7200  \cr
N_{8,3}&=21776 \quad
N_{7,4}=59424 \quad
N_{6,5}=143040  \cr
N_{5,6}&=295544 \quad
N_{4,7}=505320 \quad
N_{3,8}=699216  \cr
N_{2,9}&=783584 \quad
N_{1,10}=728160 \quad
N_{0,11}=581904  \cr
\hbox{\bf d=5 : }\ N_{14,0}&=87304 \quad
N_{13,1}=335792 \quad
N_{12,2}=1222192  \cr
N_{11,3}&=4173280 \quad
N_{10,4}=13258208 \quad
N_{9,5}=38816224  \cr
N_{8,6}&=103544272 \quad
N_{7,7}=248204432 \quad
N_{6,8}=526105120  \cr
N_{5,9}&=969325888 \quad
N_{4,10}=1532471744 \quad
N_{3,11}=2069215552  \cr}}
\medskip

Equation \tangpt\ can also be interpreted as the 
associativity condition for a deformed
ring. With $z$ as parameter, the genus $0$ free energy is now
\eqn\nouvphi{\eqalign{
\Phi(y_0,y_1,y_2;z)&=\phi_{\rm cl}(y_0,y_1,y_2;z)+\phi(y_0,y_1,y_2;z)\cr
\phi_{\rm cl}(y_0,y_1,y_2;z)&={y_0^2 y_2+y_0 y_1^2 \over 2}
-2z {y_0^2 y_1 \over 2} +4 z^2 {y_0^3 \over 6} \cr}}
It corresponds to a modified ``metric" $\eta_{ij}$ written in matrix form
$$\eqalign{\pmatrix{ 2z^2 & -2z & 1 \cr
-2 z& 1 & 0 \cr
1 & 0 & 0 \cr} &= \big( |2\rangle -z |1\rangle+{z^2 \over 2}|0\rangle \big)
\langle 0 | \cr
&+\big(|1\rangle -z |0 \rangle \big) \big(\langle 1 | -z
\langle 0 | \big) + |0 \rangle \big(\langle 2 | -z \langle 1 | +{z^2 \over
2} \langle 0 | \big) \cr}$$
and its inverse $\eta^{ij}$ which appears as a propagator (cf. 
Fig.\crossing\ and eq.\croptwo\ )
$$\eqalign{\pmatrix{ 0 & 0 & 1 \cr
0& 1 & 2z \cr
1 & 2z & 2z^2 \cr} &= \big( |0\rangle +z |1\rangle+{z^2 \over 2}|2\rangle \big)
\langle 2 | \cr
&+\big(|1\rangle +z |2 \rangle \big) \big(\langle 1 | +z
\langle 2 | \big) + |2 \rangle \big(\langle 0 | +z \langle 1 | +{z^2 \over
2} \langle 2 | \big) \cr}$$
On  the rhs of these expressions we used Dirac's bra--ket notation to emphasize
the structure, which suggests a natural generalization in higher projective spaces.

For large $\gb$ characteristic numbers can be difficult to obtain. At the very least, equation \tangpt\ gives useful constraints, 
in the form of recursion relations.

\newsec{Rational manifolds}

We consider in this section some examples of rational target manifolds and the
corresponding enumeration of rational curves. Details will be omitted except
for a few elementary comments.

\subsec{Projective spaces}

We first generalize from the plane $\IP_2$ to $\IP_n$, $n \geq 2$, 
in which case the 
classical cohomology ring may be identified with $\IC[x]/x^{n+1}$. Correspondingly,
we introduce deformation parameters $y_i$, $0 \leq i \leq n$, with weights
$[y_i]=1-i$, $i\neq 1$, and $[e^{y_1}]=n+1$, also $[F^{(g)}]=(1-g)(3-n)$, which gives
for $g=0$ ($F^{(0)}\equiv F$) $[F]=3-n$. These assignments
are consistent with the following facts 

(i) The intersection form
\eqn\intern{ \eta_{ij}=\delta_{i+j,n} = 
{\partial^3 F \over \partial y_0 \partial y_i \partial y_j} }
is dimensionless, hence $[F]=[y_0]+[y_i]+[y_{n-i}]$.

(ii) A generic rational curve of degree $d$ in $\IP_n$ depends on $(d+1)(n+1)-4$
parameters and intersection with a linear subspace of codimension $k$ imposes $k-1$
linear conditions on these parameters. Since $y_k$ is dual to cycles of codimension $k$
one expects with arguments similar to those of section 2 that the free energy 
$F$ decomposes into a classical and a quantum part $F(y)=f_{\rm cl}(y)+f(y)$, where 
as before $f_{\rm cl}$ is a cubic polynomial encoding the multiplication rules of the 
classical cohomology ring
\eqn\clan{ f_{\rm cl}(y)={ 1 \over 3!} \sum_{i+j+k=n} y_i y_j y_k }
and the quantum corrections take the form
\eqn\guessf{ f(y)= \sum N(a_2,a_3,...,a_n|d) 
{y_2^{a_2} \over a_2!}{y_3^{a_3 } \over a_3!}...{y_n^{a_n} \over a_n!} e^{d y_1} \ ,}
where the sum is running over non negative $a_i$'s such that 
\eqn\selectrule{ (n+1)d+\sum_{i=2}^n (1-i) a_i = (3-n) \ ,}
expressing the global homogeneity of $f$.
Finally $N(a_2,a_3,...,a_n|d)$ is interpreted as the number of rational curves
of degree $d$ intersecting $a_n$ points, $a_{n-1}$ lines, ... $a_{n-k}$ 
linear spaces of dimension $k$, ... in ``general position".
In particular when $d=1$, we have $\sum_{2 \leq i \leq n}(i-1)a_i=2(n-1)$ and
$N(0,0,...,0,2|1)=1$ is the number of lines through $2$ points.

The study of the quantum cohomology ring for general $n$ 
deserves a specific combinatorial treatment since the equations expressing 
associativity become quite cumbersome. 
We limit ourselves here to $\IP_3$ in which case
\eqn\inthree{ 
\eqalign{ f_{\rm cl}&={1 \over 2} y_0^2 y_3 + y_0 y_1 y_2+{1 \over 6} y_1^3 \cr
f&= \sum_{a+2b=4d \atop a,b \geq 0 d \geq 1} N_{a,b} {y_2^a \over a!} {y_3^b \over b!}
e^{d y_1} \cr}
}
where for short $N_{a,b}\equiv N(a,b|d)$.
As before, we introduce the deformed ring with basis 
$T_0$, $T_1$, $T_2$, $T_3$
\eqn\intre{ T_i T_j =\sum_{0 \leq k,l \leq n} F_{ijk} \eta^{kl} T_l }
and $T_0$ is the identity ($f$ does not depend on $y_0$).

\fig{The six duality relations for $\IP_3$.}{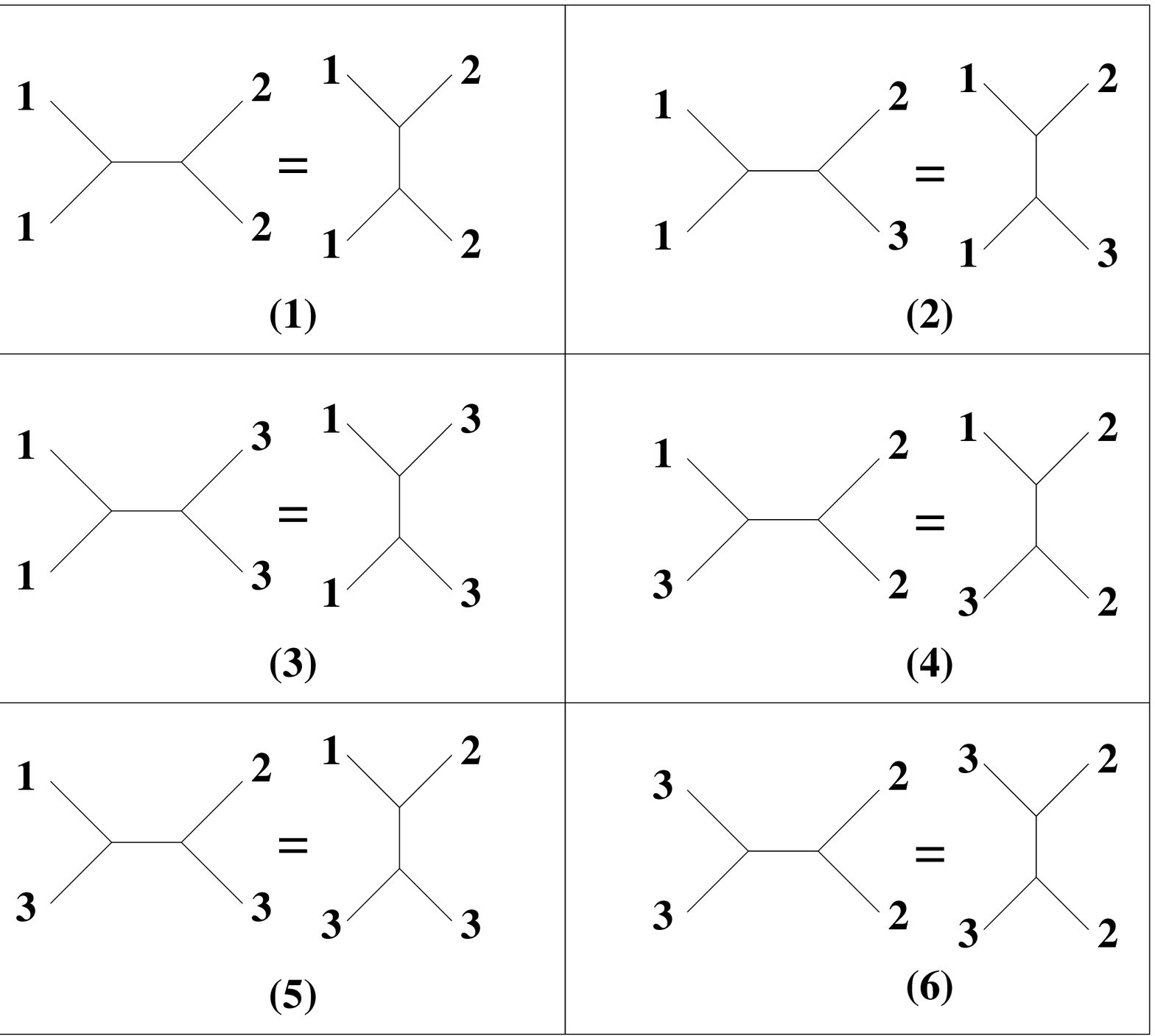}{7cm}
\figlabel\crot
 
The six associativity conditions corresponding to the ``duality" diagrams of 
Fig.\crot\ read
\eqn\crost{
\eqalign{ (1) \ \ \ & 2 f_{123} -f_{222}=f_{111}f_{222}-f_{112} f_{122} \cr
(2) \ \ \ & f_{133} -f_{223}= f_{111} f_{223}-f_{113} f_{122}\cr
(3) \ \ \ & -f_{233}=f_{111}f_{233}+f_{112}f_{133}-2f_{113}f_{123}\cr
(4) \ \ \ & f_{233}=f_{113}f_{222}-f_{112}f_{223}\cr
(5) \ \ \ &f_{333}=f_{123}^2+f_{113}f_{223}-f_{112}f_{233}-f_{122}f_{133}\cr
(6) \ \ \ & 0=f_{122}f_{233}+f_{133}f_{222}-2f_{123}f_{223} \ .\cr }
}
With $f$ of weight $0$, the equations are respectively of weight $3$, 
$4$, $5$, $6$, $7$. 
They yield (with the convention that $a+2b=4d$, $a_i+2b_i=4d_i$
for $i=1$ and $2$, $d_1$, $d_2\geq 0$, $d_1+d_2=d \geq 1$)

\bigskip
\noindent{\bf Proposition 6.} The integers $N_{a,b}$ are determined by 
$N_{0,2}=1$ and
\eqn\recurthree{
\eqalign{
(1) \ \ \ & 2d N_{a-2,b+1} -N_{a,b}= \sum N_{a_1,b_1} N_{a_2,b_2}{b \choose b_1}
\left[ d_1^3 {a-3 \choose a_1} -d_1^2 d_2 {a-3 \choose a_1-1} \right] \cr
(2) \ \ \ & d N_{a-2,b+1} -N_{a,b}= \sum N_{a_1,b_1} N_{a_2,b_2}{a-2 \choose a_1}
\left[ d_1^3 {b-1 \choose b_1} -d_1^2 d_2 {b-1 \choose b_1-1} \right] \cr
(3) \ \ \ & -N_{a,b}=\sum N_{a_1,b_1} N_{a_2,b_2}  \times \cr
&\ \ \ \ \times \left[ d_1^3{a-1 \choose a_1}
{b-2 \choose b_1}+d_1^2 d_2{a-1 \choose a_1-1}{b-2 \choose b_1}-2d_1^2d_2
{a-1 \choose a_1}{b-2 \choose b_1-1} \right] \cr
(4) \ \ \ & N_{a-2,b+1}=\sum N_{a_1,b_1} N_{a_2,b_2} d_1^2\left[
{a-3 \choose a_1}{b-1 \choose b_1-1}-{a-3 \choose a_1-1}{b-1 \choose b_1} \right] \cr
(5) \ \ \ & N_{a-2,b+1}=\sum N_{a_1,b_1} N_{a_2,b_2}\left[d_1d_2{a-2 \choose a_1-1}
{b-2 \choose b_1-1} + \right. \cr
&\ \ \ \ \ \ \left. +d_1^2{a-2 \choose a_1}{b-2 \choose b_1-1}-d_1^2
{a-2 \choose a_1-1}{b-2 \choose b_1}-d_1d_2{a-2 \choose a_1-2}{b-2 \choose b_1}
\right] \cr
(6) \ \ \ & 0= \sum N_{a_1,b_1} N_{a_2,b_2}  \times \cr
&\times d_1 \left[{a-3 \choose a_1-1}{b-2 \choose b_1}
+{a-3 \choose a_1}{b-2 \choose b_1-2}-2{a-3 \choose a_1-1}{b-2 \choose b_1-1} \right] 
\cr}
}

\noindent{}By convention,
the combinatorial factors ${n \choose n_1}$ 
are non--vanishing only for $n$, $n_1$ and $n-n_1 \geq 0$. 
The single input $N_{0,2}=1$ for the number of lines through 
$2$ points suffices to 
determine all the $N_{a,b}$. The six conditions appear consistent
although the system looks strongly overdetermined. One finds
$$
\eqalign{
\hbox{\bf d=1 : }\ 
N_{4,0}&=2 \quad
N_{2,1}=1  \quad
N_{0,2}=1  \cr
\hbox{\bf d=2 : }\ 
N_{8,0}&=92 \quad
N_{6,1}=18 \quad
N_{4,2}=4 \cr
N_{2,3}&=1  \quad
N_{0,4}=0  \cr
\hbox{\bf d=3 : }
N_{12,0}&=80160  \quad
N_{10,1}=9864 \quad
N_{8,2}=1312  \cr
N_{6,3}&=190  \quad
N_{4,4}=30  \quad
N_{2,5}=5   \cr
N_{0,6}&=1  \cr
\hbox{\bf d=4 : }
N_{16, 0}&=383306880 \quad
N_{14, 1}=34382544 \quad
N_{12, 2}=3259680 \cr
N_{10, 3}&=327888 \quad
N_{8, 4}=35104 \quad
N_{6, 5}=4000 \cr
N_{4, 6}&=480 \quad
N_{2, 7}=58 \quad
N_{0, 8}=4 \cr}$$

\eqn\respt{
\eqalign{
\hbox{\bf d=5 : }
N_{20,0}&=6089786376960 \quad
N_{18, 1}=429750191232 \quad
N_{16, 2}=31658432256 \cr
N_{14, 3}&=2440235712 \quad
N_{12, 4}=197240400 \quad
N_{10, 5}=16744080 \cr
N_{8, 6}&=1492616 \quad
N_{6, 7}=139098 \quad
N_{4, 8}=13354 \cr
N_{2, 9}&=1265 \quad
N_{0,10}=105 \cr}
}

\noindent{}In degree $1$, there is of course $N_{2,1}=1$ 
line through a
point which meets two lines, and 
$N_{4,0}=2$ lines meeting $4$ lines in general position. 
This classical result may be
obtained as follows. The set of lines intersecting three given lines
$l_1$, $l_2$, $l_3$ is a linear pencil (for instance for each point $p$ of $l_1$
there corresponds the unique line of the pencil defined by the intersection of
the planes containing $p$ and respectively $l_2$ and $l_3$) hence span a 
surface isomorphic to $\IP_1 \times \IP_1$, i.e. a quadric, 
cut by a fourth line 
$l_4$ in two points, corresponding to two lines of the pencil meeting the 
four lines $l_i$, $i=1,...,4$.
For conics there is a single
conic ($N_{2,3}=1$) through $3$ points which meets two lines. 
The three points define a plane, which is intersected by the two lines in two further points, yielding altogether the necessary $5$ points in this plane.
But to find for instance that there are $N_{8,0}=92$ 
conics meeting $8$ lines in 
general position is already non trivial.
With (twisted) cubics, we have rational normal curves in $\IP_3$ 
(i.e. not lying in a $\IP_k$, $k\leq 2$). 
In general the standard rational curve in 
$\IP_n$ of degree $n$ parametrized by $(1,t,t^2,..,t^n)$ in an appropriate 
coordinate system is uniquely stabilized by $(n+3)$ points 
(see \HAR\ for a general proof), so for $\IP_3$, $d=3$
we have $N_{0,6}=1$ and for $\IP_n$, $N(0,0,...,0,n+3|n)=1$.
These twisted cubics can be interpreted as the residual intersection of a 
linear pencil of quadrics having a line in common. 
A quadric is fixed by $9$ points, but if $3$ of them are aligned 
we only get a linear pencil of quadrics intersecting in a line and a twisted cubic
through the $6$ remaining points.  
Thus, given $6$ points in general position, if
we add $3$ other aligned points, we get the linear pencil determining uniquely 
the pencil of quadrics intersecting on the line and residually on a twisted cubic
through the $6$ points. 
The remaining enumerative numbers of twisted cubics are not so easily described.

\subsec{Quadric surfaces}

For the cases of quadric and cubic surfaces we shall be sketchy as the matter has been already reported in \CITZ. 
A smooth quadric in $\IP_3$ is isomorphic with $\IP_1 \times \IP_1$, $H^*$ is therefore 
$4$--dimensional, with Betti numbers $b^0=b^4=1$, $b^2=2$. We denote the generators
as $t_0$, $t_A$, $t_B$ and $t_2=t_At_B$. The non vanishing intersections are therefore
$\eta_{02}=\eta_{AB}=1$.
Let $y_0$, $y_A$, $y_B$, $y_2$ be the deformation parameters and $F$ the
genus zero free energy (modulo a second degree polynomial), split into
$F=f_{\rm cl}+f$ with
\eqn\spliplu{
\eqalign{ f_{\rm cl}&={1 \over 2} y_0^2 y_2 +y_0 y_A y_B \cr
f&=\sum_{a,b \geq 0 \atop a+b \geq 1} N(2(a+b)-1|a,b) 
{y_2^{2(a+b)-1} \over (2(a+b)-1)!} e^{a y_A+b y_B} \ .\cr}
}
This corresponds to the weights
\eqn\weiplu{ [y_0]=1 \ \ ; \ \ [e^{y_A}]=[e^{y_B}]=2 \ \ ; \ \ 
[y_2]=-1 \ \ ; \ \ [F]=1 }
while we will use the short hand notation $N_{a,b}=N_{b,a}\equiv N(2(a+b)-1|a,b)$
for the number of rational curves on the quadric, with bidegree $(a,b)$, 
through $2(a+b)-1$ points.

A rational curve of bidegree $(a,b)$ 
on $\IP_1 \times \IP_1$ is described by two homogeneous polynomials
of degree $a$ in a pair of parametrizing variables for the first $\IP_1$
and similarly with degree $b$ for the second $\IP_1$, 
modulo $PSL_2$ on the parameters
times $\IC^* \times \IC^*$ for homogeneity in each target $\IP_1$. 
We are left with 
$2(a+1)+2(b+1)-3-1-1=2(a+b)-1$ parameters explaining 
that a rational curve of bidegree $(a,b)$ 
is fixed by $2(a+b)-1$ points. 
Its degree as a curve in $\IP_3$ is $d=a+b$. Moreover a general curve
on $\IP_1 \times \IP_1$ with $\delta$ simple nodes has genus 
\eqn\genab{g= (a-1)(b-1) - \delta \ ,}
hence the rational curves we are counting have $\delta=(a-1)(b-1)$
simple nodes.

\fig{Duality relations for the quadric 
$\IP_1 \times \IP_1$.}{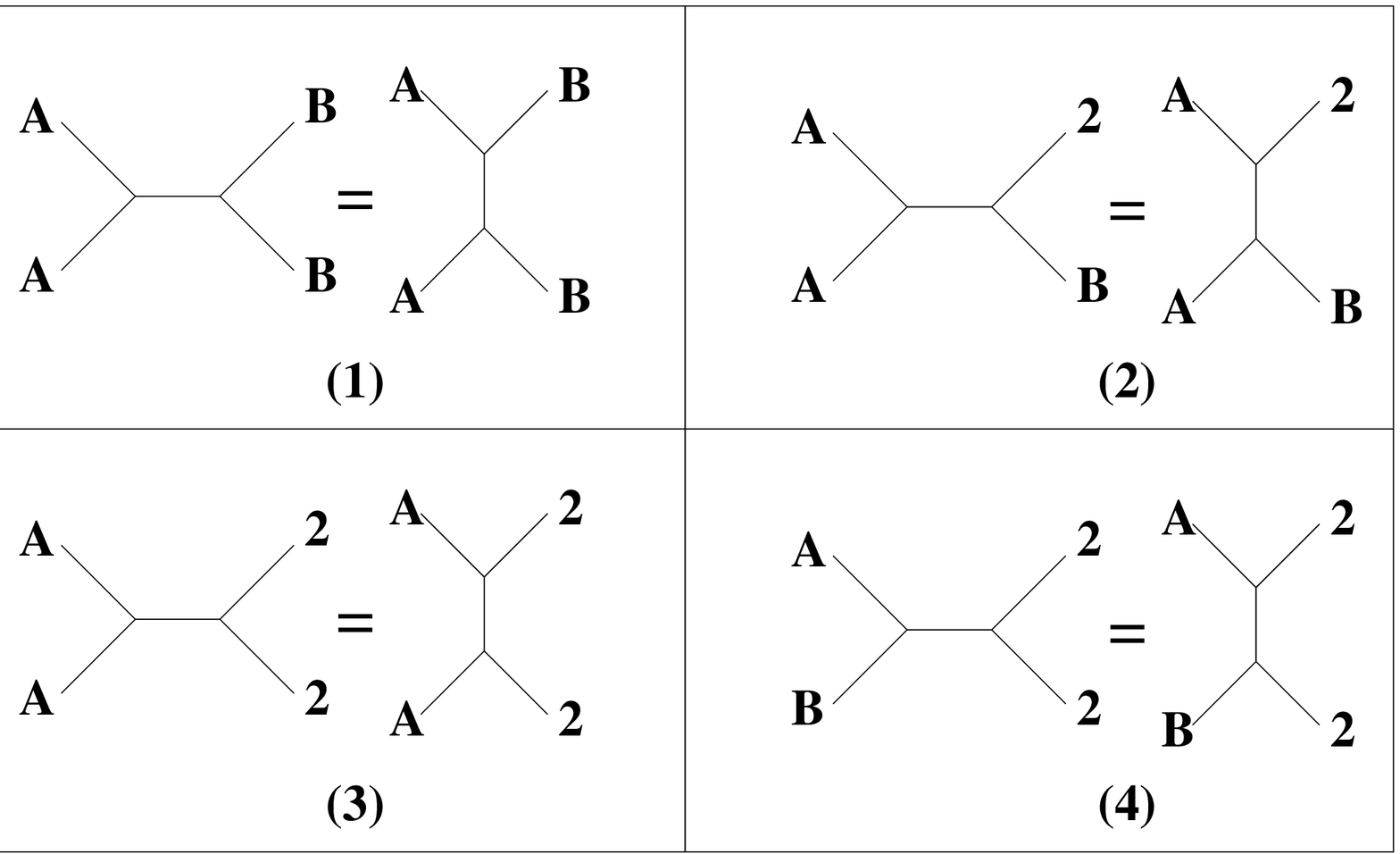}{6cm}
\figlabel\crossoo

Let $T_0$, $T_A$, $T_B$ and $T_2$ be a basis of the deformed ring with
$T_i T_j=\sum_{j,k}F_{ijk} \eta^{kl} T_l$. 
Associativity yields the $4$ conditions
depicted on Fig.\crossoo\ 
and their symmetric $A \leftrightarrow B$ counterparts 
(when the relation is not symmetric itself), leading to the $4$ recursion relations of 

\bigskip
\noindent{\bf Proposition 7.}
The integers $N_{a,b}=N_{b,a}$ are determined by the initial condition $N_{1,0}=1$
and the system
\eqn\deterplu{
\eqalign{
(1)&\ \ \  2ab N_{a,b}=\sum N_{a_1,b_1} N_{a_2,b_2} a_1^2 b_2^2 (a_1 b_2-a_2b_1)
{2(a+b)-2 \choose 2(a_1+b_1)-1} \cr
(2)&\ \ \  aN_{a,b}=\sum N_{a_1,b_1} N_{a_2,b_2} a_1(a_1^2b_2^2-a_2^2b_1^2)
{2(a+b)-3 \choose 2(a_1+b_1)-1}    \cr
(3)&\ \ \  0=\sum N_{a_1,b_1} N_{a_2,b_2} a_1^2\big[(a_2+b_2-1)(b_1a_2+b_2a_1)-
a_2b_2(2(a_1+b_1)-1)\big] {2(a+b)-3 \choose 2(a_1+b_1)-1}    \cr
(4)&\ \ \  N_{a,b}=\sum N_{a_1,b_1} N_{a_2,b_2} (a_1 b_2+a_2 b_1)b_2
\left[ a_1{2(a+b)-4 \choose 2(a_1+b_1)-2}-a_2{2(a+b)-4 \choose 2(a_1+b_1)-3} 
\right] \cr}
}
where $a_1$, $a_2\geq 0$, $b_1$, $b_2 \geq 0$, $a_1+a_2=a$, $b_1+b_2=b$, 
and the 
convention for combinatorial numbers is as before.

Up to degree $d=a+b=10$, we find
\eqn\numplu{
\eqalign{
\hbox{\bf d=1 : } 
N_{1,0}&=1 \cr
\hbox{\bf d=2 : } 
N_{1,1}&=1 \cr
\hbox{\bf d=3 : }
N_{2,1}&=1 \cr
\hbox{\bf d=4 : }
N_{3,1}&=1 \quad
N_{2,2}=12 \cr
\hbox{\bf d=5 : }
N_{4,1}&=1 \quad
N_{3,2}=96\cr
\hbox{\bf d=6 : }
N_{5,1}&=1 \quad
N_{4,2}=640 \quad
N_{3,3}=3510 \cr
\hbox{\bf d=7 : }
N_{6,1}&=1 \quad
N_{5,2}=3840 \quad
N_{4,3}=87544 \cr
\hbox{\bf d=8 : }
N_{7,1}&=1 \quad
N_{6,2}=21504 \quad
N_{5,3}=1763415 \quad
N_{4,4}=6508640\cr
\hbox{\bf d=9 : }
N_{8,1}&=1 \quad
N_{7,2}=114688 \quad
N_{6,3}=30940512 \quad
N_{5,4}=348005120 \cr
\hbox{\bf d=10 : }
N_{9,1}&=1 \quad
N_{8,2}=589824 \quad
N_{7,3}= 492675292\quad
N_{6,4}=15090252800\cr
N_{5,5}&=43628131782\cr}
}
It is readily seen that $N_{a,1}=N_{1,b}=1$ in particular 
$N_{2,1}=N_{1,2}=1$ twisted cubic through $5$ points on a quadric.
The first non trivial result is $N_{2,2}=12$ 
for the number of rational quartics
through $7$ points on a quadric. 
This number follows from the Zeuthen--Segre formula,
giving the number $n$ of singular elements in a linear 
pencil of curves of genus 
$g$ on a surface of Euler characteristic $\chi$ through $s$ base points: 
$n=\chi - 4(1-g)+s$.
Here we take the linear pencil of elliptic curves (smooth quartics, intersection
of two quadrics) through $7$ points on the quadric of characteristic $4$,
having thus an eighth base point, so $s=8$. Since the singular elements
are rational quartics, we have $N_{2,2}=4+0+8=12$.

\noindent{\bf Remark}
The recursion relations $(1)$--$(4)$ yield 
closed expressions for the first few $N_{a,b}$, with $b=0,1,2,3,4$, 
as functions of $a$
\eqn\nabclo{
\eqalign{N_{a,0}&= \delta_{a,1} \cr
N_{a,1}&=1\cr
N_{a,2}&=2^{2a} {a(a+1) \over 8}\cr
N_{a,3}&= 3^{2a}{512 a^4+1280 a^3+448 a^2- 368 a -75
\over 2^{14}}+ {32 a^2+104 a+75 \over 2^{14}}\cr
N_{a,4}&= 4^{2a}{a(a+1)(8 a^4+28 a^3 + 6 a^2 -31 a + 1) \over 768}+
2^{2a} {a(a+1)(a+2)(a+3) \over 192} \ .\cr}
}
In general, $N_{a,b}$ has the following structure as a
function of $a$ for fixed $b$
\eqn\strunab{ 
N_{a,b}= \sum_{0 \leq 2j <b} (b-2j)^{2a} P_{2b-2j-2}^{(b)}(a) \ ,}
where $P_n^{(b)}$ are degree $n$ polynomials with rational coefficients
depending on $b$ only.
It is a non trivial check that this leads to a symmetric $N_{a,b}=N_{b,a}$.
For instance $N_{4,3}=N_{3,4}=87544$ are both obtained from \nabclo.
The apparent simplicity of the structure of $N_{a,b}$ for 
$\IP_1 \times \IP_1$ may be related to the
fact that unlike the $\IP_n$, $n>1$ cases, 
$\IP_1$ has a very simple free energy
\eqn\frepone{ F= {y_0^2 \over 2} y_1 + e^{y_1} \  ,}
with the weights $[y_0]=1$ and $[e^{y_1}]=2$. It would be desirable to 
understand better the product $\IP_1 \times \IP_1$ in terms of 
free energies.

\subsec{Cubic surfaces.}

A classical description of cubic surfaces is by blowing up to lines six points of 
$\IP_2$ not lying on a conic, with any triplet not lying on a line, using the
$\IP_3$ of plane cubics through the $6$ points. These surfaces posses a system
of $27$ lines with a configuration (and intersection form) invariant under a finite group 
$G$ isomorphic to the Weyl group of $E_6$ of order $72\times 6!$  ($G$ is transitive
on the $72$ sextets of non intersecting lines). For an in--depth
study we refer to the book by Manin \YM.
In the plane, any two cubics through $6$ points intersect in $3 \times 3=6+3$ 
points. Hence a line in $\IP_3$ corresponding to a linear pencil of
cubics in the plane intersects the surface in $3$ variable points,
proving that the surface is indeed a cubic.

In the sequel we identify the second cohomology group of the surface $H^2=H^{1,1}$
and the divisor class group, isomorphic with $\IZ^7 \otimes \IC$. 
The latter is generated by the six blown lines ($l_i$, $i=1,2,...,6$) 
and the image of a line in the plane $l_0$.
A plane section of the surface (generically a genus $1$ cubic) is then equivalent to 
the divisor $3l_0-\sum_{1 \leq i \leq 6} l_i$. Generically we write a divisor (class)
\eqn\divi{ a=a^0 l_0 -a^i l_i}
with a degree 
\eqn\divdeg{ d_a=3a^0-\sum_{1 \leq i \leq 6} a^i}
and self--intersection
\eqn\selfint{ a.a\equiv (a^0)^2-\sum_{1 \leq i \leq 6} (a^i)^2 \ .}
We use physicists' notation with $\{ a^{\mu} \} = \{ a^0,a^i \}$
and a Lorentzian scalar product denoted with a dot
\eqn\cherlolo{ a.b=a^0 b^0 -\sum_{1 \leq i \leq 6} a^i b^i \ .}
Greek (resp. latin) indices run from $0$ to $6$ (resp. $1$ to $6$) and $\omega$ is
the anticanonical divisor
\eqn\antica{ \omega\equiv \{ 3,1,1,1,1,1,1\} \ \ \ , \ \ \ \omega.\omega=3 \ .}
With the interpretation in terms of curves of $\IP_2$, when $a^0>a^i \geq 0$, 
$a^0$ is the degree of the corresponding plane curve,
and $a^i$ the multiplicity 
at the $i$--th assigned point (we assume generically multiple points with distinct
tangents at the six assigned points). The apparent 
(or arithmetic) genus of a curve in the divisor class $a$, denoted $p_a$, 
is then
\eqn\papa{
\eqalign{p_a&={1 \over 2} \big[ (a^0-1)(a^0-2) - \sum_{1 \leq i \leq 6} a^i(a^i-1)
\big] \cr
&=1+{1 \over 2} (a.a-\omega.a) \ .\cr}
}
The topological genus $g$ of a representative in the class can be smaller than $p_a$
since the above formula does not take into account other singularities outside the 
assigned points. As before we assume generically that these are 
simple nodes. 
In this description, the only other effective divisor classes are given by the
six lines $l_i$ on the cubic surface.

For the quantum cohomology ring we change slightly our notations to avoid confusion. 
We use $T_x$ for the identity (instead of $T_0$) and $T_z$ instead of $T_2$, 
and keep the seven generators $T_{\mu}$.
The intersection form has non vanishing entries $\eta_{xz}=1$ and 
$\eta_{\mu \nu}=g_{\mu \nu}$ the $7$--dimensional diagonal Lorentzian metric
with $g_{00}=-g_{ii}=1$, numerically equal to its inverse. 

We are interested in counting rational irreducible curves belonging to an effective
class $a$ of degree $d=\omega.a$, satisfying extra requirements.
In the above presentation this implies that either
$a=\{0,-1,0,0,0,0,0 \}$ up to ``space" (indices $i=1,...,6$) permutations
or else
\eqn\condit{
\eqalign{
(i)&\ \ \ 0 \leq a^i < a^0 \ \ \ , \ \ \ 1 \leq i \leq 6\cr
(ii)&\ \ \ a^i+a^j \leq a^0 \ \ \ , \ \ \ 1\leq i<j \leq 6 \ ,  \ d>1 \cr
(iii)&\ \ \ \sum_{1 \leq i \leq 6} a^i \leq 2 a^0 + a^k \ \ \ ,\ \ \ 
1 \leq k \leq 6 \ ,  \ d>1 \ .\cr}
}
These conditions, necessary but not sufficient for irreducibility, mean
respectively that (i)
in the plane an irreducible curve of degree $a^0$ cannot have a point of multiplicity
as large as its degree, (ii) that no line cuts it in more that $a^0$ points and
(iii) that for $a^0>2$, a conic through $5$ points cannot cut the curve
in more that $a^0$ points. Higher conditions are subsummed under the 
inequality $p_a \geq 0$.
Let $\Delta_p\equiv \sum \Delta_{p,d}$ be the set of effective divisors
corresponding to the rational curves of arithmetic genus $p$,
split according to the degree and
$\Delta \equiv \sum_{p \geq 0} \Delta_p$. Rational curves in $\Delta_{p,d}$ have
$p$ extra double points and their divisors form invariant sets under 
the action of $G$. In particular $\Delta_0$ corresponds to smooth rational
curves on the cubic.
Plane curves of degree $a^0 \geq 1$ with assigned multiplicities $a^i$ at the base
points form a projective space of dimension $a^0(a^0+3)/2-
\sum_{1 \leq i \leq 6} a^i(a^i+1)/2=p_a+d_a-1$. Imposing $p_a$
extra conditions to reduce their topological genus to $0$ leaves a space
of rational curves of dimension $d_a-1=\omega.a-1$.
Only a finite number $N_a$ of such curves are therefore expected to 
intersect $d_a-1$ generic points. 

The action of the group $G$  which 
leaves $\omega$ invariant is generated by

(i) space permutations

(ii) reflections of the type $a \to a+(a.e)e$

\noindent{}where $e$ is one of the ${6 \choose 3}=20$ vectors of square $-2$
having only the time and $3$ space components non vanishing and equal to $1$.
If $a \in \Delta$ and $\gamma \in G$, the $\gamma(a)\in \Delta$ and 
$N_a=N_{\gamma(a)}$, a useful check on the forthcoming computation.

As usual, we write the genus $0$ free energy $F=f_{\rm cl}+f$, with
\eqn\frencu{
\eqalign{
f_{\rm cl}&= {1 \over 2} ( x^2 z + x y.y) \cr
f&= \sum_{a \in \Delta} N_a {z^{d_a-1} \over (d_a -1)!} e^{a.y} \cr}
}
corresponding to the weights
\eqn\weicu{ [x]=1 \ \ ; \ \ [z]=-1 \ \ ; \ \ [e^{y^0}]=3 \ \ ; \ \ [e^{y^i}]=1 
\ \ ; \ \  [F]=1 \ . }

\fig{Duality relations for the lines of $\IP_3$.}{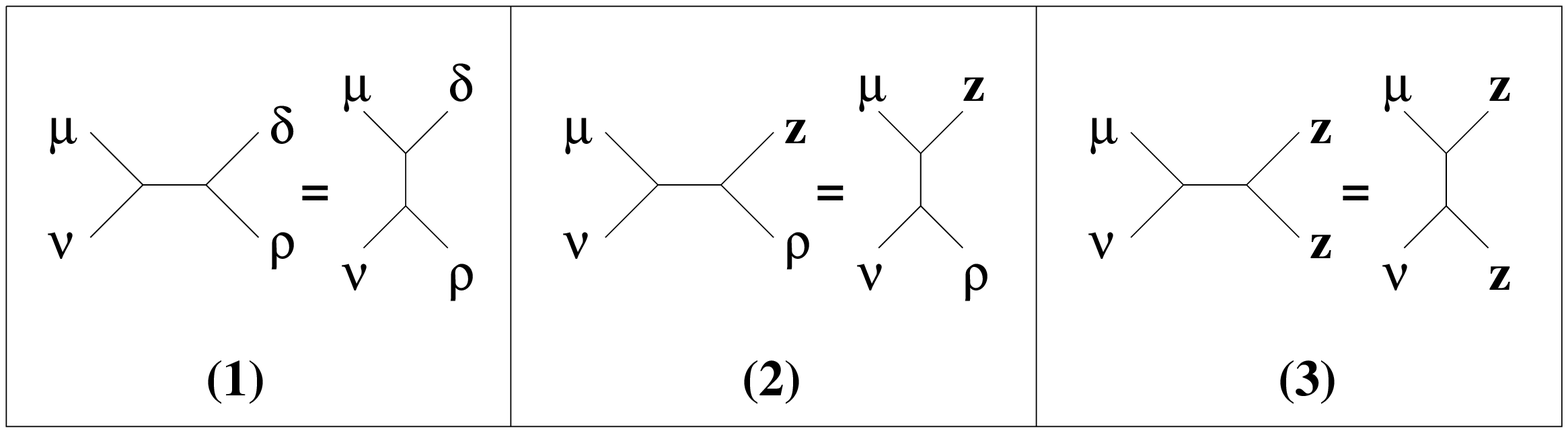}{10cm}
\figlabel\crossline

Associativity of the deformed ring leads to the ($G$--covariant) conditions
\eqn\assocu{
\eqalign{
(1)&\ \ \ g_{\mu \rho} f_{\nu \delta z} -g_{\nu \rho} f_{\mu \delta z}+g_{\nu \delta}
f_{\mu \rho z}-g_{\mu \delta} f_{\rho \nu z} = \sum_{\gs,\gs'}
g^{\gs \gs'} (f_{\mu \gd \gs}f_{\nu \gr \gs'} -f_{\nu \gd \gs}f_{\mu \gr \gs'})\cr
(2)&\ \ \ g_{\mu \gr} f_{\nu z z}-g_{\nu \gr} f_{\mu z z}=\sum_{\gs,\gs'}
g^{\gs \gs'} (f_{\mu \gs z} f_{\nu \gr \gs'} - f_{\nu \gs z} f_{\mu \gr \gs'})\cr
(3)&\ \ \ g_{\mu \nu} f_{zzz}= \sum_{\gs,\gs'} g^{\gs \gs'} (f_{\mu \gs z}
f_{\nu \gs' z}- f_{\mu \nu \gs}f_{\gs' z z})\ . \cr}
}
Substituting the expansion of $f$, we get

\bigskip
\noindent{\bf Proposition 8.}
The integers $N_a$ interpreted as numbers of rational curves on a cubic surface 
belonging to the divisor class $a$, through $d_a-1$ points are uniquely determined by
the initial condition $N_a=1$ for $d_a=1$ and, with $a,b,c \in \Delta$, $a=b+c$
\eqn\detercu{
\eqalign{
(1)&\ \ \  2(g_{\mu \gr} a_{\nu} a_\gd - g_{\nu \gr} a_\mu a_\gd+
g_{\nu \gd} a_\mu a_\gr -g_{\mu \gd} a_\gr a_\nu )N_a= \cr
&\ \ \ =\sum N_b N_c {d_a-2 \choose d_b-1} b.c(b_\mu c_\nu-b_\nu c_\mu)
(b_\gd c_\gr-b_\gr c_\gd)  \cr
(2)&\ \ \  (g_{\mu \gr} a_\nu -g_{\nu \gr} a_\mu)N_a=
\sum N_b N_c {d_a-3 \choose d_b-1} b.c(b_\mu c_\nu-b_\nu c_\mu)c_\gr \cr
(3)&\ \ \ g_{\mu \nu} N_a=\sum N_b N_c \ b.c\left[ b_\mu c_\nu {d_a-4 \choose d_b-2}-
b_\mu b_\nu {d_a-4 \choose d_b-1} \right]\ .\cr}
}
For smooth rational curves, $p_a=0$, i.e. $a \in \Delta_0$, we expect and find
that $N_a=1$. Once more the system of equations  for the 
$N_a$'s is overdetermined but consistent. For explicit calculations
it is better to get scalar equations. For instance contracting relation $(1)$
with $\omega^\mu \omega^\gr g^{\nu \gd}$, we find
\eqn\contract{
2(5d_a^2+3 a^2)N_a=\sum N_b N_c {d_a-2 \choose d_b-1} 
b.c(2 d_b d_c b.c - d_b^2 c.c-d_c^2 b.b) \ .}
Starting with the $27$ lines of the cubic (a complete orbit of $G$), 
the formula \contract\
yields a single conic through a generic point in each of the $27$ planes containing this
point and one of the $27$ lines (again a simple orbit of $G$).
We then tabulate with the same formula all $N_a$'s for $d_a \leq 3$ with  
divisors all belonging to $\Delta_0$ (for which we indeed find $N_a=1$),
except for $a = \omega$, an orbit of $G$ reduced to a single element, with 
$N_{\omega}=12$, corresponding to rational cubic curves through $2$ generic points
of the cubic surface. To proceed further it is convenient to take the trace
of relation $(3)$
\eqn\tracu{ 7 N_a=\sum N_b N_c b.c \left[ b.c {d_a-4 \choose d_b-2}-
b.b{d_a-4 \choose d_b-1} \right] \ .}
$$ \vbox{\offinterlineskip\halign{
& \vrule#& \strut\kern.3em# \kern0pt
& \vrule#& \strut\kern.3em# \kern0pt
& \vrule#& \strut\kern.3em# \kern0pt
& \vrule#& \strut\kern.3em# \kern0pt
& \vrule#& \strut\kern.3em# \kern0pt
& \vrule#& \strut\kern.3em# \kern0pt
& \vrule#& \strut\kern.3em# \kern0pt
& \vrule#& \strut\kern.3em# \kern0pt
& \vrule#& \strut\kern.3em# \kern0pt
& \vrule#& \strut\kern.3em# \kern0pt
& \vrule#& \strut\kern.3em# \kern0pt
& \vrule#& \strut\kern.3em# \kern0pt
& \vrule#& \strut\kern.3em# \kern0pt
&\vrule#\cr
\noalign{\hrule}
height2pt& \omit&& \omit&& \omit&& \omit&& \omit&& \omit&& \omit&& \omit&&
\omit&& \omit&& \omit&& \omit&& \omit&\cr
&$\displaystyle \hfill a^0 \hfill$&&$\displaystyle \hfill a^1
\hfill$&&$\displaystyle \hfill a^2 \hfill$&&$\displaystyle \hfill a^3
\hfill$&&$\displaystyle \hfill a^4 \hfill$&&$\displaystyle \hfill a^5
\hfill$&&$\displaystyle \hfill a^6 \hfill$&&$\displaystyle \hfill d
\hfill$&&$\displaystyle \hfill a.a \hfill$&&$\displaystyle \hfill p_{a} \hfill$&&$\displaystyle \hfill N_{a}
\hfill$&&$\displaystyle \hfill \ \hfill$&&$\displaystyle \hfill {\rm Orb.}
\hfill$&\cr
height2pt& \omit&& \omit&& \omit&& \omit&& \omit&& \omit&& \omit&& \omit&&
\omit&& \omit&& \omit&& \omit&& \omit&\cr
\noalign{\hrule}
height2pt& \omit&& \omit&& \omit&& \omit&& \omit&& \omit&& \omit&& \omit&&
\omit&& \omit&& \omit&& \omit&& \omit&\cr
&$\displaystyle \hfill 0 \hfill$&&$\displaystyle \hfill -1
\hfill$&&$\displaystyle \hfill 0 \hfill$&&$\displaystyle \hfill 0
\hfill$&&$\displaystyle \hfill 0 \hfill$&&$\displaystyle \hfill 0
\hfill$&&$\displaystyle \hfill 0 \hfill$&&$\displaystyle \hfill 1
\hfill$&&$\displaystyle \hfill -1 \hfill$&&$\displaystyle \hfill 0
\hfill$&&$\displaystyle \hfill 1 \hfill$&&$\displaystyle \hfill 6
\hfill$&&$\displaystyle \hfill \hfill$&\cr
height2pt& \omit&& \omit&& \omit&& \omit&& \omit&& \omit&& \omit&& \omit&&
\omit&& \omit&& \omit&& \omit&& \omit&\cr
&$\displaystyle \hfill 1 \hfill$&&$\displaystyle \hfill 1
\hfill$&&$\displaystyle \hfill 1 \hfill$&&$\displaystyle \hfill 0
\hfill$&&$\displaystyle \hfill 0 \hfill$&&$\displaystyle \hfill 0
\hfill$&&$\displaystyle \hfill 0 \hfill$&&$\displaystyle \hfill 1
\hfill$&&$\displaystyle \hfill -1 \hfill$&&$\displaystyle \hfill 0
\hfill$&&$\displaystyle \hfill 1 \hfill$&&$\displaystyle \hfill 15
\hfill$&&$\displaystyle \hfill 27 \hfill$&\cr
height2pt& \omit&& \omit&& \omit&& \omit&& \omit&& \omit&& \omit&& \omit&&
\omit&& \omit&& \omit&& \omit&& \omit&\cr
&$\displaystyle \hfill 2 \hfill$&&$\displaystyle \hfill 1
\hfill$&&$\displaystyle \hfill 1 \hfill$&&$\displaystyle \hfill 1
\hfill$&&$\displaystyle \hfill 1 \hfill$&&$\displaystyle \hfill 1
\hfill$&&$\displaystyle \hfill 0 \hfill$&&$\displaystyle \hfill 1
\hfill$&&$\displaystyle \hfill -1 \hfill$&&$\displaystyle \hfill 0
\hfill$&&$\displaystyle \hfill 1 \hfill$&&$\displaystyle \hfill 6
\hfill$&&$\displaystyle \hfill \hfill$&\cr
height2pt& \omit&& \omit&& \omit&& \omit&& \omit&& \omit&& \omit&& \omit&&
\omit&& \omit&& \omit&& \omit&& \omit&\cr
\noalign{\hrule}}}
$$
$$ \vbox{\offinterlineskip\halign{
& \vrule#& \strut\kern.3em# \kern0pt
& \vrule#& \strut\kern.3em# \kern0pt
& \vrule#& \strut\kern.3em# \kern0pt
& \vrule#& \strut\kern.3em# \kern0pt
& \vrule#& \strut\kern.3em# \kern0pt
& \vrule#& \strut\kern.3em# \kern0pt
& \vrule#& \strut\kern.3em# \kern0pt
& \vrule#& \strut\kern.3em# \kern0pt
& \vrule#& \strut\kern.3em# \kern0pt
& \vrule#& \strut\kern.3em# \kern0pt
& \vrule#& \strut\kern.3em# \kern0pt
& \vrule#& \strut\kern.3em# \kern0pt
& \vrule#& \strut\kern.3em# \kern0pt
&\vrule#\cr
\noalign{\hrule}
height2pt& \omit&& \omit&& \omit&& \omit&& \omit&& \omit&& \omit&& \omit&&
\omit&& \omit&& \omit&& \omit&& \omit&\cr
&$\displaystyle \hfill a^0 \hfill$&&$\displaystyle \hfill a^1
\hfill$&&$\displaystyle \hfill a^2 \hfill$&&$\displaystyle \hfill a^3
\hfill$&&$\displaystyle \hfill a^4 \hfill$&&$\displaystyle \hfill a^5
\hfill$&&$\displaystyle \hfill a^6 \hfill$&&$\displaystyle \hfill d
\hfill$&&$\displaystyle \hfill a.a \hfill$&&$\displaystyle \hfill p_{a} \hfill$&&$\displaystyle \hfill N_{a}
\hfill$&&$\displaystyle \hfill \ \hfill$&&$\displaystyle \hfill {\rm Orb.}
\hfill$&\cr
height2pt& \omit&& \omit&& \omit&& \omit&& \omit&& \omit&& \omit&& \omit&&
\omit&& \omit&& \omit&& \omit&& \omit&\cr
\noalign{\hrule}
height2pt& \omit&& \omit&& \omit&& \omit&& \omit&& \omit&& \omit&& \omit&&
\omit&& \omit&& \omit&& \omit&& \omit&\cr
&$\displaystyle \hfill 1 \hfill$&&$\displaystyle \hfill 1
\hfill$&&$\displaystyle \hfill 0 \hfill$&&$\displaystyle \hfill 0
\hfill$&&$\displaystyle \hfill 0 \hfill$&&$\displaystyle \hfill 0
\hfill$&&$\displaystyle \hfill 0 \hfill$&&$\displaystyle \hfill 2
\hfill$&&$\displaystyle \hfill 0 \hfill$&&$\displaystyle \hfill 0
\hfill$&&$\displaystyle \hfill 1 \hfill$&&$\displaystyle \hfill 6
\hfill$&&$\displaystyle \hfill \hfill$&\cr
height2pt& \omit&& \omit&& \omit&& \omit&& \omit&& \omit&& \omit&& \omit&&
\omit&& \omit&& \omit&& \omit&& \omit&\cr
&$\displaystyle \hfill 2 \hfill$&&$\displaystyle \hfill 1
\hfill$&&$\displaystyle \hfill 1 \hfill$&&$\displaystyle \hfill 1
\hfill$&&$\displaystyle \hfill 1 \hfill$&&$\displaystyle \hfill 0
\hfill$&&$\displaystyle \hfill 0 \hfill$&&$\displaystyle \hfill 2
\hfill$&&$\displaystyle \hfill 0 \hfill$&&$\displaystyle \hfill 0
\hfill$&&$\displaystyle \hfill 1 \hfill$&&$\displaystyle \hfill 15
\hfill$&&$\displaystyle \hfill 27 \hfill$&\cr
height2pt& \omit&& \omit&& \omit&& \omit&& \omit&& \omit&& \omit&& \omit&&
\omit&& \omit&& \omit&& \omit&& \omit&\cr
&$\displaystyle \hfill 3 \hfill$&&$\displaystyle \hfill 2
\hfill$&&$\displaystyle \hfill 1 \hfill$&&$\displaystyle \hfill 1
\hfill$&&$\displaystyle \hfill 1 \hfill$&&$\displaystyle \hfill 1
\hfill$&&$\displaystyle \hfill 1 \hfill$&&$\displaystyle \hfill 2
\hfill$&&$\displaystyle \hfill 0 \hfill$&&$\displaystyle \hfill 0
\hfill$&&$\displaystyle \hfill 1 \hfill$&&$\displaystyle \hfill 6
\hfill$&&$\displaystyle \hfill \hfill$&\cr
height2pt& \omit&& \omit&& \omit&& \omit&& \omit&& \omit&& \omit&& \omit&&
\omit&& \omit&& \omit&& \omit&& \omit&\cr
\noalign{\hrule}}}
$$
$$ \vbox{\offinterlineskip\halign{
& \vrule#& \strut\kern.3em# \kern0pt
& \vrule#& \strut\kern.3em# \kern0pt
& \vrule#& \strut\kern.3em# \kern0pt
& \vrule#& \strut\kern.3em# \kern0pt
& \vrule#& \strut\kern.3em# \kern0pt
& \vrule#& \strut\kern.3em# \kern0pt
& \vrule#& \strut\kern.3em# \kern0pt
& \vrule#& \strut\kern.3em# \kern0pt
& \vrule#& \strut\kern.3em# \kern0pt
& \vrule#& \strut\kern.3em# \kern0pt
& \vrule#& \strut\kern.3em# \kern0pt
& \vrule#& \strut\kern.3em# \kern0pt
& \vrule#& \strut\kern.3em# \kern0pt
&\vrule#\cr
\noalign{\hrule}
height2pt& \omit&& \omit&& \omit&& \omit&& \omit&& \omit&& \omit&& \omit&&
\omit&& \omit&& \omit&& \omit&& \omit&\cr
&$\displaystyle \hfill a^0 \hfill$&&$\displaystyle \hfill a^1
\hfill$&&$\displaystyle \hfill a^2 \hfill$&&$\displaystyle \hfill a^3
\hfill$&&$\displaystyle \hfill a^4 \hfill$&&$\displaystyle \hfill a^5
\hfill$&&$\displaystyle \hfill a^6 \hfill$&&$\displaystyle \hfill d
\hfill$&&$\displaystyle \hfill a.a \hfill$&&$\displaystyle \hfill p_{a} \hfill$&&$\displaystyle \hfill N_{a}
\hfill$&&$\displaystyle \hfill \ \hfill$&&$\displaystyle \hfill {\rm Orb.}
\hfill$&\cr
height2pt& \omit&& \omit&& \omit&& \omit&& \omit&& \omit&& \omit&& \omit&&
\omit&& \omit&& \omit&& \omit&& \omit&\cr
\noalign{\hrule}
height2pt& \omit&& \omit&& \omit&& \omit&& \omit&& \omit&& \omit&& \omit&&
\omit&& \omit&& \omit&& \omit&& \omit&\cr
&$\displaystyle \hfill 1 \hfill$&&$\displaystyle \hfill 0
\hfill$&&$\displaystyle \hfill 0 \hfill$&&$\displaystyle \hfill 0
\hfill$&&$\displaystyle \hfill 0 \hfill$&&$\displaystyle \hfill 0
\hfill$&&$\displaystyle \hfill 0 \hfill$&&$\displaystyle \hfill 3
\hfill$&&$\displaystyle \hfill 1 \hfill$&&$\displaystyle \hfill 0
\hfill$&&$\displaystyle \hfill 1 \hfill$&&$\displaystyle \hfill 1
\hfill$&&$\displaystyle \hfill \hfill$&\cr
height2pt& \omit&& \omit&& \omit&& \omit&& \omit&& \omit&& \omit&& \omit&&
\omit&& \omit&& \omit&& \omit&& \omit&\cr
&$\displaystyle \hfill 2 \hfill$&&$\displaystyle \hfill 1
\hfill$&&$\displaystyle \hfill 1 \hfill$&&$\displaystyle \hfill 1
\hfill$&&$\displaystyle \hfill 0 \hfill$&&$\displaystyle \hfill 0
\hfill$&&$\displaystyle \hfill 0 \hfill$&&$\displaystyle \hfill 3
\hfill$&&$\displaystyle \hfill 1 \hfill$&&$\displaystyle \hfill 0
\hfill$&&$\displaystyle \hfill 1 \hfill$&&$\displaystyle \hfill 20
\hfill$&&$\displaystyle \hfill \hfill$&\cr
height2pt& \omit&& \omit&& \omit&& \omit&& \omit&& \omit&& \omit&& \omit&&
\omit&& \omit&& \omit&& \omit&& \omit&\cr
&$\displaystyle \hfill 3 \hfill$&&$\displaystyle \hfill 2
\hfill$&&$\displaystyle \hfill 1 \hfill$&&$\displaystyle \hfill 1
\hfill$&&$\displaystyle \hfill 1 \hfill$&&$\displaystyle \hfill 1
\hfill$&&$\displaystyle \hfill 0 \hfill$&&$\displaystyle \hfill 3
\hfill$&&$\displaystyle \hfill 1 \hfill$&&$\displaystyle \hfill 0
\hfill$&&$\displaystyle \hfill 1 \hfill$&&$\displaystyle \hfill 30
\hfill$&&$\displaystyle \hfill 72 \hfill$&\cr
height2pt& \omit&& \omit&& \omit&& \omit&& \omit&& \omit&& \omit&& \omit&&
\omit&& \omit&& \omit&& \omit&& \omit&\cr
&$\displaystyle \hfill 4 \hfill$&&$\displaystyle \hfill 2
\hfill$&&$\displaystyle \hfill 2 \hfill$&&$\displaystyle \hfill 2
\hfill$&&$\displaystyle \hfill 1 \hfill$&&$\displaystyle \hfill 1
\hfill$&&$\displaystyle \hfill 1 \hfill$&&$\displaystyle \hfill 3
\hfill$&&$\displaystyle \hfill 1 \hfill$&&$\displaystyle \hfill 0
\hfill$&&$\displaystyle \hfill 1 \hfill$&&$\displaystyle \hfill 20
\hfill$&&$\displaystyle \hfill \hfill$&\cr
height2pt& \omit&& \omit&& \omit&& \omit&& \omit&& \omit&& \omit&& \omit&&
\omit&& \omit&& \omit&& \omit&& \omit&\cr
&$\displaystyle \hfill 5 \hfill$&&$\displaystyle \hfill 2
\hfill$&&$\displaystyle \hfill 2 \hfill$&&$\displaystyle \hfill 2
\hfill$&&$\displaystyle \hfill 2 \hfill$&&$\displaystyle \hfill 2
\hfill$&&$\displaystyle \hfill 2 \hfill$&&$\displaystyle \hfill 3
\hfill$&&$\displaystyle \hfill 1 \hfill$&&$\displaystyle \hfill 0
\hfill$&&$\displaystyle \hfill 1 \hfill$&&$\displaystyle \hfill 1
\hfill$&&$\displaystyle \hfill \hfill$&\cr
height2pt& \omit&& \omit&& \omit&& \omit&& \omit&& \omit&& \omit&& \omit&&
\omit&& \omit&& \omit&& \omit&& \omit&\cr
\noalign{\hrule}}}
$$
$$ \vbox{\offinterlineskip\halign{
& \vrule#& \strut\kern.3em# \kern0pt
& \vrule#& \strut\kern.3em# \kern0pt
& \vrule#& \strut\kern.3em# \kern0pt
& \vrule#& \strut\kern.3em# \kern0pt
& \vrule#& \strut\kern.3em# \kern0pt
& \vrule#& \strut\kern.3em# \kern0pt
& \vrule#& \strut\kern.3em# \kern0pt
& \vrule#& \strut\kern.3em# \kern0pt
& \vrule#& \strut\kern.3em# \kern0pt
& \vrule#& \strut\kern.3em# \kern0pt
& \vrule#& \strut\kern.3em# \kern0pt
& \vrule#& \strut\kern.3em# \kern0pt
& \vrule#& \strut\kern.3em# \kern0pt
&\vrule#\cr
\noalign{\hrule}
height2pt& \omit&& \omit&& \omit&& \omit&& \omit&& \omit&& \omit&& \omit&&
\omit&& \omit&& \omit&& \omit&& \omit&\cr
&$\displaystyle \hfill a^0 \hfill$&&$\displaystyle \hfill a^1
\hfill$&&$\displaystyle \hfill a^2 \hfill$&&$\displaystyle \hfill a^3
\hfill$&&$\displaystyle \hfill a^4 \hfill$&&$\displaystyle \hfill a^5
\hfill$&&$\displaystyle \hfill a^6 \hfill$&&$\displaystyle \hfill d
\hfill$&&$\displaystyle \hfill a.a \hfill$&&$\displaystyle \hfill p_{a} \hfill$&&$\displaystyle \hfill N_{a}
\hfill$&&$\displaystyle \hfill \ \hfill$&&$\displaystyle \hfill {\rm Orb.}
\hfill$&\cr
height2pt& \omit&& \omit&& \omit&& \omit&& \omit&& \omit&& \omit&& \omit&&
\omit&& \omit&& \omit&& \omit&& \omit&\cr
\noalign{\hrule}
height2pt& \omit&& \omit&& \omit&& \omit&& \omit&& \omit&& \omit&& \omit&&
\omit&& \omit&& \omit&& \omit&& \omit&\cr
&$\displaystyle \hfill 3 \hfill$&&$\displaystyle \hfill 1
\hfill$&&$\displaystyle \hfill 1 \hfill$&&$\displaystyle \hfill 1
\hfill$&&$\displaystyle \hfill 1 \hfill$&&$\displaystyle \hfill 1
\hfill$&&$\displaystyle \hfill 1 \hfill$&&$\displaystyle \hfill 3
\hfill$&&$\displaystyle \hfill 3 \hfill$&&$\displaystyle \hfill 1
\hfill$&&$\displaystyle \hfill 12 \hfill$&&$\displaystyle \hfill 1
\hfill$&&$\displaystyle \hfill 1 \hfill$&\cr
height2pt& \omit&& \omit&& \omit&& \omit&& \omit&& \omit&& \omit&& \omit&&
\omit&& \omit&& \omit&& \omit&& \omit&\cr
\noalign{\hrule}}}
$$
$$ \vbox{\offinterlineskip\halign{
& \vrule#& \strut\kern.3em# \kern0pt
& \vrule#& \strut\kern.3em# \kern0pt
& \vrule#& \strut\kern.3em# \kern0pt
& \vrule#& \strut\kern.3em# \kern0pt
& \vrule#& \strut\kern.3em# \kern0pt
& \vrule#& \strut\kern.3em# \kern0pt
& \vrule#& \strut\kern.3em# \kern0pt
& \vrule#& \strut\kern.3em# \kern0pt
& \vrule#& \strut\kern.3em# \kern0pt
& \vrule#& \strut\kern.3em# \kern0pt
& \vrule#& \strut\kern.3em# \kern0pt
& \vrule#& \strut\kern.3em# \kern0pt
& \vrule#& \strut\kern.3em# \kern0pt
&\vrule#\cr
\noalign{\hrule}
height2pt& \omit&& \omit&& \omit&& \omit&& \omit&& \omit&& \omit&& \omit&&
\omit&& \omit&& \omit&& \omit&& \omit&\cr
&$\displaystyle \hfill a^0 \hfill$&&$\displaystyle \hfill a^1
\hfill$&&$\displaystyle \hfill a^2 \hfill$&&$\displaystyle \hfill a^3
\hfill$&&$\displaystyle \hfill a^4 \hfill$&&$\displaystyle \hfill a^5
\hfill$&&$\displaystyle \hfill a^6 \hfill$&&$\displaystyle \hfill d
\hfill$&&$\displaystyle \hfill a.a \hfill$&&$\displaystyle \hfill p_{a} \hfill$&&$\displaystyle \hfill N_{a}
\hfill$&&$\displaystyle \hfill \  \hfill$&&$\displaystyle \hfill {\rm Orb.}
\hfill$&\cr
height2pt& \omit&& \omit&& \omit&& \omit&& \omit&& \omit&& \omit&& \omit&&
\omit&& \omit&& \omit&& \omit&& \omit&\cr
\noalign{\hrule}
height2pt& \omit&& \omit&& \omit&& \omit&& \omit&& \omit&& \omit&& \omit&&
\omit&& \omit&& \omit&& \omit&& \omit&\cr
&$\displaystyle \hfill 2 \hfill$&&$\displaystyle \hfill 1
\hfill$&&$\displaystyle \hfill 1 \hfill$&&$\displaystyle \hfill 0
\hfill$&&$\displaystyle \hfill 0 \hfill$&&$\displaystyle \hfill 0
\hfill$&&$\displaystyle \hfill 0 \hfill$&&$\displaystyle \hfill 4
\hfill$&&$\displaystyle \hfill 2 \hfill$&&$\displaystyle \hfill 0
\hfill$&&$\displaystyle \hfill 1 \hfill$&&$\displaystyle \hfill 15
\hfill$&&$\displaystyle \hfill \hfill$&\cr
height2pt& \omit&& \omit&& \omit&& \omit&& \omit&& \omit&& \omit&& \omit&&
\omit&& \omit&& \omit&& \omit&& \omit&\cr
&$\displaystyle \hfill 3 \hfill$&&$\displaystyle \hfill 2
\hfill$&&$\displaystyle \hfill 1 \hfill$&&$\displaystyle \hfill 1
\hfill$&&$\displaystyle \hfill 1 \hfill$&&$\displaystyle \hfill 0
\hfill$&&$\displaystyle \hfill 0 \hfill$&&$\displaystyle \hfill 4
\hfill$&&$\displaystyle \hfill 2 \hfill$&&$\displaystyle \hfill 0
\hfill$&&$\displaystyle \hfill 1 \hfill$&&$\displaystyle \hfill 60
\hfill$&&$\displaystyle \hfill \hfill$&\cr
height2pt& \omit&& \omit&& \omit&& \omit&& \omit&& \omit&& \omit&& \omit&&
\omit&& \omit&& \omit&& \omit&& \omit&\cr
&$\displaystyle \hfill 4 \hfill$&&$\displaystyle \hfill 3
\hfill$&&$\displaystyle \hfill 1 \hfill$&&$\displaystyle \hfill 1
\hfill$&&$\displaystyle \hfill 1 \hfill$&&$\displaystyle \hfill 1
\hfill$&&$\displaystyle \hfill 1 \hfill$&&$\displaystyle \hfill 4
\hfill$&&$\displaystyle \hfill 2 \hfill$&&$\displaystyle \hfill 0
\hfill$&&$\displaystyle \hfill 1 \hfill$&&$\displaystyle \hfill 6
\hfill$&&$\displaystyle \hfill 216 \hfill$&\cr
height2pt& \omit&& \omit&& \omit&& \omit&& \omit&& \omit&& \omit&& \omit&&
\omit&& \omit&& \omit&& \omit&& \omit&\cr
&$\displaystyle \hfill 4 \hfill$&&$\displaystyle \hfill 2
\hfill$&&$\displaystyle \hfill 2 \hfill$&&$\displaystyle \hfill 2
\hfill$&&$\displaystyle \hfill 1 \hfill$&&$\displaystyle \hfill 1
\hfill$&&$\displaystyle \hfill 0 \hfill$&&$\displaystyle \hfill 4
\hfill$&&$\displaystyle \hfill 2 \hfill$&&$\displaystyle \hfill 0
\hfill$&&$\displaystyle \hfill 1 \hfill$&&$\displaystyle \hfill 60
\hfill$&&$\displaystyle \hfill \hfill$&\cr
height2pt& \omit&& \omit&& \omit&& \omit&& \omit&& \omit&& \omit&& \omit&&
\omit&& \omit&& \omit&& \omit&& \omit&\cr
&$\displaystyle \hfill 5 \hfill$&&$\displaystyle \hfill 3
\hfill$&&$\displaystyle \hfill 2 \hfill$&&$\displaystyle \hfill 2
\hfill$&&$\displaystyle \hfill 2 \hfill$&&$\displaystyle \hfill 1
\hfill$&&$\displaystyle \hfill 1 \hfill$&&$\displaystyle \hfill 4
\hfill$&&$\displaystyle \hfill 2 \hfill$&&$\displaystyle \hfill 0
\hfill$&&$\displaystyle \hfill 1 \hfill$&&$\displaystyle \hfill 60
\hfill$&&$\displaystyle \hfill \hfill$&\cr
height2pt& \omit&& \omit&& \omit&& \omit&& \omit&& \omit&& \omit&& \omit&&
\omit&& \omit&& \omit&& \omit&& \omit&\cr
&$\displaystyle \hfill 6 \hfill$&&$\displaystyle \hfill 3
\hfill$&&$\displaystyle \hfill 3 \hfill$&&$\displaystyle \hfill 2
\hfill$&&$\displaystyle \hfill 2 \hfill$&&$\displaystyle \hfill 2
\hfill$&&$\displaystyle \hfill 2 \hfill$&&$\displaystyle \hfill 4
\hfill$&&$\displaystyle \hfill 2 \hfill$&&$\displaystyle \hfill 0
\hfill$&&$\displaystyle \hfill 1 \hfill$&&$\displaystyle \hfill 15
\hfill$&&$\displaystyle \hfill \hfill$&\cr
height2pt& \omit&& \omit&& \omit&& \omit&& \omit&& \omit&& \omit&& \omit&&
\omit&& \omit&& \omit&& \omit&& \omit&\cr
\noalign{\hrule}}}
$$
$$ \vbox{\offinterlineskip\halign{
& \vrule#& \strut\kern.3em# \kern0pt
& \vrule#& \strut\kern.3em# \kern0pt
& \vrule#& \strut\kern.3em# \kern0pt
& \vrule#& \strut\kern.3em# \kern0pt
& \vrule#& \strut\kern.3em# \kern0pt
& \vrule#& \strut\kern.3em# \kern0pt
& \vrule#& \strut\kern.3em# \kern0pt
& \vrule#& \strut\kern.3em# \kern0pt
& \vrule#& \strut\kern.3em# \kern0pt
& \vrule#& \strut\kern.3em# \kern0pt
& \vrule#& \strut\kern.3em# \kern0pt
& \vrule#& \strut\kern.3em# \kern0pt
& \vrule#& \strut\kern.3em# \kern0pt
&\vrule#\cr
\noalign{\hrule}
height2pt& \omit&& \omit&& \omit&& \omit&& \omit&& \omit&& \omit&& \omit&&
\omit&& \omit&& \omit&& \omit&& \omit&\cr
&$\displaystyle \hfill a^0 \hfill$&&$\displaystyle \hfill a^1
\hfill$&&$\displaystyle \hfill a^2 \hfill$&&$\displaystyle \hfill a^3
\hfill$&&$\displaystyle \hfill a^4 \hfill$&&$\displaystyle \hfill a^5
\hfill$&&$\displaystyle \hfill a^6 \hfill$&&$\displaystyle \hfill d
\hfill$&&$\displaystyle \hfill a.a \hfill$&&$\displaystyle \hfill p_{a} \hfill$&&$\displaystyle \hfill N_{a}
\hfill$&&$\displaystyle \hfill \  \hfill$&&$\displaystyle \hfill {\rm Orb.}
\hfill$&\cr
height2pt& \omit&& \omit&& \omit&& \omit&& \omit&& \omit&& \omit&& \omit&&
\omit&& \omit&& \omit&& \omit&& \omit&\cr
\noalign{\hrule}
height2pt& \omit&& \omit&& \omit&& \omit&& \omit&& \omit&& \omit&& \omit&&
\omit&& \omit&& \omit&& \omit&& \omit&\cr
&$\displaystyle \hfill 3 \hfill$&&$\displaystyle \hfill 1
\hfill$&&$\displaystyle \hfill 1 \hfill$&&$\displaystyle \hfill 1
\hfill$&&$\displaystyle \hfill 1 \hfill$&&$\displaystyle \hfill 1
\hfill$&&$\displaystyle \hfill 0 \hfill$&&$\displaystyle \hfill 4
\hfill$&&$\displaystyle \hfill 4 \hfill$&&$\displaystyle \hfill 1
\hfill$&&$\displaystyle \hfill 12 \hfill$&&$\displaystyle \hfill 6
\hfill$&&$\displaystyle \hfill \hfill$&\cr
height2pt& \omit&& \omit&& \omit&& \omit&& \omit&& \omit&& \omit&& \omit&&
\omit&& \omit&& \omit&& \omit&& \omit&\cr
&$\displaystyle \hfill 4 \hfill$&&$\displaystyle \hfill 2
\hfill$&&$\displaystyle \hfill 2 \hfill$&&$\displaystyle \hfill 1
\hfill$&&$\displaystyle \hfill 1 \hfill$&&$\displaystyle \hfill 1
\hfill$&&$\displaystyle \hfill 1 \hfill$&&$\displaystyle \hfill 4
\hfill$&&$\displaystyle \hfill 4 \hfill$&&$\displaystyle \hfill 1
\hfill$&&$\displaystyle \hfill 12 \hfill$&&$\displaystyle \hfill 15
\hfill$&&$\displaystyle \hfill 27 \hfill$&\cr
height2pt& \omit&& \omit&& \omit&& \omit&& \omit&& \omit&& \omit&& \omit&&
\omit&& \omit&& \omit&& \omit&& \omit&\cr
&$\displaystyle \hfill 5 \hfill$&&$\displaystyle \hfill 2
\hfill$&&$\displaystyle \hfill 2 \hfill$&&$\displaystyle \hfill 2
\hfill$&&$\displaystyle \hfill 2 \hfill$&&$\displaystyle \hfill 2
\hfill$&&$\displaystyle \hfill 1 \hfill$&&$\displaystyle \hfill 4
\hfill$&&$\displaystyle \hfill 4 \hfill$&&$\displaystyle \hfill 1
\hfill$&&$\displaystyle \hfill 12 \hfill$&&$\displaystyle \hfill 6
\hfill$&&$\displaystyle \hfill \hfill$&\cr
height2pt& \omit&& \omit&& \omit&& \omit&& \omit&& \omit&& \omit&& \omit&&
\omit&& \omit&& \omit&& \omit&& \omit&\cr
\noalign{\hrule}}}
$$
$$ \vbox{\offinterlineskip\halign{
& \vrule#& \strut\kern.3em# \kern0pt
& \vrule#& \strut\kern.3em# \kern0pt
& \vrule#& \strut\kern.3em# \kern0pt
& \vrule#& \strut\kern.3em# \kern0pt
& \vrule#& \strut\kern.3em# \kern0pt
& \vrule#& \strut\kern.3em# \kern0pt
& \vrule#& \strut\kern.3em# \kern0pt
& \vrule#& \strut\kern.3em# \kern0pt
& \vrule#& \strut\kern.3em# \kern0pt
& \vrule#& \strut\kern.3em# \kern0pt
& \vrule#& \strut\kern.3em# \kern0pt
& \vrule#& \strut\kern.3em# \kern0pt
& \vrule#& \strut\kern.3em# \kern0pt
&\vrule#\cr
\noalign{\hrule}
height2pt& \omit&& \omit&& \omit&& \omit&& \omit&& \omit&& \omit&& \omit&&
\omit&& \omit&& \omit&& \omit&& \omit&\cr
&$\displaystyle \hfill a^0 \hfill$&&$\displaystyle \hfill a^1
\hfill$&&$\displaystyle \hfill a^2 \hfill$&&$\displaystyle \hfill a^3
\hfill$&&$\displaystyle \hfill a^4 \hfill$&&$\displaystyle \hfill a^5
\hfill$&&$\displaystyle \hfill a^6 \hfill$&&$\displaystyle \hfill d
\hfill$&&$\displaystyle \hfill a.a \hfill$&&$\displaystyle \hfill p_{a} \hfill$&&$\displaystyle \hfill N_{a}
\hfill$&&$\displaystyle \hfill \  \hfill$&&$\displaystyle \hfill {\rm Orb.}
\hfill$&\cr
height2pt& \omit&& \omit&& \omit&& \omit&& \omit&& \omit&& \omit&& \omit&&
\omit&& \omit&& \omit&& \omit&& \omit&\cr
\noalign{\hrule}
height2pt& \omit&& \omit&& \omit&& \omit&& \omit&& \omit&& \omit&& \omit&&
\omit&& \omit&& \omit&& \omit&& \omit&\cr
&$\displaystyle \hfill 2 \hfill$&&$\displaystyle \hfill 1
\hfill$&&$\displaystyle \hfill 0 \hfill$&&$\displaystyle \hfill 0
\hfill$&&$\displaystyle \hfill 0 \hfill$&&$\displaystyle \hfill 0
\hfill$&&$\displaystyle \hfill 0 \hfill$&&$\displaystyle \hfill 5
\hfill$&&$\displaystyle \hfill 3 \hfill$&&$\displaystyle \hfill 0
\hfill$&&$\displaystyle \hfill 1 \hfill$&&$\displaystyle \hfill 6
\hfill$&&$\displaystyle \hfill \hfill$&\cr
height2pt& \omit&& \omit&& \omit&& \omit&& \omit&& \omit&& \omit&& \omit&&
\omit&& \omit&& \omit&& \omit&& \omit&\cr
&$\displaystyle \hfill 3 \hfill$&&$\displaystyle \hfill 2
\hfill$&&$\displaystyle \hfill 1 \hfill$&&$\displaystyle \hfill 1
\hfill$&&$\displaystyle \hfill 0 \hfill$&&$\displaystyle \hfill 0
\hfill$&&$\displaystyle \hfill 0 \hfill$&&$\displaystyle \hfill 5
\hfill$&&$\displaystyle \hfill 3 \hfill$&&$\displaystyle \hfill 0
\hfill$&&$\displaystyle \hfill 1 \hfill$&&$\displaystyle \hfill 60
\hfill$&&$\displaystyle \hfill \hfill$&\cr
height2pt& \omit&& \omit&& \omit&& \omit&& \omit&& \omit&& \omit&& \omit&&
\omit&& \omit&& \omit&& \omit&& \omit&\cr
&$\displaystyle \hfill 4 \hfill$&&$\displaystyle \hfill 3
\hfill$&&$\displaystyle \hfill 1 \hfill$&&$\displaystyle \hfill 1
\hfill$&&$\displaystyle \hfill 1 \hfill$&&$\displaystyle \hfill 1
\hfill$&&$\displaystyle \hfill 0 \hfill$&&$\displaystyle \hfill 5
\hfill$&&$\displaystyle \hfill 3 \hfill$&&$\displaystyle \hfill 0
\hfill$&&$\displaystyle \hfill 1 \hfill$&&$\displaystyle \hfill 30
\hfill$&&$\displaystyle \hfill \hfill$&\cr
height2pt& \omit&& \omit&& \omit&& \omit&& \omit&& \omit&& \omit&& \omit&&
\omit&& \omit&& \omit&& \omit&& \omit&\cr
&$\displaystyle \hfill 4 \hfill$&&$\displaystyle \hfill 2
\hfill$&&$\displaystyle \hfill 2 \hfill$&&$\displaystyle \hfill 2
\hfill$&&$\displaystyle \hfill 1 \hfill$&&$\displaystyle \hfill 0
\hfill$&&$\displaystyle \hfill 0 \hfill$&&$\displaystyle \hfill 5
\hfill$&&$\displaystyle \hfill 3 \hfill$&&$\displaystyle \hfill 0
\hfill$&&$\displaystyle \hfill 1 \hfill$&&$\displaystyle \hfill 60
\hfill$&&$\displaystyle \hfill \hfill$&\cr
height2pt& \omit&& \omit&& \omit&& \omit&& \omit&& \omit&& \omit&& \omit&&
\omit&& \omit&& \omit&& \omit&& \omit&\cr
&$\displaystyle \hfill 5 \hfill$&&$\displaystyle \hfill 3
\hfill$&&$\displaystyle \hfill 2 \hfill$&&$\displaystyle \hfill 2
\hfill$&&$\displaystyle \hfill 2 \hfill$&&$\displaystyle \hfill 1
\hfill$&&$\displaystyle \hfill 0 \hfill$&&$\displaystyle \hfill 5
\hfill$&&$\displaystyle \hfill 3 \hfill$&&$\displaystyle \hfill 0
\hfill$&&$\displaystyle \hfill 1 \hfill$&&$\displaystyle \hfill 120
\hfill$&&$\displaystyle \hfill 432 \hfill$&\cr
height2pt& \omit&& \omit&& \omit&& \omit&& \omit&& \omit&& \omit&& \omit&&
\omit&& \omit&& \omit&& \omit&& \omit&\cr
&$\displaystyle \hfill 6 \hfill$&&$\displaystyle \hfill 4
\hfill$&&$\displaystyle \hfill 2 \hfill$&&$\displaystyle \hfill 2
\hfill$&&$\displaystyle \hfill 2 \hfill$&&$\displaystyle \hfill 2
\hfill$&&$\displaystyle \hfill 1 \hfill$&&$\displaystyle \hfill 5
\hfill$&&$\displaystyle \hfill 3 \hfill$&&$\displaystyle \hfill 0
\hfill$&&$\displaystyle \hfill 1 \hfill$&&$\displaystyle \hfill 30
\hfill$&&$\displaystyle \hfill \hfill$&\cr
height2pt& \omit&& \omit&& \omit&& \omit&& \omit&& \omit&& \omit&& \omit&&
\omit&& \omit&& \omit&& \omit&& \omit&\cr
&$\displaystyle \hfill 6 \hfill$&&$\displaystyle \hfill 3
\hfill$&&$\displaystyle \hfill 3 \hfill$&&$\displaystyle \hfill 3
\hfill$&&$\displaystyle \hfill 2 \hfill$&&$\displaystyle \hfill 1
\hfill$&&$\displaystyle \hfill 1 \hfill$&&$\displaystyle \hfill 5
\hfill$&&$\displaystyle \hfill 3 \hfill$&&$\displaystyle \hfill 0
\hfill$&&$\displaystyle \hfill 1 \hfill$&&$\displaystyle \hfill 60
\hfill$&&$\displaystyle \hfill \hfill$&\cr
height2pt& \omit&& \omit&& \omit&& \omit&& \omit&& \omit&& \omit&& \omit&&
\omit&& \omit&& \omit&& \omit&& \omit&\cr
&$\displaystyle \hfill 7 \hfill$&&$\displaystyle \hfill 4
\hfill$&&$\displaystyle \hfill 3 \hfill$&&$\displaystyle \hfill 3
\hfill$&&$\displaystyle \hfill 2 \hfill$&&$\displaystyle \hfill 2
\hfill$&&$\displaystyle \hfill 2 \hfill$&&$\displaystyle \hfill 5
\hfill$&&$\displaystyle \hfill 3 \hfill$&&$\displaystyle \hfill 0
\hfill$&&$\displaystyle \hfill 1 \hfill$&&$\displaystyle \hfill 60
\hfill$&&$\displaystyle \hfill \hfill$&\cr
height2pt& \omit&& \omit&& \omit&& \omit&& \omit&& \omit&& \omit&& \omit&&
\omit&& \omit&& \omit&& \omit&& \omit&\cr
&$\displaystyle \hfill 8 \hfill$&&$\displaystyle \hfill 4
\hfill$&&$\displaystyle \hfill 3 \hfill$&&$\displaystyle \hfill 3
\hfill$&&$\displaystyle \hfill 3 \hfill$&&$\displaystyle \hfill 3
\hfill$&&$\displaystyle \hfill 3 \hfill$&&$\displaystyle \hfill 5
\hfill$&&$\displaystyle \hfill 3 \hfill$&&$\displaystyle \hfill 0
\hfill$&&$\displaystyle \hfill 1 \hfill$&&$\displaystyle \hfill 6
\hfill$&&$\displaystyle \hfill \hfill$&\cr
height2pt& \omit&& \omit&& \omit&& \omit&& \omit&& \omit&& \omit&& \omit&&
\omit&& \omit&& \omit&& \omit&& \omit&\cr
\noalign{\hrule}}}
$$
$$ \vbox{\offinterlineskip\halign{
& \vrule#& \strut\kern.3em# \kern0pt
& \vrule#& \strut\kern.3em# \kern0pt
& \vrule#& \strut\kern.3em# \kern0pt
& \vrule#& \strut\kern.3em# \kern0pt
& \vrule#& \strut\kern.3em# \kern0pt
& \vrule#& \strut\kern.3em# \kern0pt
& \vrule#& \strut\kern.3em# \kern0pt
& \vrule#& \strut\kern.3em# \kern0pt
& \vrule#& \strut\kern.3em# \kern0pt
& \vrule#& \strut\kern.3em# \kern0pt
& \vrule#& \strut\kern.3em# \kern0pt
& \vrule#& \strut\kern.3em# \kern0pt
& \vrule#& \strut\kern.3em# \kern0pt
&\vrule#\cr
\noalign{\hrule}
height2pt& \omit&& \omit&& \omit&& \omit&& \omit&& \omit&& \omit&& \omit&&
\omit&& \omit&& \omit&& \omit&& \omit&\cr
&$\displaystyle \hfill a^0 \hfill$&&$\displaystyle \hfill a^1
\hfill$&&$\displaystyle \hfill a^2 \hfill$&&$\displaystyle \hfill a^3
\hfill$&&$\displaystyle \hfill a^4 \hfill$&&$\displaystyle \hfill a^5
\hfill$&&$\displaystyle \hfill a^6 \hfill$&&$\displaystyle \hfill d
\hfill$&&$\displaystyle \hfill a.a \hfill$&&$\displaystyle \hfill p_{a} \hfill$&&$\displaystyle \hfill N_{a}
\hfill$&&$\displaystyle \hfill \  \hfill$&&$\displaystyle \hfill {\rm Orb.}
\hfill$&\cr
height2pt& \omit&& \omit&& \omit&& \omit&& \omit&& \omit&& \omit&& \omit&&
\omit&& \omit&& \omit&& \omit&& \omit&\cr
\noalign{\hrule}
height2pt& \omit&& \omit&& \omit&& \omit&& \omit&& \omit&& \omit&& \omit&&
\omit&& \omit&& \omit&& \omit&& \omit&\cr
&$\displaystyle \hfill 3 \hfill$&&$\displaystyle \hfill 1
\hfill$&&$\displaystyle \hfill 1 \hfill$&&$\displaystyle \hfill 1
\hfill$&&$\displaystyle \hfill 1 \hfill$&&$\displaystyle \hfill 0
\hfill$&&$\displaystyle \hfill 0 \hfill$&&$\displaystyle \hfill 5
\hfill$&&$\displaystyle \hfill 5 \hfill$&&$\displaystyle \hfill 1
\hfill$&&$\displaystyle \hfill 12 \hfill$&&$\displaystyle \hfill 15
\hfill$&&$\displaystyle \hfill \hfill$&\cr
height2pt& \omit&& \omit&& \omit&& \omit&& \omit&& \omit&& \omit&& \omit&&
\omit&& \omit&& \omit&& \omit&& \omit&\cr
&$\displaystyle \hfill 4 \hfill$&&$\displaystyle \hfill 2
\hfill$&&$\displaystyle \hfill 2 \hfill$&&$\displaystyle \hfill 1
\hfill$&&$\displaystyle \hfill 1 \hfill$&&$\displaystyle \hfill 1
\hfill$&&$\displaystyle \hfill 0 \hfill$&&$\displaystyle \hfill 5
\hfill$&&$\displaystyle \hfill 5 \hfill$&&$\displaystyle \hfill 1
\hfill$&&$\displaystyle \hfill 12 \hfill$&&$\displaystyle \hfill 60
\hfill$&&$\displaystyle \hfill \hfill$&\cr
height2pt& \omit&& \omit&& \omit&& \omit&& \omit&& \omit&& \omit&& \omit&&
\omit&& \omit&& \omit&& \omit&& \omit&\cr
&$\displaystyle \hfill 5 \hfill$&&$\displaystyle \hfill 3
\hfill$&&$\displaystyle \hfill 2 \hfill$&&$\displaystyle \hfill 2
\hfill$&&$\displaystyle \hfill 1 \hfill$&&$\displaystyle \hfill 1
\hfill$&&$\displaystyle \hfill 1 \hfill$&&$\displaystyle \hfill 5
\hfill$&&$\displaystyle \hfill 5 \hfill$&&$\displaystyle \hfill 1
\hfill$&&$\displaystyle \hfill 12 \hfill$&&$\displaystyle \hfill 60
\hfill$&&$\displaystyle \hfill 216 \hfill$&\cr
height2pt& \omit&& \omit&& \omit&& \omit&& \omit&& \omit&& \omit&& \omit&&
\omit&& \omit&& \omit&& \omit&& \omit&\cr
&$\displaystyle \hfill 5 \hfill$&&$\displaystyle \hfill 2
\hfill$&&$\displaystyle \hfill 2 \hfill$&&$\displaystyle \hfill 2
\hfill$&&$\displaystyle \hfill 2 \hfill$&&$\displaystyle \hfill 2
\hfill$&&$\displaystyle \hfill 0 \hfill$&&$\displaystyle \hfill 5
\hfill$&&$\displaystyle \hfill 5 \hfill$&&$\displaystyle \hfill 1
\hfill$&&$\displaystyle \hfill 12 \hfill$&&$\displaystyle \hfill 6
\hfill$&&$\displaystyle \hfill \hfill$&\cr
height2pt& \omit&& \omit&& \omit&& \omit&& \omit&& \omit&& \omit&& \omit&&
\omit&& \omit&& \omit&& \omit&& \omit&\cr
&$\displaystyle \hfill 6 \hfill$&&$\displaystyle \hfill 3
\hfill$&&$\displaystyle \hfill 3 \hfill$&&$\displaystyle \hfill 2
\hfill$&&$\displaystyle \hfill 2 \hfill$&&$\displaystyle \hfill 2
\hfill$&&$\displaystyle \hfill 1 \hfill$&&$\displaystyle \hfill 5
\hfill$&&$\displaystyle \hfill 5 \hfill$&&$\displaystyle \hfill 1
\hfill$&&$\displaystyle \hfill 12 \hfill$&&$\displaystyle \hfill 60
\hfill$&&$\displaystyle \hfill \hfill$&\cr
height2pt& \omit&& \omit&& \omit&& \omit&& \omit&& \omit&& \omit&& \omit&&
\omit&& \omit&& \omit&& \omit&& \omit&\cr
&$\displaystyle \hfill 7 \hfill$&&$\displaystyle \hfill 3
\hfill$&&$\displaystyle \hfill 3 \hfill$&&$\displaystyle \hfill 3
\hfill$&&$\displaystyle \hfill 3 \hfill$&&$\displaystyle \hfill 2
\hfill$&&$\displaystyle \hfill 2 \hfill$&&$\displaystyle \hfill 5
\hfill$&&$\displaystyle \hfill 5 \hfill$&&$\displaystyle \hfill 1
\hfill$&&$\displaystyle \hfill 12 \hfill$&&$\displaystyle \hfill 15
\hfill$&&$\displaystyle \hfill \hfill$&\cr
height2pt& \omit&& \omit&& \omit&& \omit&& \omit&& \omit&& \omit&& \omit&&
\omit&& \omit&& \omit&& \omit&& \omit&\cr
\noalign{\hrule}}}
$$
$$ \vbox{\offinterlineskip\halign{
& \vrule#& \strut\kern.3em# \kern0pt
& \vrule#& \strut\kern.3em# \kern0pt
& \vrule#& \strut\kern.3em# \kern0pt
& \vrule#& \strut\kern.3em# \kern0pt
& \vrule#& \strut\kern.3em# \kern0pt
& \vrule#& \strut\kern.3em# \kern0pt
& \vrule#& \strut\kern.3em# \kern0pt
& \vrule#& \strut\kern.3em# \kern0pt
& \vrule#& \strut\kern.3em# \kern0pt
& \vrule#& \strut\kern.3em# \kern0pt
& \vrule#& \strut\kern.3em# \kern0pt
& \vrule#& \strut\kern.3em# \kern0pt
& \vrule#& \strut\kern.3em# \kern0pt
&\vrule#\cr
\noalign{\hrule}
height2pt& \omit&& \omit&& \omit&& \omit&& \omit&& \omit&& \omit&& \omit&&
\omit&& \omit&& \omit&& \omit&& \omit&\cr
&$\displaystyle \hfill a^0 \hfill$&&$\displaystyle \hfill a^1
\hfill$&&$\displaystyle \hfill a^2 \hfill$&&$\displaystyle \hfill a^3
\hfill$&&$\displaystyle \hfill a^4 \hfill$&&$\displaystyle \hfill a^5
\hfill$&&$\displaystyle \hfill a^6 \hfill$&&$\displaystyle \hfill d
\hfill$&&$\displaystyle \hfill a.a \hfill$&&$\displaystyle \hfill p_{a} \hfill$&&$\displaystyle \hfill N_{a}
\hfill$&&$\displaystyle \hfill \  \hfill$&&$\displaystyle \hfill {\rm Orb.}
\hfill$&\cr
height2pt& \omit&& \omit&& \omit&& \omit&& \omit&& \omit&& \omit&& \omit&&
\omit&& \omit&& \omit&& \omit&& \omit&\cr
\noalign{\hrule}
height2pt& \omit&& \omit&& \omit&& \omit&& \omit&& \omit&& \omit&& \omit&&
\omit&& \omit&& \omit&& \omit&& \omit&\cr
&$\displaystyle \hfill 4 \hfill$&&$\displaystyle \hfill 2
\hfill$&&$\displaystyle \hfill 1 \hfill$&&$\displaystyle \hfill 1
\hfill$&&$\displaystyle \hfill 1 \hfill$&&$\displaystyle \hfill 1
\hfill$&&$\displaystyle \hfill 1 \hfill$&&$\displaystyle \hfill 5
\hfill$&&$\displaystyle \hfill 7 \hfill$&&$\displaystyle \hfill 2
\hfill$&&$\displaystyle \hfill 96 \hfill$&&$\displaystyle \hfill 6
\hfill$&&$\displaystyle \hfill \hfill$&\cr
height2pt& \omit&& \omit&& \omit&& \omit&& \omit&& \omit&& \omit&& \omit&&
\omit&& \omit&& \omit&& \omit&& \omit&\cr
&$\displaystyle \hfill 5 \hfill$&&$\displaystyle \hfill 2
\hfill$&&$\displaystyle \hfill 2 \hfill$&&$\displaystyle \hfill 2
\hfill$&&$\displaystyle \hfill 2 \hfill$&&$\displaystyle \hfill 1
\hfill$&&$\displaystyle \hfill 1 \hfill$&&$\displaystyle \hfill 5
\hfill$&&$\displaystyle \hfill 7 \hfill$&&$\displaystyle \hfill 2
\hfill$&&$\displaystyle \hfill 96 \hfill$&&$\displaystyle \hfill 15
\hfill$&&$\displaystyle \hfill 27 \hfill$&\cr
height2pt& \omit&& \omit&& \omit&& \omit&& \omit&& \omit&& \omit&& \omit&&
\omit&& \omit&& \omit&& \omit&& \omit&\cr
&$\displaystyle \hfill 6 \hfill$&&$\displaystyle \hfill 3
\hfill$&&$\displaystyle \hfill 2 \hfill$&&$\displaystyle \hfill 2
\hfill$&&$\displaystyle \hfill 2 \hfill$&&$\displaystyle \hfill 2
\hfill$&&$\displaystyle \hfill 2 \hfill$&&$\displaystyle \hfill 5
\hfill$&&$\displaystyle \hfill 7 \hfill$&&$\displaystyle \hfill 2
\hfill$&&$\displaystyle \hfill 96 \hfill$&&$\displaystyle \hfill 6
\hfill$&&$\displaystyle \hfill \hfill$&\cr
height2pt& \omit&& \omit&& \omit&& \omit&& \omit&& \omit&& \omit&& \omit&&
\omit&& \omit&& \omit&& \omit&& \omit&\cr
\noalign{\hrule}}}
$$
$$ \vbox{\offinterlineskip\halign{
& \vrule#& \strut\kern.3em# \kern0pt
& \vrule#& \strut\kern.3em# \kern0pt
& \vrule#& \strut\kern.3em# \kern0pt
& \vrule#& \strut\kern.3em# \kern0pt
& \vrule#& \strut\kern.3em# \kern0pt
& \vrule#& \strut\kern.3em# \kern0pt
& \vrule#& \strut\kern.3em# \kern0pt
& \vrule#& \strut\kern.3em# \kern0pt
& \vrule#& \strut\kern.3em# \kern0pt
& \vrule#& \strut\kern.3em# \kern0pt
& \vrule#& \strut\kern.3em# \kern0pt
& \vrule#& \strut\kern.3em# \kern0pt
& \vrule#& \strut\kern.3em# \kern0pt
&\vrule#\cr
\noalign{\hrule}
height2pt& \omit&& \omit&& \omit&& \omit&& \omit&& \omit&& \omit&& \omit&&
\omit&& \omit&& \omit&& \omit&& \omit&& \cr
&$\displaystyle \hfill a^0 \hfill$&&$\displaystyle \hfill a^1
\hfill$&&$\displaystyle \hfill a^2 \hfill$&&$\displaystyle \hfill a^3
\hfill$&&$\displaystyle \hfill a^4 \hfill$&&$\displaystyle \hfill a^5
\hfill$&&$\displaystyle \hfill a^6 \hfill$&&$\displaystyle \hfill d
\hfill$&&$\displaystyle \hfill a.a \hfill$&&$\displaystyle \hfill p_{a} \hfill$&&$\displaystyle \hfill N_{a}
\hfill$&&$\displaystyle \hfill \  \hfill$&&$\displaystyle \hfill {\rm Orb.}
\hfill$&\cr
height2pt& \omit&& \omit&& \omit&& \omit&& \omit&& \omit&& \omit&& \omit&&
\omit&& \omit&& \omit&& \omit&& \omit&\cr
\noalign{\hrule}
height2pt& \omit&& \omit&& \omit&& \omit&& \omit&& \omit&& \omit&& \omit&&
\omit&& \omit&& \omit&& \omit&& \omit&\cr
&$\displaystyle \hfill 2 \hfill$&&$\displaystyle \hfill 0
\hfill$&&$\displaystyle \hfill 0 \hfill$&&$\displaystyle \hfill 0
\hfill$&&$\displaystyle \hfill 0 \hfill$&&$\displaystyle \hfill 0
\hfill$&&$\displaystyle \hfill 0 \hfill$&&$\displaystyle \hfill 6
\hfill$&&$\displaystyle \hfill 4 \hfill$&&$\displaystyle \hfill 0
\hfill$&&$\displaystyle \hfill 1 \hfill$&&$\displaystyle \hfill 1
\hfill$&&$\displaystyle \hfill \hfill$&\cr
height2pt& \omit&& \omit&& \omit&& \omit&& \omit&& \omit&& \omit&& \omit&&
\omit&& \omit&& \omit&& \omit&& \omit&\cr
&$\displaystyle \hfill 4 \hfill$&&$\displaystyle \hfill 2
\hfill$&&$\displaystyle \hfill 2 \hfill$&&$\displaystyle \hfill 2
\hfill$&&$\displaystyle \hfill 0 \hfill$&&$\displaystyle \hfill 0
\hfill$&&$\displaystyle \hfill 0 \hfill$&&$\displaystyle \hfill 6
\hfill$&&$\displaystyle \hfill 4 \hfill$&&$\displaystyle \hfill 0
\hfill$&&$\displaystyle \hfill 1 \hfill$&&$\displaystyle \hfill 20
\hfill$&&$\displaystyle \hfill 72 \hfill$&\cr
height2pt& \omit&& \omit&& \omit&& \omit&& \omit&& \omit&& \omit&& \omit&&
\omit&& \omit&& \omit&& \omit&& \omit&\cr
&$\displaystyle \hfill 6 \hfill$&&$\displaystyle \hfill 4
\hfill$&&$\displaystyle \hfill 2 \hfill$&&$\displaystyle \hfill 2
\hfill$&&$\displaystyle \hfill 2 \hfill$&&$\displaystyle \hfill 0
\hfill$&&$\displaystyle \hfill 0 \hfill$&&$\displaystyle \hfill 6
\hfill$&&$\displaystyle \hfill 4 \hfill$&&$\displaystyle \hfill 0
\hfill$&&$\displaystyle \hfill 1 \hfill$&&$\displaystyle \hfill 30
\hfill$&&$\displaystyle \hfill \hfill$&\cr
height2pt& \omit&& \omit&& \omit&& \omit&& \omit&& \omit&& \omit&& \omit&&
\omit&& \omit&& \omit&& \omit&& \omit&\cr
&$\displaystyle \hfill 8 \hfill$&&$\displaystyle \hfill 4
\hfill$&&$\displaystyle \hfill 4 \hfill$&&$\displaystyle \hfill 4
\hfill$&&$\displaystyle \hfill 2 \hfill$&&$\displaystyle \hfill 2
\hfill$&&$\displaystyle \hfill 2 \hfill$&&$\displaystyle \hfill 6
\hfill$&&$\displaystyle \hfill 4 \hfill$&&$\displaystyle \hfill 0
\hfill$&&$\displaystyle \hfill 1 \hfill$&&$\displaystyle \hfill 20
\hfill$&&$\displaystyle \hfill \hfill$&\cr
height2pt& \omit&& \omit&& \omit&& \omit&& \omit&& \omit&& \omit&& \omit&&
\omit&& \omit&& \omit&& \omit&& \omit&\cr
&$\displaystyle \hfill 10 \hfill$&&$\displaystyle \hfill 4
\hfill$&&$\displaystyle \hfill 4 \hfill$&&$\displaystyle \hfill 4
\hfill$&&$\displaystyle \hfill 4 \hfill$&&$\displaystyle \hfill 4
\hfill$&&$\displaystyle \hfill 4 \hfill$&&$\displaystyle \hfill 6
\hfill$&&$\displaystyle \hfill 4 \hfill$&&$\displaystyle \hfill 0
\hfill$&&$\displaystyle \hfill 1 \hfill$&&$\displaystyle \hfill 1
\hfill$&&$\displaystyle \hfill \hfill$&\cr
height2pt& \omit&& \omit&& \omit&& \omit&& \omit&& \omit&& \omit&& \omit&&
\omit&& \omit&& \omit&& \omit&& \omit&\cr
\noalign{\hrule}}}
$$
$$ \vbox{\offinterlineskip\halign{
& \vrule#& \strut\kern.3em# \kern0pt
& \vrule#& \strut\kern.3em# \kern0pt
& \vrule#& \strut\kern.3em# \kern0pt
& \vrule#& \strut\kern.3em# \kern0pt
& \vrule#& \strut\kern.3em# \kern0pt
& \vrule#& \strut\kern.3em# \kern0pt
& \vrule#& \strut\kern.3em# \kern0pt
& \vrule#& \strut\kern.3em# \kern0pt
& \vrule#& \strut\kern.3em# \kern0pt
& \vrule#& \strut\kern.3em# \kern0pt
& \vrule#& \strut\kern.3em# \kern0pt
& \vrule#& \strut\kern.3em# \kern0pt
& \vrule#& \strut\kern.3em# \kern0pt
&\vrule#\cr
\noalign{\hrule}
height2pt& \omit&& \omit&& \omit&& \omit&& \omit&& \omit&& \omit&& \omit&&
\omit&& \omit&& \omit&& \omit&& \omit&\cr
&$\displaystyle \hfill a^0 \hfill$&&$\displaystyle \hfill a^1
\hfill$&&$\displaystyle \hfill a^2 \hfill$&&$\displaystyle \hfill a^3
\hfill$&&$\displaystyle \hfill a^4 \hfill$&&$\displaystyle \hfill a^5
\hfill$&&$\displaystyle \hfill a^6 \hfill$&&$\displaystyle \hfill d
\hfill$&&$\displaystyle \hfill a.a \hfill$&&$\displaystyle \hfill p_{a} \hfill$&&$\displaystyle \hfill N_{a}
\hfill$&&$\displaystyle \hfill \  \hfill$&&$\displaystyle \hfill {\rm Orb.}
\hfill$&\cr
height2pt& \omit&& \omit&& \omit&& \omit&& \omit&& \omit&& \omit&& \omit&&
\omit&& \omit&& \omit&& \omit&& \omit&\cr
\noalign{\hrule}
height2pt& \omit&& \omit&& \omit&& \omit&& \omit&& \omit&& \omit&& \omit&&
\omit&& \omit&& \omit&& \omit&& \omit&\cr
&$\displaystyle \hfill 3 \hfill$&&$\displaystyle \hfill 2
\hfill$&&$\displaystyle \hfill 1 \hfill$&&$\displaystyle \hfill 0
\hfill$&&$\displaystyle \hfill 0 \hfill$&&$\displaystyle \hfill 0
\hfill$&&$\displaystyle \hfill 0 \hfill$&&$\displaystyle \hfill 6
\hfill$&&$\displaystyle \hfill 4 \hfill$&&$\displaystyle \hfill 0
\hfill$&&$\displaystyle \hfill 1 \hfill$&&$\displaystyle \hfill 30
\hfill$&&$\displaystyle \hfill \hfill$&\cr
height2pt& \omit&& \omit&& \omit&& \omit&& \omit&& \omit&& \omit&& \omit&&
\omit&& \omit&& \omit&& \omit&& \omit&\cr
&$\displaystyle \hfill 4 \hfill$&&$\displaystyle \hfill 3
\hfill$&&$\displaystyle \hfill 1 \hfill$&&$\displaystyle \hfill 1
\hfill$&&$\displaystyle \hfill 1 \hfill$&&$\displaystyle \hfill 0
\hfill$&&$\displaystyle \hfill 0 \hfill$&&$\displaystyle \hfill 6
\hfill$&&$\displaystyle \hfill 4 \hfill$&&$\displaystyle \hfill 0
\hfill$&&$\displaystyle \hfill 1 \hfill$&&$\displaystyle \hfill 60
\hfill$&&$\displaystyle \hfill \hfill$&\cr
height2pt& \omit&& \omit&& \omit&& \omit&& \omit&& \omit&& \omit&& \omit&&
\omit&& \omit&& \omit&& \omit&& \omit&\cr
&$\displaystyle \hfill 5 \hfill$&&$\displaystyle \hfill 4
\hfill$&&$\displaystyle \hfill 1 \hfill$&&$\displaystyle \hfill 1
\hfill$&&$\displaystyle \hfill 1 \hfill$&&$\displaystyle \hfill 1
\hfill$&&$\displaystyle \hfill 1 \hfill$&&$\displaystyle \hfill 6
\hfill$&&$\displaystyle \hfill 4 \hfill$&&$\displaystyle \hfill 0
\hfill$&&$\displaystyle \hfill 1 \hfill$&&$\displaystyle \hfill 6
\hfill$&&$\displaystyle \hfill \hfill$&\cr
height2pt& \omit&& \omit&& \omit&& \omit&& \omit&& \omit&& \omit&& \omit&&
\omit&& \omit&& \omit&& \omit&& \omit&\cr
&$\displaystyle \hfill 5 \hfill$&&$\displaystyle \hfill 3
\hfill$&&$\displaystyle \hfill 2 \hfill$&&$\displaystyle \hfill 2
\hfill$&&$\displaystyle \hfill 2 \hfill$&&$\displaystyle \hfill 0
\hfill$&&$\displaystyle \hfill 0 \hfill$&&$\displaystyle \hfill 6
\hfill$&&$\displaystyle \hfill 4 \hfill$&&$\displaystyle \hfill 0
\hfill$&&$\displaystyle \hfill 1 \hfill$&&$\displaystyle \hfill 60
\hfill$&&$\displaystyle \hfill 432 \hfill$&\cr
height2pt& \omit&& \omit&& \omit&& \omit&& \omit&& \omit&& \omit&& \omit&&
\omit&& \omit&& \omit&& \omit&& \omit&\cr
&$\displaystyle \hfill 6 \hfill$&&$\displaystyle \hfill 3
\hfill$&&$\displaystyle \hfill 3 \hfill$&&$\displaystyle \hfill 3
\hfill$&&$\displaystyle \hfill 2 \hfill$&&$\displaystyle \hfill 1
\hfill$&&$\displaystyle \hfill 0 \hfill$&&$\displaystyle \hfill 6
\hfill$&&$\displaystyle \hfill 4 \hfill$&&$\displaystyle \hfill 0
\hfill$&&$\displaystyle \hfill 1 \hfill$&&$\displaystyle \hfill 120
\hfill$&&$\displaystyle \hfill \hfill$&\cr
height2pt& \omit&& \omit&& \omit&& \omit&& \omit&& \omit&& \omit&& \omit&&
\omit&& \omit&& \omit&& \omit&& \omit&\cr
&$\displaystyle \hfill 7 \hfill$&&$\displaystyle \hfill 5
\hfill$&&$\displaystyle \hfill 2 \hfill$&&$\displaystyle \hfill 2
\hfill$&&$\displaystyle \hfill 2 \hfill$&&$\displaystyle \hfill 2
\hfill$&&$\displaystyle \hfill 2 \hfill$&&$\displaystyle \hfill 6
\hfill$&&$\displaystyle \hfill 4 \hfill$&&$\displaystyle \hfill 0
\hfill$&&$\displaystyle \hfill 1 \hfill$&&$\displaystyle \hfill 6
\hfill$&&$\displaystyle \hfill \hfill$&\cr
height2pt& \omit&& \omit&& \omit&& \omit&& \omit&& \omit&& \omit&& \omit&&
\omit&& \omit&& \omit&& \omit&& \omit&\cr
&$\displaystyle \hfill 7 \hfill$&&$\displaystyle \hfill 4
\hfill$&&$\displaystyle \hfill 3 \hfill$&&$\displaystyle \hfill 3
\hfill$&&$\displaystyle \hfill 3 \hfill$&&$\displaystyle \hfill 1
\hfill$&&$\displaystyle \hfill 1 \hfill$&&$\displaystyle \hfill 6
\hfill$&&$\displaystyle \hfill 4 \hfill$&&$\displaystyle \hfill 0
\hfill$&&$\displaystyle \hfill 1 \hfill$&&$\displaystyle \hfill 60
\hfill$&&$\displaystyle \hfill \hfill$&\cr
height2pt& \omit&& \omit&& \omit&& \omit&& \omit&& \omit&& \omit&& \omit&&
\omit&& \omit&& \omit&& \omit&& \omit&\cr
&$\displaystyle \hfill 8 \hfill$&&$\displaystyle \hfill 5
\hfill$&&$\displaystyle \hfill 3 \hfill$&&$\displaystyle \hfill 3
\hfill$&&$\displaystyle \hfill 3 \hfill$&&$\displaystyle \hfill 2
\hfill$&&$\displaystyle \hfill 2 \hfill$&&$\displaystyle \hfill 6
\hfill$&&$\displaystyle \hfill 4 \hfill$&&$\displaystyle \hfill 0
\hfill$&&$\displaystyle \hfill 1 \hfill$&&$\displaystyle \hfill 60
\hfill$&&$\displaystyle \hfill \hfill$&\cr
height2pt& \omit&& \omit&& \omit&& \omit&& \omit&& \omit&& \omit&& \omit&&
\omit&& \omit&& \omit&& \omit&& \omit&\cr
&$\displaystyle \hfill 9 \hfill$&&$\displaystyle \hfill 5
\hfill$&&$\displaystyle \hfill 4 \hfill$&&$\displaystyle \hfill 3
\hfill$&&$\displaystyle \hfill 3 \hfill$&&$\displaystyle \hfill 3
\hfill$&&$\displaystyle \hfill 3 \hfill$&&$\displaystyle \hfill 6
\hfill$&&$\displaystyle \hfill 4 \hfill$&&$\displaystyle \hfill 0
\hfill$&&$\displaystyle \hfill 1 \hfill$&&$\displaystyle \hfill 30
\hfill$&&$\displaystyle \hfill \hfill$&\cr
height2pt& \omit&& \omit&& \omit&& \omit&& \omit&& \omit&& \omit&& \omit&&
\omit&& \omit&& \omit&& \omit&& \omit&\cr
\noalign{\hrule}}}
$$
$$ \vbox{\offinterlineskip\halign{
& \vrule#& \strut\kern.3em# \kern0pt
& \vrule#& \strut\kern.3em# \kern0pt
& \vrule#& \strut\kern.3em# \kern0pt
& \vrule#& \strut\kern.3em# \kern0pt
& \vrule#& \strut\kern.3em# \kern0pt
& \vrule#& \strut\kern.3em# \kern0pt
& \vrule#& \strut\kern.3em# \kern0pt
& \vrule#& \strut\kern.3em# \kern0pt
& \vrule#& \strut\kern.3em# \kern0pt
& \vrule#& \strut\kern.3em# \kern0pt
& \vrule#& \strut\kern.3em# \kern0pt
& \vrule#& \strut\kern.3em# \kern0pt
& \vrule#& \strut\kern.3em# \kern0pt
&\vrule#\cr
\noalign{\hrule}
height2pt& \omit&& \omit&& \omit&& \omit&& \omit&& \omit&& \omit&& \omit&&
\omit&& \omit&& \omit&& \omit&& \omit&\cr
&$\displaystyle \hfill a^0 \hfill$&&$\displaystyle \hfill a^1
\hfill$&&$\displaystyle \hfill a^2 \hfill$&&$\displaystyle \hfill a^3
\hfill$&&$\displaystyle \hfill a^4 \hfill$&&$\displaystyle \hfill a^5
\hfill$&&$\displaystyle \hfill a^6 \hfill$&&$\displaystyle \hfill d
\hfill$&&$\displaystyle \hfill a.a \hfill$&&$\displaystyle \hfill p_{a} \hfill$&&$\displaystyle \hfill N_{a}
\hfill$&&$\displaystyle \hfill \  \hfill$&&$\displaystyle \hfill {\rm Orb.}
\hfill$&\cr
height2pt& \omit&& \omit&& \omit&& \omit&& \omit&& \omit&& \omit&& \omit&&
\omit&& \omit&& \omit&& \omit&& \omit&\cr
\noalign{\hrule}
height2pt& \omit&& \omit&& \omit&& \omit&& \omit&& \omit&& \omit&& \omit&&
\omit&& \omit&& \omit&& \omit&& \omit&\cr
&$\displaystyle \hfill 3 \hfill$&&$\displaystyle \hfill 1
\hfill$&&$\displaystyle \hfill 1 \hfill$&&$\displaystyle \hfill 1
\hfill$&&$\displaystyle \hfill 0 \hfill$&&$\displaystyle \hfill 0
\hfill$&&$\displaystyle \hfill 0 \hfill$&&$\displaystyle \hfill 6
\hfill$&&$\displaystyle \hfill 6 \hfill$&&$\displaystyle \hfill 1
\hfill$&&$\displaystyle \hfill 12 \hfill$&&$\displaystyle \hfill 20
\hfill$&&$\displaystyle \hfill \hfill$&\cr
height2pt& \omit&& \omit&& \omit&& \omit&& \omit&& \omit&& \omit&& \omit&&
\omit&& \omit&& \omit&& \omit&& \omit&\cr
&$\displaystyle \hfill 4 \hfill$&&$\displaystyle \hfill 2
\hfill$&&$\displaystyle \hfill 2 \hfill$&&$\displaystyle \hfill 1
\hfill$&&$\displaystyle \hfill 1 \hfill$&&$\displaystyle \hfill 0
\hfill$&&$\displaystyle \hfill 0 \hfill$&&$\displaystyle \hfill 6
\hfill$&&$\displaystyle \hfill 6 \hfill$&&$\displaystyle \hfill 1
\hfill$&&$\displaystyle \hfill 12 \hfill$&&$\displaystyle \hfill 90
\hfill$&&$\displaystyle \hfill \hfill$&\cr
height2pt& \omit&& \omit&& \omit&& \omit&& \omit&& \omit&& \omit&& \omit&&
\omit&& \omit&& \omit&& \omit&& \omit&\cr
&$\displaystyle \hfill 5 \hfill$&&$\displaystyle \hfill 3
\hfill$&&$\displaystyle \hfill 2 \hfill$&&$\displaystyle \hfill 2
\hfill$&&$\displaystyle \hfill 1 \hfill$&&$\displaystyle \hfill 1
\hfill$&&$\displaystyle \hfill 0 \hfill$&&$\displaystyle \hfill 6
\hfill$&&$\displaystyle \hfill 6 \hfill$&&$\displaystyle \hfill 1
\hfill$&&$\displaystyle \hfill 12 \hfill$&&$\displaystyle \hfill 180
\hfill$&&$\displaystyle \hfill \hfill$&\cr
height2pt& \omit&& \omit&& \omit&& \omit&& \omit&& \omit&& \omit&& \omit&&
\omit&& \omit&& \omit&& \omit&& \omit&\cr
&$\displaystyle \hfill 6 \hfill$&&$\displaystyle \hfill 3
\hfill$&&$\displaystyle \hfill 3 \hfill$&&$\displaystyle \hfill 3
\hfill$&&$\displaystyle \hfill 1 \hfill$&&$\displaystyle \hfill 1
\hfill$&&$\displaystyle \hfill 1 \hfill$&&$\displaystyle \hfill 6
\hfill$&&$\displaystyle \hfill 6 \hfill$&&$\displaystyle \hfill 1
\hfill$&&$\displaystyle \hfill 12 \hfill$&&$\displaystyle \hfill 20
\hfill$&&$\displaystyle \hfill \hfill$&\cr
height2pt& \omit&& \omit&& \omit&& \omit&& \omit&& \omit&& \omit&& \omit&&
\omit&& \omit&& \omit&& \omit&& \omit&\cr
&$\displaystyle \hfill 6 \hfill$&&$\displaystyle \hfill 3
\hfill$&&$\displaystyle \hfill 3 \hfill$&&$\displaystyle \hfill 2
\hfill$&&$\displaystyle \hfill 2 \hfill$&&$\displaystyle \hfill 2
\hfill$&&$\displaystyle \hfill 0 \hfill$&&$\displaystyle \hfill 6
\hfill$&&$\displaystyle \hfill 6 \hfill$&&$\displaystyle \hfill 1
\hfill$&&$\displaystyle \hfill 12 \hfill$&&$\displaystyle \hfill 60
\hfill$&&$\displaystyle \hfill 720 \hfill$&\cr
height2pt& \omit&& \omit&& \omit&& \omit&& \omit&& \omit&& \omit&& \omit&&
\omit&& \omit&& \omit&& \omit&& \omit&\cr
&$\displaystyle \hfill 6 \hfill$&&$\displaystyle \hfill 4
\hfill$&&$\displaystyle \hfill 2 \hfill$&&$\displaystyle \hfill 2
\hfill$&&$\displaystyle \hfill 2 \hfill$&&$\displaystyle \hfill 1
\hfill$&&$\displaystyle \hfill 1 \hfill$&&$\displaystyle \hfill 6
\hfill$&&$\displaystyle \hfill 6 \hfill$&&$\displaystyle \hfill 1
\hfill$&&$\displaystyle \hfill 12 \hfill$&&$\displaystyle \hfill 60
\hfill$&&$\displaystyle \hfill \hfill$&\cr
height2pt& \omit&& \omit&& \omit&& \omit&& \omit&& \omit&& \omit&& \omit&&
\omit&& \omit&& \omit&& \omit&& \omit&\cr
&$\displaystyle \hfill 7 \hfill$&&$\displaystyle \hfill 4
\hfill$&&$\displaystyle \hfill 3 \hfill$&&$\displaystyle \hfill 3
\hfill$&&$\displaystyle \hfill 2 \hfill$&&$\displaystyle \hfill 2
\hfill$&&$\displaystyle \hfill 1 \hfill$&&$\displaystyle \hfill 6
\hfill$&&$\displaystyle \hfill 6 \hfill$&&$\displaystyle \hfill 1
\hfill$&&$\displaystyle \hfill 12 \hfill$&&$\displaystyle \hfill 180
\hfill$&&$\displaystyle \hfill \hfill$&\cr
height2pt& \omit&& \omit&& \omit&& \omit&& \omit&& \omit&& \omit&& \omit&&
\omit&& \omit&& \omit&& \omit&& \omit&\cr
&$\displaystyle \hfill 8 \hfill$&&$\displaystyle \hfill 4
\hfill$&&$\displaystyle \hfill 4 \hfill$&&$\displaystyle \hfill 3
\hfill$&&$\displaystyle \hfill 3 \hfill$&&$\displaystyle \hfill 2
\hfill$&&$\displaystyle \hfill 2 \hfill$&&$\displaystyle \hfill 6
\hfill$&&$\displaystyle \hfill 6 \hfill$&&$\displaystyle \hfill 1
\hfill$&&$\displaystyle \hfill 12 \hfill$&&$\displaystyle \hfill 90
\hfill$&&$\displaystyle \hfill \hfill$&\cr
height2pt& \omit&& \omit&& \omit&& \omit&& \omit&& \omit&& \omit&& \omit&&
\omit&& \omit&& \omit&& \omit&& \omit&\cr
&$\displaystyle \hfill 9 \hfill$&&$\displaystyle \hfill 4
\hfill$&&$\displaystyle \hfill 4 \hfill$&&$\displaystyle \hfill 4
\hfill$&&$\displaystyle \hfill 3 \hfill$&&$\displaystyle \hfill 3
\hfill$&&$\displaystyle \hfill 3 \hfill$&&$\displaystyle \hfill 6
\hfill$&&$\displaystyle \hfill 6 \hfill$&&$\displaystyle \hfill 1
\hfill$&&$\displaystyle \hfill 12 \hfill$&&$\displaystyle \hfill 20
\hfill$&&$\displaystyle \hfill \hfill$&\cr
height2pt& \omit&& \omit&& \omit&& \omit&& \omit&& \omit&& \omit&& \omit&&
\omit&& \omit&& \omit&& \omit&& \omit&\cr
\noalign{\hrule}}}
$$
$$ \vbox{\offinterlineskip\halign{
& \vrule#& \strut\kern.3em# \kern0pt
& \vrule#& \strut\kern.3em# \kern0pt
& \vrule#& \strut\kern.3em# \kern0pt
& \vrule#& \strut\kern.3em# \kern0pt
& \vrule#& \strut\kern.3em# \kern0pt
& \vrule#& \strut\kern.3em# \kern0pt
& \vrule#& \strut\kern.3em# \kern0pt
& \vrule#& \strut\kern.3em# \kern0pt
& \vrule#& \strut\kern.3em# \kern0pt
& \vrule#& \strut\kern.3em# \kern0pt
& \vrule#& \strut\kern.3em# \kern0pt
& \vrule#& \strut\kern.3em# \kern0pt
& \vrule#& \strut\kern.3em# \kern0pt
&\vrule#\cr
\noalign{\hrule}
height2pt& \omit&& \omit&& \omit&& \omit&& \omit&& \omit&& \omit&& \omit&&
\omit&& \omit&& \omit&& \omit&& \omit&\cr
&$\displaystyle \hfill a^0 \hfill$&&$\displaystyle \hfill a^1
\hfill$&&$\displaystyle \hfill a^2 \hfill$&&$\displaystyle \hfill a^3
\hfill$&&$\displaystyle \hfill a^4 \hfill$&&$\displaystyle \hfill a^5
\hfill$&&$\displaystyle \hfill a^6 \hfill$&&$\displaystyle \hfill d
\hfill$&&$\displaystyle \hfill a.a \hfill$&&$\displaystyle \hfill p_{a} \hfill$&&$\displaystyle \hfill N_{a}
\hfill$&&$\displaystyle \hfill \  \hfill$&&$\displaystyle \hfill {\rm Orb.}
\hfill$&\cr
height2pt& \omit&& \omit&& \omit&& \omit&& \omit&& \omit&& \omit&& \omit&&
\omit&& \omit&& \omit&& \omit&& \omit&\cr
\noalign{\hrule}
height2pt& \omit&& \omit&& \omit&& \omit&& \omit&& \omit&& \omit&& \omit&&
\omit&& \omit&& \omit&& \omit&& \omit&\cr
&$\displaystyle \hfill 4 \hfill$&&$\displaystyle \hfill 2
\hfill$&&$\displaystyle \hfill 1 \hfill$&&$\displaystyle \hfill 1
\hfill$&&$\displaystyle \hfill 1 \hfill$&&$\displaystyle \hfill 1
\hfill$&&$\displaystyle \hfill 0 \hfill$&&$\displaystyle \hfill 6
\hfill$&&$\displaystyle \hfill 8 \hfill$&&$\displaystyle \hfill 2
\hfill$&&$\displaystyle \hfill 96 \hfill$&&$\displaystyle \hfill 30
\hfill$&&$\displaystyle \hfill \hfill$&\cr
height2pt& \omit&& \omit&& \omit&& \omit&& \omit&& \omit&& \omit&& \omit&&
\omit&& \omit&& \omit&& \omit&& \omit&\cr
&$\displaystyle \hfill 5 \hfill$&&$\displaystyle \hfill 2
\hfill$&&$\displaystyle \hfill 2 \hfill$&&$\displaystyle \hfill 2
\hfill$&&$\displaystyle \hfill 2 \hfill$&&$\displaystyle \hfill 1
\hfill$&&$\displaystyle \hfill 0 \hfill$&&$\displaystyle \hfill 6
\hfill$&&$\displaystyle \hfill 8 \hfill$&&$\displaystyle \hfill 2
\hfill$&&$\displaystyle \hfill 96 \hfill$&&$\displaystyle \hfill 30
\hfill$&&$\displaystyle \hfill \hfill$&\cr
height2pt& \omit&& \omit&& \omit&& \omit&& \omit&& \omit&& \omit&& \omit&&
\omit&& \omit&& \omit&& \omit&& \omit&\cr
&$\displaystyle \hfill 5 \hfill$&&$\displaystyle \hfill 3
\hfill$&&$\displaystyle \hfill 2 \hfill$&&$\displaystyle \hfill 1
\hfill$&&$\displaystyle \hfill 1 \hfill$&&$\displaystyle \hfill 1
\hfill$&&$\displaystyle \hfill 1 \hfill$&&$\displaystyle \hfill 6
\hfill$&&$\displaystyle \hfill 8 \hfill$&&$\displaystyle \hfill 2
\hfill$&&$\displaystyle \hfill 96 \hfill$&&$\displaystyle \hfill 30
\hfill$&&$\displaystyle \hfill \hfill$&\cr
height2pt& \omit&& \omit&& \omit&& \omit&& \omit&& \omit&& \omit&& \omit&&
\omit&& \omit&& \omit&& \omit&& \omit&\cr
&$\displaystyle \hfill 6 \hfill$&&$\displaystyle \hfill 3
\hfill$&&$\displaystyle \hfill 3 \hfill$&&$\displaystyle \hfill 2
\hfill$&&$\displaystyle \hfill 2 \hfill$&&$\displaystyle \hfill 1
\hfill$&&$\displaystyle \hfill 1 \hfill$&&$\displaystyle \hfill 6
\hfill$&&$\displaystyle \hfill 8 \hfill$&&$\displaystyle \hfill 2
\hfill$&&$\displaystyle \hfill 96 \hfill$&&$\displaystyle \hfill 90
\hfill$&&$\displaystyle \hfill 270 \hfill$&\cr
height2pt& \omit&& \omit&& \omit&& \omit&& \omit&& \omit&& \omit&& \omit&&
\omit&& \omit&& \omit&& \omit&& \omit&\cr
&$\displaystyle \hfill 7 \hfill$&&$\displaystyle \hfill 4
\hfill$&&$\displaystyle \hfill 3 \hfill$&&$\displaystyle \hfill 2
\hfill$&&$\displaystyle \hfill 2 \hfill$&&$\displaystyle \hfill 2
\hfill$&&$\displaystyle \hfill 2 \hfill$&&$\displaystyle \hfill 6
\hfill$&&$\displaystyle \hfill 8 \hfill$&&$\displaystyle \hfill 2
\hfill$&&$\displaystyle \hfill 96 \hfill$&&$\displaystyle \hfill 30
\hfill$&&$\displaystyle \hfill \hfill$&\cr
height2pt& \omit&& \omit&& \omit&& \omit&& \omit&& \omit&& \omit&& \omit&&
\omit&& \omit&& \omit&& \omit&& \omit&\cr
&$\displaystyle \hfill 7 \hfill$&&$\displaystyle \hfill 3
\hfill$&&$\displaystyle \hfill 3 \hfill$&&$\displaystyle \hfill 3
\hfill$&&$\displaystyle \hfill 3 \hfill$&&$\displaystyle \hfill 2
\hfill$&&$\displaystyle \hfill 1 \hfill$&&$\displaystyle \hfill 6
\hfill$&&$\displaystyle \hfill 8 \hfill$&&$\displaystyle \hfill 2
\hfill$&&$\displaystyle \hfill 96 \hfill$&&$\displaystyle \hfill 30
\hfill$&&$\displaystyle \hfill \hfill$&\cr
height2pt& \omit&& \omit&& \omit&& \omit&& \omit&& \omit&& \omit&& \omit&&
\omit&& \omit&& \omit&& \omit&& \omit&\cr
&$\displaystyle \hfill 8 \hfill$&&$\displaystyle \hfill 4
\hfill$&&$\displaystyle \hfill 3 \hfill$&&$\displaystyle \hfill 3
\hfill$&&$\displaystyle \hfill 3 \hfill$&&$\displaystyle \hfill 3
\hfill$&&$\displaystyle \hfill 2 \hfill$&&$\displaystyle \hfill 6
\hfill$&&$\displaystyle \hfill 8 \hfill$&&$\displaystyle \hfill 2
\hfill$&&$\displaystyle \hfill 96 \hfill$&&$\displaystyle \hfill 30
\hfill$&&$\displaystyle \hfill \hfill$&\cr
height2pt& \omit&& \omit&& \omit&& \omit&& \omit&& \omit&& \omit&& \omit&&
\omit&& \omit&& \omit&& \omit&& \omit&\cr
\noalign{\hrule}}}
$$
$$ \vbox{\offinterlineskip\halign{
& \vrule#& \strut\kern.3em# \kern0pt
& \vrule#& \strut\kern.3em# \kern0pt
& \vrule#& \strut\kern.3em# \kern0pt
& \vrule#& \strut\kern.3em# \kern0pt
& \vrule#& \strut\kern.3em# \kern0pt
& \vrule#& \strut\kern.3em# \kern0pt
& \vrule#& \strut\kern.3em# \kern0pt
& \vrule#& \strut\kern.3em# \kern0pt
& \vrule#& \strut\kern.3em# \kern0pt
& \vrule#& \strut\kern.3em# \kern0pt
& \vrule#& \strut\kern.3em# \kern0pt
& \vrule#& \strut\kern.3em# \kern0pt
& \vrule#& \strut\kern.3em# \kern0pt
&\vrule#\cr
\noalign{\hrule}
height2pt& \omit&& \omit&& \omit&& \omit&& \omit&& \omit&& \omit&& \omit&&
\omit&& \omit&& \omit&& \omit&& \omit&\cr
&$\displaystyle \hfill a^0 \hfill$&&$\displaystyle \hfill a^1
\hfill$&&$\displaystyle \hfill a^2 \hfill$&&$\displaystyle \hfill a^3
\hfill$&&$\displaystyle \hfill a^4 \hfill$&&$\displaystyle \hfill a^5
\hfill$&&$\displaystyle \hfill a^6 \hfill$&&$\displaystyle \hfill d
\hfill$&&$\displaystyle \hfill a.a \hfill$&&$\displaystyle \hfill p_{a} \hfill$&&$\displaystyle \hfill N_{a}
\hfill$&&$\displaystyle \hfill \  \hfill$&&$\displaystyle \hfill {\rm Orb.}
\hfill$&\cr
height2pt& \omit&& \omit&& \omit&& \omit&& \omit&& \omit&& \omit&& \omit&&
\omit&& \omit&& \omit&& \omit&& \omit&\cr
\noalign{\hrule}
height2pt& \omit&& \omit&& \omit&& \omit&& \omit&& \omit&& \omit&& \omit&&
\omit&& \omit&& \omit&& \omit&& \omit&\cr
&$\displaystyle \hfill 4 \hfill$&&$\displaystyle \hfill 1
\hfill$&&$\displaystyle \hfill 1 \hfill$&&$\displaystyle \hfill 1
\hfill$&&$\displaystyle \hfill 1 \hfill$&&$\displaystyle \hfill 1
\hfill$&&$\displaystyle \hfill 1 \hfill$&&$\displaystyle \hfill 6
\hfill$&&$\displaystyle \hfill 10 \hfill$&&$\displaystyle \hfill 3
\hfill$&&$\displaystyle \hfill 620 \hfill$&&$\displaystyle \hfill 1
\hfill$&&$\displaystyle \hfill \hfill$&\cr
height2pt& \omit&& \omit&& \omit&& \omit&& \omit&& \omit&& \omit&& \omit&&
\omit&& \omit&& \omit&& \omit&& \omit&\cr
&$\displaystyle \hfill 5 \hfill$&&$\displaystyle \hfill 2
\hfill$&&$\displaystyle \hfill 2 \hfill$&&$\displaystyle \hfill 2
\hfill$&&$\displaystyle \hfill 1 \hfill$&&$\displaystyle \hfill 1
\hfill$&&$\displaystyle \hfill 1 \hfill$&&$\displaystyle \hfill 6
\hfill$&&$\displaystyle \hfill 10 \hfill$&&$\displaystyle \hfill 3
\hfill$&&$\displaystyle \hfill 620 \hfill$&&$\displaystyle \hfill 20
\hfill$&&$\displaystyle \hfill \hfill$&\cr
height2pt& \omit&& \omit&& \omit&& \omit&& \omit&& \omit&& \omit&& \omit&&
\omit&& \omit&& \omit&& \omit&& \omit&\cr
&$\displaystyle \hfill 6 \hfill$&&$\displaystyle \hfill 3
\hfill$&&$\displaystyle \hfill 2 \hfill$&&$\displaystyle \hfill 2
\hfill$&&$\displaystyle \hfill 2 \hfill$&&$\displaystyle \hfill 2
\hfill$&&$\displaystyle \hfill 1 \hfill$&&$\displaystyle \hfill 6
\hfill$&&$\displaystyle \hfill 10 \hfill$&&$\displaystyle \hfill 3
\hfill$&&$\displaystyle \hfill 620 \hfill$&&$\displaystyle \hfill 30
\hfill$&&$\displaystyle \hfill 72 \hfill$&\cr
height2pt& \omit&& \omit&& \omit&& \omit&& \omit&& \omit&& \omit&& \omit&&
\omit&& \omit&& \omit&& \omit&& \omit&\cr
&$\displaystyle \hfill 7 \hfill$&&$\displaystyle \hfill 3
\hfill$&&$\displaystyle \hfill 3 \hfill$&&$\displaystyle \hfill 3
\hfill$&&$\displaystyle \hfill 2 \hfill$&&$\displaystyle \hfill 2
\hfill$&&$\displaystyle \hfill 2 \hfill$&&$\displaystyle \hfill 6
\hfill$&&$\displaystyle \hfill 10 \hfill$&&$\displaystyle \hfill 3
\hfill$&&$\displaystyle \hfill 620 \hfill$&&$\displaystyle \hfill 20
\hfill$&&$\displaystyle \hfill \hfill$&\cr
height2pt& \omit&& \omit&& \omit&& \omit&& \omit&& \omit&& \omit&& \omit&&
\omit&& \omit&& \omit&& \omit&& \omit&\cr
&$\displaystyle \hfill 8 \hfill$&&$\displaystyle \hfill 3
\hfill$&&$\displaystyle \hfill 3 \hfill$&&$\displaystyle \hfill 3
\hfill$&&$\displaystyle \hfill 3 \hfill$&&$\displaystyle \hfill 3
\hfill$&&$\displaystyle \hfill 3 \hfill$&&$\displaystyle \hfill 6
\hfill$&&$\displaystyle \hfill 10 \hfill$&&$\displaystyle \hfill 3
\hfill$&&$\displaystyle \hfill 620 \hfill$&&$\displaystyle \hfill 1
\hfill$&&$\displaystyle \hfill \hfill$&\cr
height2pt& \omit&& \omit&& \omit&& \omit&& \omit&& \omit&& \omit&& \omit&&
\omit&& \omit&& \omit&& \omit&& \omit&\cr
\noalign{\hrule}}}
$$
$$ \vbox{\offinterlineskip\halign{
& \vrule#& \strut\kern.3em# \kern0pt
& \vrule#& \strut\kern.3em# \kern0pt
& \vrule#& \strut\kern.3em# \kern0pt
& \vrule#& \strut\kern.3em# \kern0pt
& \vrule#& \strut\kern.3em# \kern0pt
& \vrule#& \strut\kern.3em# \kern0pt
& \vrule#& \strut\kern.3em# \kern0pt
& \vrule#& \strut\kern.3em# \kern0pt
& \vrule#& \strut\kern.3em# \kern0pt
& \vrule#& \strut\kern.3em# \kern0pt
& \vrule#& \strut\kern.3em# \kern0pt
& \vrule#& \strut\kern.3em# \kern0pt
& \vrule#& \strut\kern.3em# \kern0pt
&\vrule#\cr
\noalign{\hrule}
height2pt& \omit&& \omit&& \omit&& \omit&& \omit&& \omit&& \omit&& \omit&&
\omit&& \omit&& \omit&& \omit&& \omit&\cr
&$\displaystyle \hfill a^0 \hfill$&&$\displaystyle \hfill a^1
\hfill$&&$\displaystyle \hfill a^2 \hfill$&&$\displaystyle \hfill a^3
\hfill$&&$\displaystyle \hfill a^4 \hfill$&&$\displaystyle \hfill a^5
\hfill$&&$\displaystyle \hfill a^6 \hfill$&&$\displaystyle \hfill d
\hfill$&&$\displaystyle \hfill a.a \hfill$&&$\displaystyle \hfill p_{a} \hfill$&&$\displaystyle \hfill N_{a}
\hfill$&&$\displaystyle \hfill \  \hfill$&&$\displaystyle \hfill {\rm Orb.}
\hfill$&\cr
height2pt& \omit&& \omit&& \omit&& \omit&& \omit&& \omit&& \omit&& \omit&&
\omit&& \omit&& \omit&& \omit&& \omit&\cr
\noalign{\hrule}
height2pt& \omit&& \omit&& \omit&& \omit&& \omit&& \omit&& \omit&& \omit&&
\omit&& \omit&& \omit&& \omit&& \omit&\cr
&$\displaystyle \hfill 6 \hfill$&&$\displaystyle \hfill 2
\hfill$&&$\displaystyle \hfill 2 \hfill$&&$\displaystyle \hfill 2
\hfill$&&$\displaystyle \hfill 2 \hfill$&&$\displaystyle \hfill 2
\hfill$&&$\displaystyle \hfill 2 \hfill$&&$\displaystyle \hfill 6
\hfill$&&$\displaystyle \hfill 12 \hfill$&&$\displaystyle \hfill 4
\hfill$&&$\displaystyle \hfill 2376 \hfill$&&$\displaystyle \hfill 1
\hfill$&&$\displaystyle \hfill 1 \hfill$&\cr
height2pt& \omit&& \omit&& \omit&& \omit&& \omit&& \omit&& \omit&& \omit&&
\omit&& \omit&& \omit&& \omit&& \omit&\cr
\noalign{\hrule}}}
$$
\vskip 15pt
\noindent{\bf Table II:} Rational curves on a cubic surface.
We display respectively the divisor classes, degree, self--intersection,
arithmetic genus, $N_a$, number of conjugates under space permutation,
order of the $G$--orbit.
\vskip 17pt
Table II exhibits $N_a$ for divisors in the form
$a^0>a^1\geq a^2\geq ...\geq a^6\geq 0$ up to degree $6$, 
with the number of distinct 
equivalent ones under space permutations recorded 
in the column before the last and
grouped in complete orbits under $G$. The last column yields 
the order of the orbit.
We observe that the $72$ divisors of degree $3$ and $p_a=0$, 
$N_a=1$ form
a single orbit of $G$. Indeed \CITZ\ they are in one to one 
correspondence with the
system $\cal R$ of $72$ non zero roots of $E_6$ ($\dim(E_6)=72+6=78$)
${\cal R}=\{ a \in \IZ^7 | \omega.a=0, a.a=-2\}$, and
$\Delta_{3,0}={\cal R}+\omega$.
Some numbers are recognizable.
For instance
for $a=\omega$, $d_a=3$ interpreted in the plane as uninodal cubics through 
$6+(3-1)=8$ points, we find $N_a=12$ as in section 2. Similarly for 
$a=\{4,1,1,1,1,1,1\}$, $d_a=6$ interpreted in the plane as $3$--nodal quartics
through $6+(6-1)=11$ points, we recover $N_a=620$ again as in section 2.

One could similarly discuss other Del Pezzo surfaces in particular those
with intersection form on divisor classes invariant under the Weyl
groups of $E_7$ and $E_8$.

\subsec{Lines in $\IP_3$.}

As an example of Grassmannian we shall content ourselves to
consider the $4$ dimensional Pl\"ucker quadric in $\IP_5$, 
corresponding to the variety of lines in $\IP_3$. In physicist's 
notation, this is also the (complex) electromagnetic fields up to
scale $F\equiv (\vec{E},\vec{B})$, such that $\vec{E}.\vec{B}=0$.
If a line in $\IP_3$ is given by $2$ points with homogeneous coordinates
$(x_0,x_1,x_2,x_3)$ and $(y_0,y_1,y_2,y_3)$, the coordinates of the line are
$F_{\mu \nu}=x_\mu y_\nu -x_\nu y_\mu$, $0\leq \mu < \nu \leq 3$, i.e.
$E_i=F_{0i}$ and $B_i=F_{jk}$, where $i,j,k$ is a cyclic permutation
of $1,2,3$. One checks that $\vec{E}.\vec{B}=0$ is the necessary and sufficient
condition that
$F$ lies on the Pl\"ucker quadric $Q$ in $\IP_5$ \HAR.
The cohomology ring is spanned by $t_0$, $t_1$, $t_{2a}$, $t_{2b}$, $t_3$ 
and $t_4$ dual to divisors given in the table below, where we first give an 
interpretation of the cycles as lines in $\IP_3$, then as points on the quadric $Q$.

%
$$\vbox{\offinterlineskip
\halign{\tv\quad # & \quad\tv \quad 
# & \quad \tv \quad  # & \quad \tv \quad # & \quad \tv #\cr 
\noalign{\hrule}
\tvi \  & dim. & \hfill lines in $\IP_3$ \hfill & \hfill points on $Q$
\hfill &\cr
\noalign{\hrule}
\tvi $t_0$ &$4$ & \hfill {all lines}\hfill 
&\hfill { $Q$  }\hfill &\cr
\tvi  $t_1$ &$3$ &  \hfill { lines meeting a line}\hfill  
&\hfill { hyperplane sect. of $Q$ }\hfill &\cr
\tvi $t_{2a}$ &$2$ &  \hfill { lines through a point}\hfill  
& \hfill { $\alpha$--plane $\subset Q$}\hfill &\cr
\tvi $t_{2b}$ &$2$ &   
\hfill {lines in a plane}\hfill  
& \hfill {$\beta$--plane $\subset Q$}\hfill &\cr
\tvi  $t_3$ &$1$ & 
\hfill {lines in a plane through a point}\hfill  
&  \hfill {line $\subset Q$}\hfill &\cr
\tvi  $t_4$ &$0$ &  
\hfill { fixed line }\hfill &  \hfill {point $\in Q$}\hfill &\cr
\noalign{\hrule} }} $$
\noindent{\bf Table III:} Homology cycles for the lines of $\IP_3$. We display 
the cohomology basis, dimension of the divisor,
and the dual cycles respectively as lines in $\IP_3$, and on
the Pl\"ucker quadric in $\IP_5$.
\vskip 1cm
%
The classical ring with unit $t_0$ is given by the relations
\eqn\claco{
\eqalign{t_1^2&=t_{2a}+t_{2b} \cr
t_1t_{2a}&=t_1t_{2b}=t_3\cr
t_1 t_3&=t_{2a}^2=t_{2b}^2=t_4 \ ,\cr}
}
all other products being $0$. Consequently the non vanishing elements of the intersection
form are
$\eta_{04}=\eta_{13}=\eta_{2a,2a}=\eta_{2b,2b}=1$. We introduce deformation 
parameters $y_0$, $y_1$, $y_{2a}$, $y_{2b}$, $y_3$ and $y_4$ and the
genus $0$ free energy $F$, with weights
\eqn\weipl{[y_0]=1 \ ; \ [e^{y_1}]=4 \ ; \ [y_{2a}]=[y_{2b}]=-1 \ ; \ 
[y_3]=-2 \ ; \ [y_4]=-3 \ ; \ [F]=-1 \ .}
As usual the free energy is split into $F=f_{\rm cl}+f$ with
\eqn\splipl{
\eqalign{
f_{\rm cl}&= {1 \over 2} y_0(y_0 y_4+y_{2a}^2+y_{2b}^2)+y_0y_1y_3+
{1 \over 2}y_1^2(y_{2a}+y_{2b}) \cr
f&= \sum_{\ga,\gb,\gc,\gd \geq 0 \atop \ga+\gb+2\gc+3\gd=4d+1}
N(\ga,\gb,\gc,\gd|d) {y_{2a}^\ga \over \ga !}{y_{2b}^\gb \over \gb !}
{y_3^\gc \over \gc !}{y_4^\gd \over \gd !} e^{d y_1} \ .\cr}
}
To confirm this assignment of weights, we
note that there is a single linear pencil of lines ($d=1$), i.e. a single
set of lines in a plane through a point, containing a fixed line $l_0$ 
and intersecting another generic such pencil in one line.
Indeed the former is the set of lines through a point on $l_0$ contained
in a plane through $l_0$. To intersect a pencil given by the set of lines 
in a plane $\pi_0$ through a point $p_0 \in \pi_0$, in a single line $l_1$,
we consider the point $p_1=\pi_0.l_0$, the line $l_1=[p_0,p_1]$ and
the plane $\pi_1=[l_0,l_1]$, then the pencil looked for is the set of lines 
through $p_1$ in $\pi_1$. Hence we find $N(0,0,1,1|1)=1$, and 
obtain a first relation on the weights
$[F]=[y_3]+[y_4]+[e^{y_1}]$.

Curves on $Q$ correspond to ruled surfaces in $\IP_3$. The 
integers $N(\ga,\gb,\gc,\gd|d)$ are interpreted as counting the number of rational
ruled surfaces of degree $d$ subject to auxiliary relations dictated by the
exponents $\ga$, $\gb$, $\gc$, $\gd$ in terms of their ruling.
For instance we expect a single quadric ($\IP_1 \times \IP_1$) through 
$3$ lines in general position, i.e. $N(0,0,0,3|2)=1$ which implies
that $[F]=3[y_4]+2 [e^{y_1}]$. With the assignments $[y_0]=1$, $[y_{2a,b}]=-1$,
$[y_3]=-2$ and $[y_4]=-3$, these relations specify the weights
of $F$ and $e^{y_1}$ to be $-1$ and $4$ respectively.
We also have the obvious symmetry
$N(\ga,\gb,\gc,\gd|d)=N(\gb,\ga,\gc,\gd|d)$. 

We will not write explicitly the $15$
relations expressing the associativity of the 
deformed ring and their symmetric counterparts.
It is interesting to note however than only $5$ of them are sufficient to 
completely specify the numbers $N(\ga,\gb,\gc,\gd|d)$. We content ourselves with
the following

\bigskip
\noindent{\bf Proposition 9.}
For the Grassmannian of lines in $\IP_3$ the associativity conditions, together with 
the initial condition $N(0,0,1,1|1)=1$ determine recursively all $N(\ga,\gb,\gc,\gd|d)$.
The first few $N$'s for $d \leq 2$ read (we denote for short
$N(\ga,\gb,\gc,\gd|d)\equiv N(\ga,\gb,\gc,\gd)$)
\eqn\respl{
\eqalign{
\hbox{\bf d=1 : }\ \ 
N(5,0,0,0)&=0 \quad
N(4,1,0,0)=0 \quad
N(3,2,0,0)=1 \quad
N(3,0,1,0)=0 \cr
N(2,1,1,0)&=1 \quad
N(1,0,2,0)=1 \quad
N(2,0,0,1)=0 \quad
N(1,1,0,1)=1 \cr
N(0,0,1,1)&=1 \cr
\hbox{\bf d=2 : }\ \
N(9,0,0,0)&=2 \quad
N(8,1,0,0)=6 \quad
N(7,2,0,0)=18 \quad
N(6,3,0,0)=34\cr
N(5,4,0,0)&=42 \quad
N(7,0,1,0)=3 \quad
N(6,1,1,0)=9 \quad
N(5,2,1,0)=17\cr
N(4,3,1,0)&=21 \quad
N(5,0,2,0)=5 \quad
N(4,1,2,0)=9 \quad
N(3,2,2,0)=11\cr
N(3,0,3,0)&=5 \quad
N(2,1,3,0)=6 \quad
N(1,0,4,0)=3 \quad
N(6,0,0,1)=1\cr
N(5,1,0,1)&=3 \quad
N(4,2,0,1)=5 \quad
N(3,3,0,1)=5 \quad
N(4,0,1,1)=2\cr
N(3,1,1,1)&=3 \quad
N(2,2,1,1)=3 \quad
N(2,0,2,1)=2 \quad
N(1,1,2,1)=2\cr
N(0,0,3,1)&=1 \quad
N(3,0,0,2)=1 \quad
N(2,1,0,2)=1 \quad
N(1,0,1,2)=1\cr
N(0,0,0,3)&=1\cr}
}
We get a number of enumerative data on quadrics. We recover $N(0,0,0,3)=1$.
Also $N(9,0,0,0)=N(0,9,0,0)=2$ correspond to the fact that a quadric is 
uniquely fixed by $9$ points or dually by $9$ tangent planes, and that 
there are two rulings on such a quadric. Similarly there are $3$ quadrics
through $8$ points tangent to a plane. Indeed through $8$ points a linear
pencil of quadrics cuts a plane in a linear pencil of conics through $4$ points,
which degenerates in $3$ ways in a pair of lines corresponding to a 
quadric tangent to the plane. Since again a quadric has two rulings, this 
explains $N(8,1,0,0)=N(1,8,0,0)=2 \times 3=6$.
Similarly $N(7,0,1,0)=N(0,7,1,0)=3$ since now 
we have quadrics through $8$ points
tangent to a plane but one of the rulings is selected. One can similarly 
check part of our results for $d=2$ against table VI on page 329 of the
book by Semple and Roth \SR, namely there are
$9$ quadrics through $7$ points and tangent to $2$ planes 
($N(7,2,0,0)=2N(6,1,1,0)=18$), $17$ quadrics through $6$ points and tangent 
to $3$ planes ($N(6,3,0,0)=2N(5,2,1,0)=34$), and $21$ quadrics through $5$ 
points and tangent to $4$ planes ($N(5,4,0,0)=2N(4,3,1,0)=42$).

\bigskip
\noindent{\bf Remark.}
Consider the deformed ring when 
$y_0=y_{2a}=y_{2b}=y_3=y_4=0$. Setting $e^{y_1}=q^4$, 
the multiplication table 
of the deformed ring, with identity $T_0$, reduces to
\eqn\defmult{
\eqalign{
T_1^2&=T_{2a}+T_{2b} \quad , \quad T_1 T_{2a}=T_1 T_{2b}=T_3 \cr
T_1 T_3&= T_4+q^4 T_0 \quad , \quad T_1 T_4=q^4 T_1\cr
T_{2a}^2&=T_{2b}^2= T_4 \quad , \quad T_{2a} T_{2b}= q^4 T_0 \cr
T_{2a}T_3&=T_{2b} T_3=q^4 T_1 \quad , \quad T_{2a}  T_4=q^4 T_{2b} \cr
T_{2b}T_4&=q^4 T_{2a} \quad , \quad T_3^2=q^4 (T_{2a}+T_{2b}) \cr
T_3 T_4&= q^4 T_3 \quad , \quad T_4^2=q^8 T_0 \ ,\cr}
}
since the only surviving third derivatives of $f$ are 
$f_{134}=f_{2a,2b,4}=q^4$ and $f_{444}=q^8$.
Indeed in the sum over non negative exponents the condition $\ga+\gb+2\gc+3\gd=4d+1$
entails that for $d>2$, $\ga+\gb+\gc+\gd>3$, and for $d=2$ the only possibility
for $\ga+\gb+\gc+\gd \leq 3$ is $\ga=\gb=\gc=0$, and $\gd=3$.

The intermediate ring \defmult\ for the Grassmannian $G(2,4)$ of lines
in $\IP_3$ was considered by Witten \WG, elaborating 
earlier work by Gepner, in relation to Verlinde's formula. 
It is identified as the ring of 
symmetric functions in two variables quotiented by the gradient of a symmetric
polynomial as follows. Let the two variables be $\gl_1$, $\gl_2$, and 
$s=\gl_1+\gl_2$, $p=\gl_1 \gl_2$.
The polynomial ring of symmetric functions in $\gl_1$, $\gl_2$ is $\IC[s,p]$.
With $W \in \IC[s,p]$ defined as
\eqn\potgra{ W(s,p)= {\gl_1^5+\gl_2^5 \over 5} +q^4 (\gl_1+\gl_2) \ .}
the corresponding Gepner--Witten ring is $\IC[s,p]/{\rm grad} W(s,p)$.
The gradient conditions read
\eqn\gracond{
\eqalign{
{\partial W \over \partial s}&= {\gl_1^5 -\gl_2^5 \over \gl_1 -\gl_2}+q^4=
s^4-3ps^2+p^2+q^4\cr
{\partial W \over \partial p}&={\gl_1^4 -\gl_2^4 \over \gl_1 -\gl_2}=
s^3-2ps \cr}
}
So we can also write $\IC[s,p]/\{s^4-3ps^2+p^2+q^4,s^3-2ps\}$. When $q=0$, the
Hilbert polynomial for this ring is 
\eqn\poin{ {(1-t^3)(1-t^4) \over (1-t)(1-t^2)}=1+t+2t^2+t^3+t^4}
indeed the same as for the classical Grassmannian. The correspondence
\eqn\corresp{
\eqalign{
T_0 \leftrightarrow 1 \ \ ; \ \ &T_1\leftrightarrow s \ \ ; \ \ 
T_{2a}\leftrightarrow s^2-p \cr
T_{2b}\leftrightarrow p \ \ ; \ \ &T_3\leftrightarrow sp \ \ ; \ \ 
T_4\leftrightarrow p^4 \cr}
}
identifies the two rings as the reader can easily check (the roles
of $T_{2a}$ and $T_{2b}$ could be permuted).

\subsec{Flags.}

The flag manifold ${\cal  F}_n$ in $\IP_n$ is the set 
of sequences formed by a point on a line in a plane ... Viewing $\IP_n$ as
$\IC^{n+1}-\{ 0\} / \IC^*$, it may be identified either as $GL_{n+1}/B_{n+1}$,
quotient of the linear group by the Borel subgroup 
of upper triangular matrices,
which exhibits ${\cal F}_n$ as a complex variety, or equivalently as
$U_{n+1}/U_1^{n+1}$ proving that it is compact, with complex dimension
dim$({\cal F}_n)=n(n+1)/2$. Its intersection ring is generated by divisors $u_k$,
$k=1,...,n$, corresponding to the $n$ cycles of codimension $1$,
${\cal C}_k \equiv$ {\it the linear $(k-1)$ subvariety in the flag intersects a fixed linear subvariety of codimension $k$}.
For the following presentation see the book of Macdonald \MAC, 
and references therein.
Set $u_0=u_{n+1}=0$.
It is convenient to introduce $x_r=u_{r+1}-u_r$ for $r=0,1,...,n$ with
the relation $\sum_{0 \leq r \leq n} x_r=0$, or 
equivalently $u_k=\sum_{0 \leq r \leq k-1} x_r$ for $k=1,2,...,n$.
Let $\IC[x_0,...,x_n]$ be the (graded) polynomial ring on $x_0$, ..., $x_n$,
with $x_k$ of grade $1$, and ${\cal S}[x_0,...,x_n]$ the ideal 
generated by the elementary symmetric functions in $x_0$, ..., $x_n$, 
\eqn\cohoflag{H^*({\cal F}_n)=\IC[x_0,...,x_n]/{\cal S}[x_0,...,x_n] \ .}
As a $\IZ$--module a basis is given by the so--called ``Schubert polynomials"
indexed by permutations of $(n+1)$ objects (group $S_{n+1}$). 
The Hilbert polynomial
counting the number of elements in the ring of given degree is
\eqn\hilfla{
\eqalign{
P(x)&= {1 \over (1-x)^{n+1}} (1-x)(1-x^2)...(1-x^{n+1}) \cr
&=(1+x)(1+x+x^2)...(1+x+x^2+....+x^n) \ ,\cr}
}
expressing the fact that we have one relation in each degree from $1$
to $(n+1)$ (given by the elementary symmetric functions in $x_0$, ..., $x_n$),
and showing that the dimension of the ring is $(n+1)!$.
Let $p(x_0,...,x_n)$ and $q(x_0,...,x_n)$ be two 
polynomials representative of classes in $H^*$ of eq.\cohoflag.
The group $S_{n+1}$ acts by permuting the arguments, 
and this action is still well  
defined on the quotient \cohoflag. The intersection
form on the ring is then obtained as 
\eqn\intfla{
\langle p,q \rangle=\oint {dx_0 \over 2i \pi x_0} ... \oint
{dx_n \over 2i \pi x_n}
{A[pq](x_0,...,x_n) \over \prod_{0 \leq i <j \leq n} (x_i -x_j)} \ , }
where, for any polynomial $r(x_0,...,x_n)$, $A[r]$ denotes
the antisymmetrized polynomial 
$A[r](x_0,...,x_n)\equiv \sum_{\gs \in S_{n+1}} \ge(\gs) 
r(x_{\gs(0)},...,x_{\gs(n)})$, $\ge(\gs)$ the signature of the
permutation $\gs$. The integrals in \intfla\ are over the unit circle.
For instance, a representative of top degree $n(n+1)/2$ is
$p_{\rm max}=x_0^n x_1^{n-1} x_2^{n-2} ...x_n^0$, while a representative of lowest 
($0$) degree is $1$, with $\langle 1,p_{max}\rangle=1$.
Rather than developing the general theory let us now concentrate on 
the first non--trivial case i.e. the flags of $\IP_2$, in which case 
we have the generators

$u_1=x_0$ : {\it the point lies on a fixed line},

$u_2=x_0+x_1$ : {\it the line goes through a fixed point},

\noindent{}and one adds $x_2$ such that $x_0+x_1+x_2=0$. 

\fig{Cycles in the flags of $\IP_2$. Each drawing is
indexed by the corresponding dual cohomology class. Arrows indicate
the freedom of the flag cycles.}{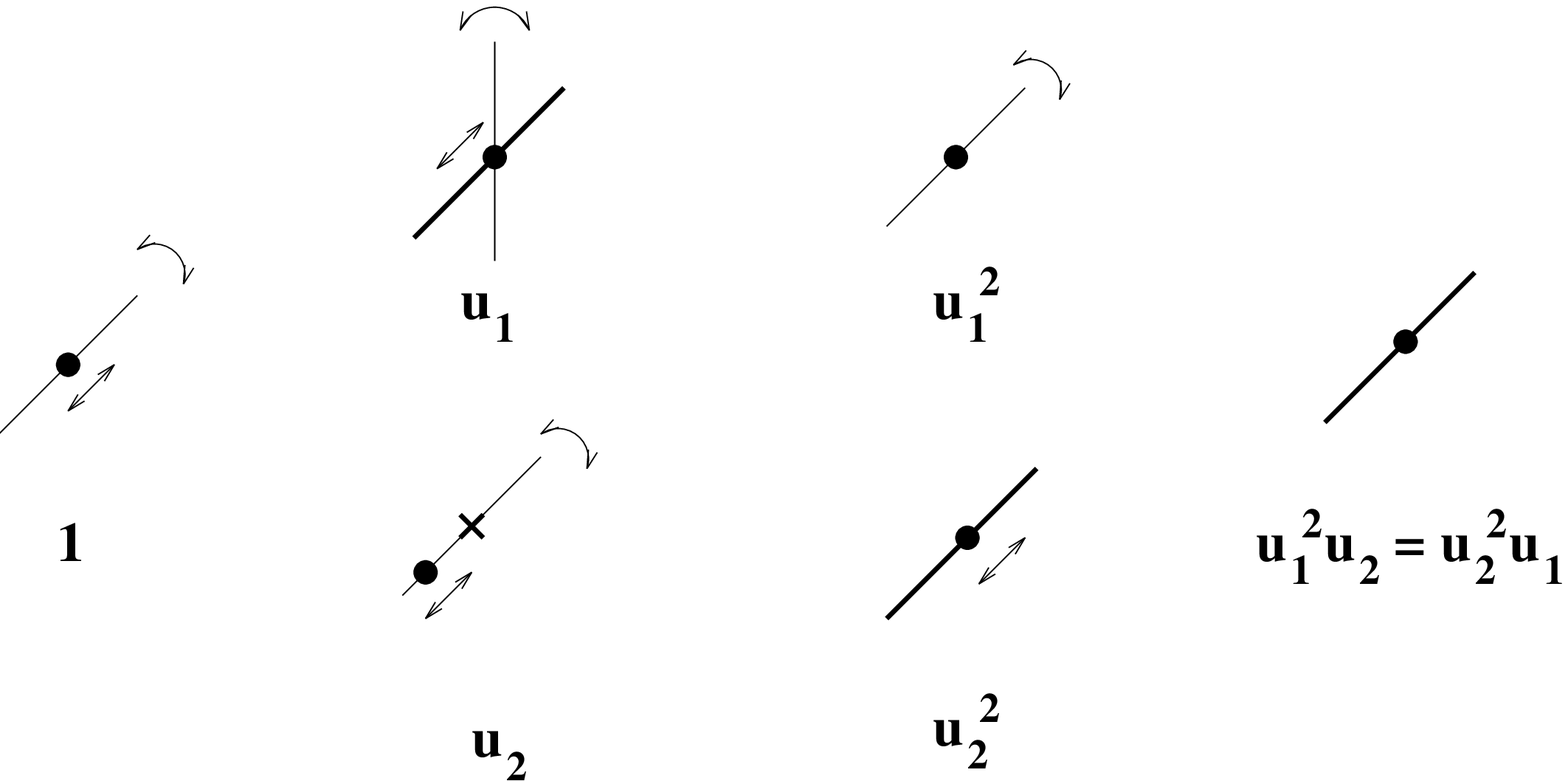}{10cm}
\figlabel\flags

The number of elements
in the ring $\IC[x_0,x_1,x_2]/{\cal S}[x_0,x_1,x_2]$ is $6$,
with basis
\eqn\basfla{
1  \qquad \matrix{x_0 \cr x_0+x_1} \qquad \matrix{x_0^2\cr x_0x_1\cr}
\qquad x_0^2 x_1 }
i.e., as depicted on Fig.\flags,
\eqn\flabas{
1 \left\{\matrix{{\it any}\cr {\it flag} \cr}\right\}\quad 
\matrix{ u_1 & 
\left\{\matrix{\hbox{\it the point lies}\cr 
\hbox{\it on a line}\cr}\right\} 
&u_1^2 & \{ {\it the}\ {\it point}\ {\it is}\ {\it fixed} \ \}\cr
u_2 & \left\{\matrix{\hbox{\it the line goes}\cr 
\hbox{\it through a point}\cr}\right\} 
& u_2^2 & \{ {\it the}\ {\it line} \ {\it is}\  {\it fixed} \} \cr} \quad
u_1^2 u_2 \ \{ {\it the}\  {\it flag}\ {\it is}\ {\it fixed} \} }
The relations read
\eqn\relaax{ x_0+x_1+x_2=0 \ \ ; \ \ x_0x_1+x_1x_2+x_2x_0=0 \ \ ; \ \ 
x_0x_1x_2=0 }
or eliminating $x_2$
\eqn\aaxrel{x_0^2+x_0x_1+x_1^2=0 \ \ \ \hbox{and} \ \ \ x_0^2x_1+x_1^2 x_0=0,}
which entails $x_0^3=x_1^3=x_2^3=0$. Expressed in $u_1$, $u_2$,
this yields
\eqn\relau{ u_1 u_2=u_1^2+u_2^2 \qquad u_1^2u_2=u_2^2 u_1 }
and therefore $u_1^3=u_2^3=0$. The non--vanishing intersections are
\eqn\nvint{
\eqalign{
\langle 1 , x_0^2 x_1 \rangle =&\langle x_0 , x_0 x_1 \rangle =
\langle x_0+x_1 , x_0^2 \rangle =1 \cr
\langle 1 , u_1^2 u_2 \rangle &=\langle u_2 , u_1^2 \rangle =
\langle u_1 , u_2^2 \rangle =1\ ,\cr}
}
which are readily checked geometrically, 
using the representations of Fig.\flags.

The flag manifold of $\IP_2$ is $3$ dimensional and can also be
described by the incidence relation of points and lines in
$\IP_2\times {}^*\IP_2$ (where ${}^*\IP_2$ stands for the dual plane).
If $x,y,z$ (resp. $X,Y,Z$) stand for the coordinates in $\IP_2$
(resp. ${}^*\IP_2$), we have a ``quadric" in $\IP_2\times {}^*\IP_2$
of equation $xX+yY+zZ=0$.
Rational curves in this space have a bidegree $(a,b)$ for their
projections on the two planes. Such a curve is described by three
homogeneous polynomials of degree $a$ (resp. $b$) in two variables $(u,v)$
parametrizing $\IP_1$, namely $x(u,v)$, 
$y(u,v)$ and $z(u,v)$ (resp. $X(u,v)$, $Y(u,v)$, $Z(u,v)$), modulo the 
action of $PSL_2$ on each triple of polynomials, and subject to the
incidence condition $xX+yY+zZ=0$. The number of free parameters is therefore
\eqn\flapa{[3(a+1)-3]+[3(b+1)-3]-[(a+b+1)-1]=2(a+b) \ .}
To fix the curve we therefore need $2(a+b)$ conditions.
Intersection with a codimension $2$ cycle in the classes 
$u_1^2$ or $u_2^2$ counts as one condition, 
whereas intersection with a codimension $3$ cycle (in the class $u_1^2 u_2$)
i.e. a fixed flag counts for two conditions. 
Using these remarks, one can generate the deformed ring. With obvious 
notations, let  
$T_0$, $T_{1a}$, $T_{1b}$, $T_{2a}$, $T_{2b}$ 
and $T_3$ correspond to the deformations of $1$, $u_1$, $u_2$, 
$u_1^2$, $u_2^2$ and $u_1^2 u_2=u_2^2 u_1$ respectively. 
The associated parameters $y_0$, $y_{1a}$,
$y_{1b}$, $y_{2a}$, $y_{2b}$ and $y_3$ and the genus $0$ free
energy $F$ have weights
\eqn\weifla{[y_0]=1 \ \ ; \ \ [e^{y_{1a}}]=[e^{y_{1b}}]=2 \ \ ; \ \ 
[y_{2a}]=[y_{2b}]=-1 \ \ ; \ \ [y_3]=-2 \ \ ; \ \ [F]=0 \ . }
{}From the preceding, we have $F=f_{\rm cl}+f$, with
\eqn\splifla{
\eqalign{
f_{\rm cl}&= {y_0^2 \over 2} y_3 + y_0(y_{1a}y_{2b}+y_{1b}y_{2a})+
{1 \over 2} y_{1a}y_{1b}(y_{1a}+y_{1b}) \cr
f&= \sum_{\gl+\mu+2 \nu=2(a+b) \atop
\gl,\mu,\nu,a,b \geq 0 ; a+b \geq 1} N(\gl,\mu,\nu|a,b)
{y_{2a}^\gl \over \gl !}{y_{2b}^\mu \over \mu !}{y_3^\nu \over \nu !}
e^{a y_{1a}+b y_{1b}} \cr}
}
Obviously $N(\gl,\mu,\nu|a,b)=N(\mu,\gl,\nu|b,a)$, reflecting the duality 
between point and line. 

A curve in the flag space such that the line
is fixed intersects a generic cycle in the class $u_1$ in a single 
flag, and does not intersect a generic cycle in $u_2$. Hence its bidegree is
$(1,0)$. It is fixed if we require that it contains a fixed flag (which
fixes the line), and obviously does not intersect generic cycles in the classes 
$u_1^2$ or $u_2^2$, hence
\eqn\initfla{ N(0,0,1|1,0)=N(0,0,1|0,1)=1 \ .}
Similarly a curve of bidegree $(1,0)$ required to intersect two cycles in 
the class $u_1^2$, which fixes $2$ distinct points, hence the line,
is uniquely determined, hence
\eqn\flainit{ N(2,0,0|1,0)=N(0,2,0|0,1)=1 \ .}
We refrain again from writing explicitly all associativity conditions for the 
deformed intersection ring. We checked up to $a+b \leq 5$ that they 
were all consistent and claim

\bigskip
\noindent{\bf Proposition 10.}
Either initial condition \initfla\ or \flainit, together with the
associativity conditions, determine uniquely the 
integers $N(\gl,\mu,\nu|a,b)$, and one finds ($d=a+b$)
\eqn\resfla{
\eqalign{
\hbox{\bf d=1 : }
N(2,0,0\vert 1,0)&=1 \quad 
N(0,0,1\vert 1,0)=1 \cr
\hbox{\bf d=2 : }
N(0,0,2\vert 1,1)&=1 \quad 
N(1,1,1\vert 1,1)=1 \quad
N(2,2,0\vert 1,1)=1\cr
\hbox{\bf d=3 : }
N(2,0,2\vert 2,1)&=1 \quad 
N(4,0,1\vert 2,1)=1  \quad 
N(3,1,1\vert 2,1)=1 \quad 
N(5,1,0\vert 2,1)=2  \cr
N(4,2,0\vert 2,1)&=1\cr
\hbox{\bf d=4 : }
N(4,0,2\vert 3,1)&=1 \quad 
N(6,0,1\vert 3,1)=4  \quad
N(5,1,1\vert 3,1)=1 \quad 
N(8,0,0\vert 3,1)=12 \cr
N(7,1,0\vert 3,1)&=6 \quad 
N(6,2,0\vert 3,1)=1  \quad
N(0,0,4\vert 2,2)=1 \quad 
N(2,0,3\vert 2,2)=1 \cr
N(1,1,3\vert 2,2)&=2 \quad 
N(3,1,2\vert 2,2)=2  \quad
N(2,2,2\vert 2,2)=3 \quad 
N(4,2,1\vert 2,2)=4 \cr
N(3,3,1\vert 2,2)&=5\quad 
N(5,3,0\vert 2,2)=8  \quad
N(4,4,0\vert 2,2)=10\cr
\hbox{\bf d=5 : }
N(10,0,0\vert 4,1)&=60 \quad
N(9,1,0\vert 4,1)=12  \quad
N(8,2,0\vert 4,1)=1 \quad
N(8,0,1\vert 4,1)=9 \cr
N(7,1,1\vert 4,1)&=1 \quad
N(6,0,2\vert 4,1)=1  \quad
N(8,2,0\vert 3,2)=108 \quad
N(7,3,0\vert 3,2)=150 \cr
N(6,4,0\vert 3,2)&=96 \quad
N(5,5,0\vert 3,2)=32  \quad
N(7,1,1\vert 3,2)=36 \quad
N(6,2,1\vert 3,2)=60 \cr
N(5,3,1\vert 3,2)&=40 \quad
N(4,4,1\vert 3,2)=16  \quad
N(6,0,2\vert 3,2)=12 \quad
N(5,1,2\vert 3,2)=26 \cr
N(4,2,2\vert 3,2)&=18 \quad
N(3,3,2\vert 3,2)=8  \quad
N(4,0,3\vert 3,2)=12 \quad
N(3,1,3\vert 3,2)=9 \cr
N(2,2,3\vert 3,2)&=4 \quad
N(2,0,4\vert 3,2)=5  \quad
N(1,1,4\vert 3,2)=2  \quad
N(0,0,5\vert 3,2)=1 \cr }
}
Apart from the two simple sets of initial conditions \initfla\--\flainit,
we recognize a few numbers. Take for instance $N(8,0,0|3,1)=12$.
We deal with a curve in flag space intersecting in $3$ flags a cycle in the class $u_1$. Clearly the projection from our curve to 
the first $\IP_2$ should be a uninodal cubic (since our curves are rational).
Moreover, this cubic should pass through $8$ points ($\gl=8$).
Now the curve in flag space should be such that as the point varies on
the cubic with accompanying line, it intersects in a single flag a cycle
in the class $u_2$. This forces the line to pass through the node of
the cubic. Indeed given a cycle in the class $u_2$, with lines through a
fixed point of the plane, we join this point to the node to obtain the
line of the desired flag, its point being the only
other intersection of this line with the cubic.
Finally it is readily seen that such a family does not intersect
a generic cycle in either $u_1^2$ (the point is fixed) or $u_2^2$ (the line
is fixed), hence $\gl=\mu=0$, a fortiori it does not pass through an
arbitrarily fixed flag, $\gc=0$. Now we know from
section 2 that there are $N_3=12$ uninodal cubics
through $8$ generic points of the plane, confirming $N(8,0,0|3,1)=12$.

Finally, following the remark at the end of the previous subsection,
we exhibit the deformed ring multiplication table, when restricted 
to the plane $y_0=y_{2a}=y_{2b}=y_3=0$, setting
$e^{y_{1a}}=q_a^2$ and $e^{y_{1b}}=q_b^2$, and with unit $T_0$
\eqn\lawfla{
\eqalign{
T_{1a}^2&=T_{2a}+q_a^2 T_0 \ \ ; \ \ T_{1a} T_{1b}=T_{2a}+T_{2b} \ \ 
; \ \ T_{1a}T_{2a}=q_a^2 T_{1b} \cr
T_{1a}T_{2b}&=T_{1b}T_{2a}= T_3 \ \ ; 
\ \ T_{1a} T_3=q_a^2 T_{2b}+q_a^2q_b^2 T_0\ \ 
; \ \ T_{1b}^2=T_{2b}+q_b^2 T_0\cr
T_{1b}T_{2b}&=q_b^2 T_{1a} \ \ ; \ \ T_{1b}T_3=q_b^2 T_{2a}+q_a^2q_b^2 T_0
\ \ ; \ \ T_{2a}^2=q_a^2 T_{2b}\cr 
T_{2a}T_{2b}&=q_a^2 q_b^2 T_0\ \ ; \ \ 
T_{2a}T_3=q_a^2 q_b^2 T_{1a} \ \ ; \ \ T_{2b}^2=q_b^2 T_{2a}\cr
T_{2b}T_3&=q_a^2 q_b^2 T_{1b} \ \ ; \ \ 
T_3^2=q_a^2 q_b^2(T_{2b}+T_{2a}) \ .\cr}
}
This reduced ring may be interpreted as $\IC[u_1,u_2]/{\cal I}$,
where the ideal ${\cal I}$ is generated by the polynomials
\eqn\consi{ 
\eqalign{
&u_1^2+u_2^2-u_1u_2-q_a^2-q_b^2 \cr
&u_1u_2^2-u_2u_1^2-q_b^2 u_1+q_a^2 u_2 \cr}
}
with the identifications
\eqn\idenfla{
\eqalign{
T_{1a} &\leftrightarrow u_1 \cr
T_{1b} &\leftrightarrow u_2  \cr
T_{2a} &\leftrightarrow u_1^2-q_a^2 \cr
T_{2b} &\leftrightarrow u_2^2-q_b^2\cr
T_3 &\leftrightarrow u_1u_2^2-q_b^2 u_1\equiv u_2u_1^2 -q_a^2 u_2 \ .\cr}
}
Unfortunately $T_{1a}$ and $T_{1b}$ satisfy now sixth degree equations 
(instead of a third degree one in the classical case $u_1^3=u_2^3=0$)
\eqn\sixfla{ (u_1^2-q_a^2)^3=q_a^4 q_b^2 \qquad (u_2^2-q_b^2)^3=q_a^2 q_b^4}
and therefore it is impossible 
(even for $q_a=q_b$) to write it as a deformation 
of $\IC[x_0,x_1,x_2]/{\cal S}[x_0,x_1,x_2]$, insisting that the
elementary symmetric functions of $x_0,x_1,x_2$ take assigned values.
Rather the ideal $\cal I$ of \consi\ is the one of the six points
of intersection of a conic and a cubic in the affine plane $(u_1,u_2)$,
with coordinates
\eqn\cococu{ u_1=q_a \big[ 1+ \big({q_b^2/q_a^2}\big)^{1/3} \big]^{1/2} \quad
u_2=u_1\big( {u_1^2 \over q_a^2} -1 \big) .}
The six determinations arise from the cubic and square roots, while
$u_2$ is rational in $u_1$. Thus the reduced ring is also of the form
$\IC[x]/P(x)$ with
\eqn\ideap{ P(x)=(x^2-q_a^2)^3 -q_a^4 q_b^2 }
and
\eqn\subi{ \eqalign{ u_1&\to x \cr
u_2 &\to x \big( {x^2 \over q_a^2} -1 \big) \cr}}
or equivalently
\eqn\subit{ 
\eqalign{
x_0&\to x \cr
x_1&\to x\big( {x^2 \over q_a^2} -2 \big) \cr
x_2&\to x \big({x^2 \over q_a^2} -1 \big) \ , \cr}}
i.e. a specific deformation of the $A_5$ singularity.
The six points in \cococu\ are not in general position
since they lie on a conic, i.e. a rational curve, explaining
why the ring can be expressed in terms of the roots of a sixth--degree
polynomial in a single variable.

\newsec{Comments and questions}

In this (in)conclusive section, we add a few remarks, questions and results
from the literature.

\subsec{Manifolds with rational curves}

In a systematic discussion we should have commented on the type of generic
projective varieties for which one would expect to enumerate rational curves.
For simplicity we restrict ourselves to irreducible 
hypersurfaces $X$ of degree $d$ in $\IP_n$.
It is known \HAR\ that the variety ${\cal F}_k(X)$ of $k$--planes
in $X$ has dimension
\eqn\hyperdim{ dim {\cal F}_k(X) = (n-k)(k+1) -{ d+k \choose k}} 
More precisely for $d>2$, when this number is non negative,
it is the correct dimension. There are exceptions for $d=2$. Applying this to lines,
$k=1$, we get 
\eqn\superdim{ dim {\cal F}_1(X) = 2n-3-d} 
So one expects to find lines up to degree $d=2n-3$, i.e. up to cubics
($27$ lines) in $\IP_3$, or quintics ($2875$ lines) for 
threefolds in $\IP_4$, the famous example for mirror symmetry, etc...
A general formula for the number $L_d$ of lines on a generic hypersurface of
degree $2n-3$ in $\IP_n$ is due to Harris \HH\ 
\eqn\harifor{L_d=d \times  
d! \sum_{k=0}^{n-2} {(2k)! \over (k+1)! k!} \sum_{I_k}
\prod_{i \in I_k} {(d-2i)^2 \over i(d-i)}} 
where $I_k$ runs over subsets of $\{1,2,...,n-2\}$ with $n-k-2$ elements
and an empty product is equal to $1$.

The above does not mean that smooth hypersurfaces in $\IP_n$ of
degree larger than $2n-3$ do not possess lines, rather such
hypersurfaces form a positive codimension submanifold in the space of 
degree $2n-3$ hypersurfaces (i.e. they are not generic).
On the other hand generic hypersurfaces of higher degree 
might still possess rational curves (but these will not be lines).

As an example take the generic smooth quartic surface in $\IP_3$, a typical
case of a $K_3$ surface. It is known that all curves on such a surface
are complete intersections with another surface --  hence, by Bezout, have
degrees multiple of $4$. The dimension of the space of curves of 
degree $4d$ is then expected to be $h_d-1$, where $h_d$ is the 
dimension of the space of homogeneous polynomials in $4$ variables modulo
those that vanish on the quartic $X$. A generating function is
\eqn\formular{
\sum_{d=0}^\infty h_d \lambda^d = {1-\lambda^4 \over (1-\lambda)^4} 
= 1+2 \sum_{d=0}^\infty (1+d^2) \lambda^d} 
For $d \geq 1$ we then have a space of curves of degree $4d$ on $X$ of 
dimension $2d^2+1$. The arithmetic genus of such a curve $C$ 
(i.e. disregarding possible singularities) is
\eqn\genusp{ p = 1+{C.C +C.K \over 2}}
with the canonical divisor $K$ vanishing for a $K_3$ surface, while
$C.C=4d^2$ (the value $C.C'$, where $C$ and $C'$ are intersections
with two surfaces $Y$ and $Y'$ of degree $d$, reads $C.C'=4\times d\times d$ 
by Bezout).
Consequently $p=1+2d^2$. We conclude (with Kontsevich) that requiring $C$
to be rational, i.e. imposing $1+2d^2$ conditions (generically to have
$1+2d^2$ double points) in the space of the same dimension, leads to
expect finitely many rational curves in each degree $4d$.
In the first non trivial instance on a smooth quartic surface in $\IP_3$ there
are $3200$ rational (plane, trinodal)  quartics cut by $3$--tangent 
planes according to the classical formula of Salmon for the number of tritangent 
planes to a smooth surface of degree $r$
\eqn\salmontrigo{
tritg(r)={1 \over 6} (r^9-6r^8+15r^7-59r^6+204r^5-339r^4+770r^3
-2056r^2+1920r)} 
evaluated at $r=4$. To extend this to curves of higher degree on a quartic
is apparently an open problem. For instance in degree $8$ one is required 
to count $9$--fold tangent quadrics to $X$!

To emphasize the point, non generic quartic surfaces
(indeed a codimension $1$ in a $19$ dimensional space) do possess lines,
as examplified by the Fermat quartic, $\sum_{0 \leq i \leq 3} x_i^4=0$,
on which we find at least $48$ lines of the type 
$\lambda(1,\omega,0,0)+\mu (0,0,1,\omega')$ with $\omega^4=\omega'^4=-1$.
In general the most naive reasoning goes as follows. Using homogeneous
coordinates in $\IP_1$ a parametrized curve of degree $k$ in $\IP_n$
depends on $(n+1)(k+1)$ parameters.
The constraint to lie on a degree $d$ hypersurface implies the vanishing 
of a homogeneous polynomial in the two coordinates, of degree $kd$,
hence $kd+1$ conditions. If the difference
\eqn\peticalc{ (n+1)(k+1) -(kd+1) = k(n+1-d) +n}
is larger or equal to $4$ (the arbitrariness in the parametrization)
one expects rational curves (this extends readily to complete intersections).
A marginal case occurs when $d=n+1$ (so that $k$ does not matter).
If moreover $n$ is precisely equal to $4$, i.e. for the 
famous quintic threefold in $\IP_4$, the expectation 
(Clemens) is the existence of finitely many rational curves 
of any degree. Conjecturally mirror symmetry enables one to compute
these numbers in any degree $k$ (and more).

On the other hand for $n=3$ and $d=4$ the naive reasoning fails for 
$k$ a multiple of $4$ in which case one of the conditions must be redundant, while if $d=2$ the counting of degrees of freedom 
(i.e. subtracting $4$ from \peticalc\ ) yields $2k-1$ in agreement with 
equation \spliplu\ where $k=a+b$, and when $d=3$ one gets
$k-1$, again in agreement 
with equation \frencu\ where $k=d_a$.

\subsec{Parallel with matrix models}

Topological field theories assumed to underlie this paper are generalizations
of those arising from matrix models of $2D$--quantum gravity as interpreted
and elaborated in references \Wun\ \DVV\ \KM\ \KQG\ \IZQG.
Similar formulas as those presented above also hold in these cases for
the genus zero ``little phase space" free energies.
Take for instance the generalized Airy matrix integrals which we denote, 
as models, by $W_{n+2}$, $n \geq 0$, the
original case being $W_2$ with reference to the corresponding 
$W$--algebra constraints satisfied by the partition 
function\foot{
One also finds the notation $A_{n+1}$ from singularity theory.
After the parallel we will make between $W_{n+2}$ models and 
enumerations in $\IP_n$,
this denomination raises the following question: what are the 
``natural" families of targets in the enumerative context, corresponding to 
the other $D$ and $E$ series of singularities and matrix models if any?}

Using for $W_{n+2}$ models similar notations as for the $\IP_n$ case
of this paper, let us denote the little phase space variables $y_i$,
$i=0,1,...,n$, and the genus zero free energy $F$, with weights
\eqn\weighw{ [y_i]=1-{i \over n+2} \qquad [F]=3-{n \over n+2}} 
to be compared with those in the $\IP_n$ case
\eqn\weighip{ [y_i]=1-i \qquad [F]=3-n} 
That all weights are positive implies that $F$ is a polynomial in the 
$W_{n+2}$ case.
It splits as
$F= f_{\rm cl}+f$
where $f_{\rm cl}$ is the same cubic polynomial as in the $\IP_n$ case
\eqn\claf{ f_{\rm cl} = {1 \over 3!} 
\sum_{i_1+i_2+i_3=n} y_{i_1} y_{i_2} y_{i_3} }
The quantum part $f$ contains finitely many higher degree terms in the $y$'s.
{}From unpublished work by J.-B. Zuber (whom we take this opportunity to thank)
we extract the following table up to $W_6$
\eqn\zubeq{
\eqalign{ W_2 (n=0) \ &: \  f=0 \cr
W_3 (n=1) \ &: \ f={1 \over 3} {y_1^4 \over 4!} \cr
W_4 (n=2) \ &: \ f={1 \over 4} {y_1^2 \over 2!}{y_2^2 \over 2!}+
{1 \over 8}{y_2^5 \over 5!} \cr
W_5 (n=3) \ &: \ f={1 \over 5} \big[ {y_1^2 \over 2!}{y_3^2 \over 2!} 
+y_1 {y_2^2 \over 2!}y_3 +2 {y_2^4 \over 4!} \big]
+{2 \over 5^2}{y_2^2 \over 2!}{y_3^3 \over 3!}+{6 \over 5^3}{y_3^6 \over 6!}\cr
W_6 (n=4) \ &: \ f={1 \over 6} \big[ {y_1^2 \over 2!}{y_4^2 \over 2!}+
y_1 y_2 y_3 y_4+y_1 {y_3^3 \over 3!}+2{y_2^2 \over 2!}{y_3^2 \over 2!}+
{y_2^3 \over 3!}y_4 \big] \cr
&\qquad +{2 \over 6^2}\big[ {y_2^2 \over 2!}{y_4^3 \over 3!}
+y_2{y_3^2 \over 2!}{y_4^2 \over 2!}+2 {y_3^4 \over 4!}y_4 \big]+{3! \over 6^3}
{y_3^2 \over 2!}{y_4^4 \over 4!}+{4! \over 6^4}{y_4^7 \over 7!}\cr} }
One checks that the equations for the associativity of the 
quantum ring are indeed 
satisfied. Take for instance $W_4$ in parallel to target $\IP_2$,
with $[y_0]=1$, $[y_1]=3/4$, $[y_2]=1/2$, $[F]=5/2$. Writing
\eqn\alacon{
f=\sum_{3a+2b=10} \nu_{a,b} {y_1^a \over a!}{y_2^b \over b!}=
\nu_{2,2} {y_1^2 \over 2!}{y_2^2 \over 2!}+\nu_{0,5}{y_2^5 \over 5!}}
the familiar equation $f_{222}=f_{112}^2 -f_{111}f_{122}$ yields
\eqn\familiere{
\nu_{0,5}=2 \nu_{2,2}^2 }
and the ``initial condition" $\nu_{2,2}=1/4$ gives agreement with the
above table, for $W_4$.

There also exist polynomial solutions for the same associativity conditions
but with different weights. For instance with three generators
as in $W_4$ but with weights $[y_0]=1$, $[y_1]=2/3$, $[y_2]=1/3$ and
$[F]=7/3$, with the same $f_{\rm cl}$, we find
\eqn\pafamil{
f=\ga {y_1^3 \over 3!}y_2 +2 \ga^2 {y_1^2 \over 2!}{y_2^3 \over 3!}+24 \ga^4
{y_2^7 \over 7!} }
where the remaining arbitrary coefficient $\ga$ would be determined by an
hypothetic geometric interpretation. That the cubic part $f_{\rm cl}$ is common
to $W_{n+2}$ and $\IP_n$ remains slightly mysterious to the 
authors. Is there some geometric connection, or could one ``twist" the 
matrix model to yield the same results as for $\IP_n$?

\subsec{Partition function}

There are indications \KLEI\ that formulas generalizing those in proposition 
2, as well as formula \conjnd, hold for quadrics.
This points to some system of equations satisfied by $\exp \sum F^{(g)}$,
where the free energy is decomposed according to the genus $g$,
analogous to the Virasoro constraints and their $W$--generalizations for
matrix models, which remain to be found.

To elaborate the case for $\IP_2$, assume that for $g>0$,
$F^{(g)}$ has no polynomial part and define a partition function
\eqn\partif{
\eqalign{
Z&= e^{- \big[ {y_0^2 y_2 +y_0 y_1^2 \over 2} \big]
+\sum_{g \geq 0} F^{(g)}}
\cr
&=\sum_{d \geq 0} \sum_{0 \leq \delta \leq d(d-1)/2}  z_{d,\delta}
{y_2^{d(d+3)/2-\delta} \over (d(d+3)/2-\delta)!} e^{d y_1} \cr}}
Here $z_{d,\delta}$ are for fixed $\delta$ the polynomials in $d$ of degree 
$2 \delta$ introduced in \conjnd\ and given explicitly \VAIN\ \KLEI\ up
to $\delta=6$ by the polynomial parts of proposition 2.
These numbers count reducible as well as irreducible degree $d$ plane 
curves with $\delta$ simple nodes through $d(d+3)/2-\delta$ generic points.
{}From homogeneity, if $g_i$ stands for the genus
of the $i$--th irreducible part 
(the same irreducible part may occur several times)
then
\eqn\movegenre{ \sum_i (g_i -1) ={d(d-3)\over 2} -\delta}
The sum over $\delta$ extends up to $d(d-1)/2$ (and not $(d-1)(d-2)/2$
for irreducible curves) corresponding to systems of $d$ lines through
$2d$ points in which case
\eqn\vivic{z_{d,d(d-1)/2} = {(2d)! \over 2^d d!} = (2d-1)!!}
the number of pairings as in Wick's theorem.
{}From section 2
\eqn\merde{
\eqalign{
Z&=\sum_{d \geq 0} {y_2^{d(d+3)/2} \over (d(d+3)/2)!} e^{d y_1}
+\sum_{d \geq 2} 3(d-1)^2 {y_2^{d(d+3)/2-1} \over (d(d+3)/2-1)!}
e^{d y_1} \cr
&+ \sum_{d \geq 3} {3 \over 2}(d-1)(d-2)(3d^2-3d-11) 
{y_2^{d(d+3)/2-2} \over (d(d+3)/2-2)!} e^{d y_1} \cr
&+ \cdots + \sum_{2 \delta \leq d(d-1)}
({(3d^2)^\delta \over \delta!}+\cdots) 
{y_2^{d(d+3)/2-\delta} \over (d(d+3)/2-\delta)!} e^{d y_1}+ \cdots \cr}}

\subsec{Higher genus}

As already stressed the challenge is to find a compact means 
-- at least as a conjecture -- to generate the contributions of higher genera. Focussing on the $\IP_2$ case, we have unsuccessfully tried a number
of possibilities with disappointing results. Although this is
not the usual practice, let us briefly mention some.

Perhaps too naively one might think that equation \assop\ is some genus zero restriction of an equation valid
for the full free energy. To check this idea we have once more tried to 
imitate the reasoning of section 2.3 by looking at a one 
dimensional family of elliptic curves $C_\lambda$ of degree $d \geq 3$ through 
$3d-1$ points using a projection 
$$C_\lambda \to \IP_1$$
where $\IP_1$ is the linear pencil of degree $d$ 
smooth adjoints of degree $d-2$ 
(i.e. curves through the nodes of $C_\lambda$) through $d-2$ of the fixed
points of the family $C_\lambda$.
Such adjoints cut residually $C_\lambda$ in two points. The projection
therefore gives a presentation of the elliptic curves as double covers
of $\IP_1$ ramified at $4$ points (corresponding to the $4$ tangent adjoints in the
family).
Then repeating the argument of section 2.3 one obtains, with $N_d^{(0)}$
($N_d^{(1)}$) the number of rational (elliptic)
 degree $d$ curves through $3d-1$ ($3d$) points we find a relation
\eqn\effort{
\eqalign{
N_d^{(1)}&= \sum_{d_1+d_2=d} \bigg[ 
N_{d_1}^{(1)} N_{d_2}^{(0)}
\big[2 d_1^2 d_2^2 { 3d-3\choose 3 d_1 -1 }-d_1 d_2^3 {3d-3 \choose 3d_1 -2}
\big] \cr
&- N_{d_1}^{(0)} N_{d_2}^{(1)} d_1 d_2^3{3d-3 \choose 3d_1 -3}
\bigg] +2 \nu_d^{(0,1)} -\nu_d^{(0,0)} -\nu_d^{(1,1)} \cr}}
where the numbers $\nu_d^{(0,0)}$ stand for the numbers of elliptic curves
through $3d-1$ points which have two extra assigned points with 
equal projections (i.e. the number of elliptic curves in such a family 
such that an adjoint goes through $d$ assigned points), 
$\nu_d^{(0,1)}$ a similar number when  one of the 
extra assigned points and a point of $C_\lambda \cap l$
(where $l$ is a fixed line) have equal projection.
Finally $\nu_d^{(1,1)}$ counts the curves when two points, one from 
$C_\lambda \cap l$ and one from $C_\lambda \cap l'$ have equal projections,
$l$ and $l'$ two fixed lines.

Unfortunately we do not know how to evaluate the 
$\nu_d^{(i,j)}$ but at least it gives a feeling, and a geometric interpretation, of the
corrections to a polarized form of equation \assop\ to evaluate the contributions in higher genus.

Another direction is to try to generalize the 
topological relations in a way similar
to those resulting from Fig.\crossing, including $1$ loop for genus $1$ 
as an example, in this way yielding several equations. Apparently we failed, perhaps for not having introduced ``gravitational descendants"
in the calculation.

\subsec{Miscellany}

\noindent{(i)} We have tentatively attributed the simplifications that 
seem to occur for quadrics (isomorphic to $\IP_1 \times \IP_1$) to the fact
that the genus $0$ free energy for $\IP_1$ is very simple, assuming that 
some relations exist between $F_{M_1}$, $F_{M_2}$ and $F_{M_1 \times M_2}$.
What is such a relation?

\noindent{(ii)} There are some analogies between counting rational curves 
over $\IC$ and counting rational points for varieties over $\IQ$ and even counting the integral points (i.e. points with integer coordinates)
inside a dilated integral polytope
(i.e. a polytope with vertices at integral points).
It would be highly interesting to discover a common framework.

\noindent{(iii)} Last but not least it remains of course to ascertain 
properly the status of the various relations in this paper. 

In short, the subject remains wide open and the above presentation had no
other pretence than to encourage further investigations.

\vfill\eject

\appendix{A}{}

Following Dubrovin \DUB\ let us show that the differential equation 
\diffG, taking into account homogeneity, 
and governing the quantum ring for $\IP_2$, is equivalent to a 
Painlev\'e VI equation. The argument amounts to recognize that for 
generic values of the deformation parameters the corresponding commutative
algebra over $\IC$ is semi--simple.
A reparametrization expresses it in the form 
\eqn\idempa{ \CT_{\ga} \circ \CT_{\gb} \ = \ \delta_{\ga,\gb} \ \CT_{\ga} }
as noticed in \idempo\ (we use the symbol $\circ$ for the ring multiplication
to avoid confusion in the sequel). In this basis the equivalent intersection 
form is interpreted as a metric on the tangent space to the 
parameter space
$H^*$. Expressing that the metric is flat 
(constant in the original coordinates) 
and taking into account homogeneity, 
one obtains, after some tedious calculations, the required Painlev\'e VI 
equation.  As far as possible we try to keep general notations for an expected 
generalization to other quantum rings.

\bigskip

\noindent{\bf 1.} We revert to upper indices for the variables $y^i$, 
$0 \leq i \leq 2$, and identify the ring structure as one on the tangent 
space to $H^*$ through the correspondence
\eqn\tang{ T_i \leftrightarrow {\partial \over \partial y^i} }
The ``metric"
\eqn\mettan{ \eta_{ij} = \langle T_i , T_j \rangle}
is constant in these coordinates (the only non vanishing components 
being $\eta_{02}=\eta_{11}=1$) and numerically equal to its inverse
\eqn\intan{ \eta^{ij} = (\eta^{-1})_{ij} }
It enjoys the fundamental property
\eqn\funtan{ \langle T_i \circ T_j , T_k \rangle 
=\langle T_i, T_j \circ T_k \rangle= {\partial^3 F \over \partial y^i
\partial y^j \partial y^k} }
Latin indices will label the initial {\it flat coordinates} $y^i$,
while greek indices will be used for canonical ones to be introduced
below.

As noticed at the end of section 2, eqns \potring\--\polpot,
the ring for $\IP_2$ is identified with $\IC[x] / P(x)$, with
\eqn\polquo{ P(x)=x^3 - f_{111} x^2 -2 f_{112} x -f_{122}=
\prod_{\ga=1}^3 (x-q^{\ga}) }
through
\eqn\idenappa{ T_0 \leftrightarrow 1 \quad T_1 \leftrightarrow x \quad T_2
\leftrightarrow x^2 -f_{111}x -f_{112} }
where $q^{\ga}$ denote the roots of the polynomial $P(x)$.
For generic $y$'s one introduces new {\it canonical coordinates} $u^{\ga}$
and a corresponding basis $\CT_{\ga} \leftrightarrow \partial / \partial u^{\ga}$ such that the ring structure takes its canonical form \idempa.
Using the identification with the tangent space (and summation over 
dummy indices) we have on the one hand
\eqn\handone{ \CT_{\ga} \ = \ {\partial y^i \over  \partial u^\ga}\
\ T_i }
while on the other hand in $\IC[x]/ P(x)$
\eqn\otherhand{ \CT_\ga \leftrightarrow \prod_{\gb, \gb \neq \ga}
{x - q^\ga \over q^\ga - q^\gb} }
Hence 
\eqn\basch{ \eqalign{
T_0&= \sum_{\ga} \CT_\ga \cr
T_1&= \sum_{\ga} q^\ga \ \CT_\ga \cr
T_2&= \sum_\ga \big[ (q^\ga)^2 -f_{111} q^\ga -f_{112} \big] \ \CT_\ga \cr}}

\bigskip

\noindent{\bf 2.} In canonical coordinates the intersection form,
or metric, reads
\eqn\metint{ \eta_{\ga \gb} = \langle \CT_\ga, \CT_{\gb} \rangle =
{\partial y^i \over \partial u^\ga} \eta_{ij} {\partial y^j \over 
\partial u^\gb} }
The unit being $T_0= \sum_\ga \CT_\ga$, we have from \funtan\
\eqn\funant{\eta_{\ga \gb} = 
\langle \CT_\ga \circ \sum_\gc \CT_\gc,\CT_\gb \rangle
=\langle \sum_\gc \CT_\gc, \CT_\ga \circ \CT_\gb \rangle=
\delta_{\ga,\gb} \langle \sum_\gc \CT_\gc, \CT_\ga \rangle}
and the metric reduces to its diagonal elements $\eta_{\ga \ga}$.
With the notation $( \ , \ )$ for the duality between tangent and
cotangent spaces
\eqn\duatct{ (dy^i , {\partial \over \partial y^j})\  = \ \delta^i_j }
we have
\eqn\notethat{ \langle T_i, T_j \rangle = \eta_{ik} (dy^k, {\partial \over \partial y^j})}
Correspondingly
\eqn\corap{\eqalign{
\eta_{\ga \ga}=\langle \sum_\gb \CT_\gb, \CT_\ga \rangle&= \langle 
T_0, \CT_\ga \rangle=\eta_{0k} (dy^k,{\partial \over \partial u^\ga})\cr
\eta_{\ga \ga}= {\partial y_0 \over \partial u^\ga} &\qquad 
y_0= \eta_{0k} y^k \ \hbox{(here $y^2$)}. }}
\bigskip

\noindent{\bf 3.} Inserting in the structure equations 
$T_i \circ T_j=F_{ijl} \eta^{lk} T_k$
the expressions $T_i=(\partial u^\ga / \partial y^i) \CT_\ga$, and
taking into account \idempa, one finds for any $u^\ga$, denoted $u$,
the differential equations
\eqn\difu{\eqalign{
{\partial u \over \partial y^0} &=1 \quad \left( 
{\partial u \over \partial y^1} \right)^2={\partial u \over \partial y^2} 
+f_{111}{\partial u \over \partial y^1}+f_{112} \cr
{\partial u \over \partial y^1}{\partial u \over \partial y^2}&=f_{112}
{\partial u \over \partial y^1}+f_{122} \quad \left( 
{\partial u \over \partial y^2} \right)^2=f_{122}
{\partial u \over \partial y^1}+f_{222} \cr}}
Therefore each $u^\ga - y^0$ is a function of $y^1,y^2$ only. In terms
of the roots $q^\ga$ we can identify the derivatives as 
\eqn\derident{ \eqalign{
{\partial u^\ga \over \partial y^0}&=1 \cr
{\partial u^\ga \over \partial y^1}&= q^\ga \cr
{\partial u^\ga \over \partial y^2}&={ P'(q^\ga)-(q^\ga)^2 \over 2}=
(q^\ga)^2 - f_{111} q^\ga -f_{112} .\cr}}
Changing coordinates from flat ($y$) to canonical ($u$) is highly 
non trivial, it not only requires the knowledge of the function $f$,
but also of the roots of $P(x)$.

\bigskip

\noindent{\bf 4.} Using homogeneity properties it is possible to find 
expressions for the coordinates $u^\ga$ rather than their derivatives.
Recall that the free energy satisfies
\eqn\pronefe{ F= f_{\rm cl} + f }
\eqn\protfe{\eqalign{
Ef_{\rm cl}&=3y^0 y^1+f_{\rm cl} \cr
Ef&=f \cr
{\partial f \over \partial y^0}&=0 \cr}}
where $E$ is the vector field
\eqn\vfe{
E=y^0 {\partial \over \partial y^0}+3 {\partial \over \partial y^1}-y^2
{\partial \over \partial y^2}=\sum_\ga u^\ga {\partial \over \partial u^\ga}}
The second equality stems from the fact that the roots $q^\ga$ as
well as the canonical coordinates are of weight $1$
\eqn\weican{ E q^\ga = {\partial q^\ga \over  \partial y^0} = q^\ga \qquad
E u^\ga =u^\ga }
and the vector field $E$ can be interpreted as an element of the 
ring\foot{From eqs.\basch\ and (A.22) it follows that setting
$L_{-1}=T_0$, $L_n=E^{\circ (n+1)}$, $n \geq 0$, and considering the
$L$'s as vector
fields, they form a Lie algebra $\{ L_n,L_m \}= (n-m) L_{n+m}$,
isomorphic to the Lie algebra on the affine line (part of a Virasoro algebra)
as noticed by Kontsevich. It is not known at present what to do with
this remark.}
\eqn\interE{ E \leftrightarrow y^0 T_0 + 3 T_1 -y^2 T_2=\sum_\ga u^\ga
\CT_\ga}
Since the metric identifies the tangent space with its dual, the product
structure can be transfered to the cotangent space. With $\eta^{\ga \gb}=
(\eta^{-1})_{\ga \gb}$, we have
\eqn\transfapp{ \sum_\gc \eta^{\ga \gc} \langle 
{\partial \over \partial u^\ga},{\partial \over \partial u^\gb}\rangle=
\delta^\ga_\gb=(du^\ga,{\partial \over \partial u^\gb}) }
Moreover $\eta_{\ga \gb}$ is diagonal, so is $\eta^{\ga \gb}$, hence the 
differentials $du^\ga$ behave under multiplication as $\eta^{\ga \ga} 
\partial / \partial u^\ga$ (no summation), i.e.
\eqn\multiapp{\eqalign{
du^\ga \circ du^\gb &\leftrightarrow \eta^{\ga \ga }
{\partial \over \partial u^\ga} \circ  \eta^{\gb \gb }
{\partial \over \partial u^\gb}=\eta^{\ga \ga} \eta^{\gb \gb} \delta_{\ga,\gb}
{\partial \over \partial u^\ga} \cr
du^\ga \circ du^\gb &=\eta^{\ga \ga} \delta_{\ga,\gb} du^\ga={\delta_{\ga,\gb}
\over \eta_{\ga \ga}} du^\ga \cr}}
Define 
\eqn\defscap{ \eqalign{
g^{\ga \gb}&= \ll du^\ga, du^\gb \gg\equiv (du^\ga \circ du^\gb,E ) \cr
&=\sum_\gc u^\gc (du^\ga \circ du^\gb, {\partial \over \partial u^\gc})=
{\delta_{\ga,\gb} \over \eta_{\ga \ga} } u^\ga \cr}}
i.e.
\eqn\defscaie{ \sum_{\gc} \eta_{\ga \gc} g^{\gc \gb} \ = \ u^\ga \ \delta_{\ga,\gb} }
and 
\eqn\scadefie{ g^{ij} = \ll dy^i,dy^j \gg = \sum_{\ga} 
{\partial y^i\over \partial u^\ga}{\partial y^j \over \partial u^\ga}
{u^\ga \over \eta_{\ga \ga }} }
The canonical coordinates $u^\ga$ appear as eigenvalues of the matrix 
$\sum_\gc \eta_{\ga \gc} g^{\gc \gb}$ or equivalently of 
$\sum_k \eta_{ik} g^{kj}$, thus as solutions of
\eqn\deteqap{ \det ( g^{ij} - u \eta^{ij} ) = 0}
Now $g^{ij} = (dy^i \circ dy^j,E)$, and since the differentials 
$dy^i$ behave under multiplication as $\eta^{ik} \partial / \partial y^k$
we have
\eqn\wehave{ g^{ij}=\eta^{ik} \eta^{jl} E F_{kl} \qquad 
g_{ij} = E F_{ij} }
{}From \protfe\ \vfe\
\eqn\calapp{ g_{ij}=F_{ij}-(\delta_{0i}F_{0j}-\delta_{2i}F_{2j})-
(\delta_{0j} F_{0i} - \delta_{2j} F_{2i})+3(\delta_{0i} \delta_{1j}+
\delta_{0j} \delta_{1i}) }
In a more transparent notation, denote the weights as
\eqn\denow{ d_F=d_f=1 \quad , \quad d_i = \left\{ \matrix{ 1 & {\rm if} & i=0 \cr
0 & {\rm if} & i=1 \cr -1 & {\rm if} & i=2} \right. }
then
\eqn\thenapp{ g_{ij}=(d_F-d_i-d_j)F_{ij}+3 (\delta_{0i} \delta_{1j}+
\delta_{0j} \delta_{1i}) }
The equation determining the coordinates $u^\ga$ (rather than their 
derivatives) takes the form
\eqn\taketheapp{ \det ( g_{ij} - u \eta_{ij}) =0 }
This is again given in terms of roots of a cubic polynomial depending
on the second derivatives of $F$. Explicitly this is 
\eqn\expliap{
\eqalign{ 
&\left\vert\matrix{ -y^2 & 3 & y^0 - u  \cr
3 & f_{11} +y^0 - u & 2 f_{12} \cr
y^0-u & 2 f_{12} & 3 f_{22} } \right\vert  \cr 
= (u-y^0)^3-f_{11}(u-y^0)^2+
3 &(y^2 f_{22}-4 f_{12})(u-y^0)+y^2(4f_{12}^2-3 f_{11} f_{22}) -27 
f_{22}=0 \cr}}
Once more we see that $(u^\ga - y^0)$ are functions of $y^1,y^2$ only.

\bigskip

\noindent{\bf 5.} The metric is flat
and the curvature vanishes.  In the coordinates $u^\ga$, the metric
is diagonal. This simplifies the corresponding (Darboux) equations which
in terms of the (symmetric) ``rotation coefficients", defined for 
$\ga \neq \gb$ through
\eqn\darmet{\eqalign{
\Gamma_{\ga \gb}&= {1 \over \sqrt{ \eta_{\gb \gb}}} {\partial 
\sqrt{ \eta_{\ga \ga}} \over \partial u^\gb}= 
{1 \over 2 \sqrt{ \eta_{\ga \ga} \eta_{\gb \gb}}} {\partial 
\eta_{\ga \ga} \over \partial u^\gb} \cr
&={1 \over 2 \sqrt{ \eta_{\ga \ga} \eta_{\gb \gb}}} {\partial^2 
y_0 \over \partial u^\ga \partial u^\gb} \cr}}
where we used \corap, take the form
\eqn\finga{ 
\encadremath{\eqalign{ \ga \neq \gb \neq \gc \ \ \ \ \ \ \ 
\ {\partial \Gamma_{\gb \gc}
\over \partial u^\ga} \ &= \ \Gamma_{\gb \ga} \ \Gamma_{\gc \ga} \cr
\ga \neq \gb \ \ \ \ \ \ \ \sum_\gc {\partial \Gamma_{\ga \gb}
\over \partial u^\gc}&=0 }}}
{}From their definition the $\Gamma$'s are of weight $-1$, hence
\eqn\fingaap{ \encadremath{
\big( \sum_\gc u^\gc {\partial \over \partial u^\gc} +1 \big) 
\Gamma_{\ga \gb}=0 }}
In deriving \finga\ we shall also obtain translation invariance of the 
metric, by an equal shift on all $u$'s
\eqn\shiftu{\sum_\gc
{\partial \over \partial u^\gamma}\eta_{\ga \ga}=0}
\noindent{\bf Proof:}
With a metric of the form $\sum_\ga \eta_{\ga \ga} (du_\ga)^2$, the
covariant derivative of a vector field $\sum_\ga w^\ga 
{\partial / \partial u^\ga}$ reads
\eqn\codervf{\eqalign{
D_\gb w^\ga &= {\partial w^\ga \over \partial u^\gb} +{1 \over 2\eta_{\ga\ga}}
\bigg({\partial \eta_{\ga \ga}\over \partial u^\gb} w^\ga -
{\partial \eta_{\gb \gb}\over \partial u^\ga}w^\gb+
{1 \over 2} \delta_\gb^\ga \sum_\gc 
{\partial \eta_{\ga\ga} \over 2 \eta_{\ga\ga} \partial u^\gc}w^\gc\bigg) \cr
&={1 \over \sqrt{\eta_{\ga\ga}}} \bigg( 
{\partial \psi^\ga \over \partial u^\gb} -\Gamma_{\gb \ga} \psi^\gc +
\delta_\gb^\ga \sum_\gc \Gamma_{\ga \gc}\psi^\gc \bigg) \cr}}
where we have set
\eqn\setpsia{ \psi^\ga = \sqrt{\eta_{\ga \ga}} w^\ga}
and $\Gamma_{\ga \gb}$ is as above (note that only $\Gamma_{\ga\gb}$
for $\ga\neq\gb$ enters in \codervf\ ).

In particular the vector 
$\sum_\ga \partial / \partial u^\ga=\partial /\partial y^0$ with constant
components $w^\ga=1$ for each $\ga$, is covariantly constant, i.e.
\eqn\covcons{ 0= \bigg( {\partial \eta_{\ga\ga} \over \partial u^\gb}-
{\partial \eta_{\gb\gb} \over \partial u^\ga} \bigg) +\sum_\gc 
{\partial \eta_{\ga\ga} \over \partial u^\gc} }
The first bracket vanishes by virtue of \corap\ i.e. $\eta_{\ga\ga}$
is a gradient, proving \shiftu.  From \covdef\ it follows that the zero
curvature conditions are equivalent to the integrability conditions
of the system
\eqn\intecond{ {\partial \psi^\ga \over \partial u^\gb} = \Gamma_{\ga \gb}
\psi^\gb - \delta_\gb^\ga \sum_\gc \Gamma_{\ga\gc} \psi^\gc}
In matrix notation, with $\psi$ a vector of component $\psi^\ga$,
$[\Gamma]_{\ga\gb}=\Gamma_{\ga\gb}=\Gamma_{\gb\ga}$ for $\ga\neq \gb$
and $[\Gamma]_{\ga\ga}=0$, and $E_\ga$ the matrix with elements
$[E_\ga]_{\gb\gc}=\delta_{\ga\gb}\delta_{\ga\gc}$, this reads
\eqn\matcondint{{\partial \psi \over \partial u^\gb} = 
\big[ \Gamma,E_\gb \big] \psi}
{}From the definition of $\Gamma$ and \shiftu\ a first natural solution of this
system is
\eqn\natsol{ \psi^\ga_0=\sqrt{\eta_{\ga\ga}} ={1 \over  \sqrt{\eta_{\ga\ga}}}
{\partial y_0 \over \partial u^\ga}}
One readily sees that the compatibility conditions of
\matcondint\ are the Darboux equations \finga\ as claimed. The  latter
together with the homogeneity constraint \fingaap\ can be summarized in a 
single equation as follows. Define the antisymmetric matrix $V$
through
\eqn\defVmat{ V=[\Gamma,U] \qquad V^{\ga\gb}=-V^{\gb\ga}=
\Gamma_{\ga\gb}(u^\gb -u^\ga)}
where 
\eqn\defUmat{ U={\rm diag}\{ u^\ga\} }
Then a little calculation shows that \finga\--\fingaap\ are equivalent to
\eqn\reformat{ \encadremath{ {\partial V \over \partial u^\ga}
+\bigg[ V,\big[ \Gamma,E_\ga \big] \bigg] =0}}
The matrix $V$ enjoys the following properties

\noindent{(i)} If $\psi$ is a solution of \matcondint\ it follows from 
\reformat\ that so is $V\psi$, and generically $V$ can be diagonalized 
in the space of solutions of \matcondint\ with constant 
eigenvalues. Indeed let $\psi$ be an eigenvector of $V$ with 
eigenvalue $\mu$
\eqn\eigenpsi{ V \psi = \mu \psi}
then
\eqn\thenpsi{\eqalign{
{\partial(V\psi) \over \partial u^\ga}&=
{\partial \mu \over u^\ga} \psi + \mu [\Gamma,E_\ga] \psi \cr
&={\partial \mu \over u^\ga} \psi +[\Gamma,E_\ga] (V\psi) \cr}}
and from $\partial (V\psi)/\partial u^\ga=[\Gamma,E_\ga](V\psi)$ we
conclude that
\eqn\concmu{ {\partial \mu \over \partial  u^\ga}=0}

\noindent{(ii)} The eigenvalues of $V$ are the weights of the corresponding
eigenvectors, indeed from $\sum_\ga u^\ga E_\ga=U$ and the fact that 
$\psi$ is an eigenvector of $V$, we get
\eqn\alltomat{ \sum_\ga u^\ga {\partial \psi \over \partial u^\ga}=
\sum_\ga u^\ga [\Gamma,E_\ga]\psi=[\Gamma,U]\psi=V\psi=\mu \psi}

\noindent{(iii)} We noted in \natsol\ that 
$\psi^\ga_0=\partial y_0 / \sqrt{\eta_{\ga\ga}} \partial u^\ga$ 
is a solution of
\matcondint. It is also an eigenvector of
$V$ with eigenvalue $\mu_0=-1$ since $\psi^\ga_0=\sqrt{\eta_{\ga\ga}}$, and
the weight of $\eta_{\ga\ga}$ is 
$[\eta_{\ga\ga}]=[y_0]-[u^\ga]=[y^2]-[u^\ga]=-2$.
More generally if $\psi^\ga$ is of the form $\sqrt{\eta_{\ga\ga}}w^\ga$,
where $w^\ga$ are the components of a covariantly constant vector field,
it satisfies eq.\intecond\ from its very derivation.
Now for each index $i$ the vector field $\partial/\partial y^i$ is
covariantly constant, hence yields a solution
\eqn\solcov{\eqalign{\psi^\ga_i&= \sqrt{\eta_{\ga\ga}} 
{\partial u^\ga \over \partial y^i}=\sum_\gb{\eta_{\ga\gb} \over \sqrt{\eta_{\ga\ga}}}{\partial u^\gb \over \partial y^i} \cr
&=\sum_j {\eta_{ij} \over \sqrt{\eta_{\ga\ga}}} 
{\partial y^j \over \partial u^\ga}={1 \over \sqrt{\eta_{\ga\ga}}}
{\partial y_i \over \partial u^\ga} \cr}}
In this derivation we have used the fact that $\eta_{\ga\gb}$ is diagonal 
and equal to 
$(\partial y^i / \partial u^\ga)\eta_{ij} (\partial y^j / \partial u^\gb)$,
with $\eta_{ij}$ constant, and we have set
\eqn\wehaveset{ y_i=\sum_j \eta_{ij} y^j }
The solutions $\psi_i$ are a complete set of homogeneous eigenvectors
of $V$ and in the present case their weights are
\eqn\weighpres{ \mu_0=-1 \qquad \mu_1=0 \qquad \mu_2=1 }
Seen as a matrix, $\psi^\ga_i$ is, up to a multiplicative
constant, nothing but the Jacobian of the change of coordinates.
It follows from \solcov\ that the $\psi$'s satisfy the orthogonality
relations
\eqn\psiorto{ \eta_{ij}= \sum_\ga {\partial u^\ga \over \partial y^i}
\eta_{\ga\ga} {\partial u^\ga \over \partial y^j}=\sum_\ga
\psi^\ga_i \psi^\ga_j}
and the completeness relation
\eqn\compsi{ \delta^{\ga\gb}=\sum_{i,j} \psi^\ga_i \eta^{ij} \psi^\gb_j}
Furthermore from \natsol\ and \solcov\  we have
\eqn\derder{ {\partial y_i \over \partial u^\ga}=\psi^\ga_0 \psi^\ga_i}
Finally we obtain the structure constants of the ring as follows.
Since
$$ {\partial u^\ga \over \partial y^i}=
{1 \over \sqrt{\eta_{\ga\ga}}} \psi^\ga_i={\psi^\ga_i \over
\psi^\ga_0}$$
and
$$\eqalign{ {\cal T}_\ga \circ {\cal T}_\gb &= \delta_{\ga\gb} {\cal T}_\ga
\cr  T_i &= \sum_\ga {\partial u^\ga \over \partial y^i} {\cal T}_\ga \cr
{\cal T}_\ga &=\sum_\ga {\partial y^i \over \partial u^\ga} T_i \cr}$$
we have
$$\eqalign{T_i \circ T_j &= \sum_{\ga\gb} {\partial u^\ga \over \partial y^i}
{\partial u^\gb \over \partial y^j}  {\cal T}_\ga \circ {\cal T}_\gb \cr
&=\sum_{\ga} {\partial u^\ga \over \partial y^i}
{\partial u^\ga \over \partial y^j} {\cal T}_\ga \cr
&=\sum_{\ga,l}  {\partial u^\ga \over \partial y^i}
{\partial u^\ga \over \partial y^j}{\partial y^l \over \partial u^\ga} T_l\cr
&=\sum_{\ga,k,l}  {\partial u^\ga \over \partial y^i}
{\partial u^\ga \over \partial y^j}{\partial y_k \over \partial u^\ga}
\eta^{kl} T_l\cr
&=\sum_{\ga,k,l}{\psi^\ga_i \over \psi^\ga_0} {\psi^\ga_j \over
\psi^\ga_0} \psi^\ga_0 \psi^\ga_k \eta^{kl} T_l\cr
&= \sum_{k,l} F_{ijk} \eta^{kl} T_l \cr}$$
We finally deduce that
\eqn\deducF{
\encadremath{ F_{ijk} \equiv {\partial^3 F \over \partial y^i \partial y^j 
\partial y^k }= \sum_\ga {\psi^\ga_i \psi^\ga_j 
\psi^\ga_k 
 \over \psi^\ga_0}}}
a formula reminiscent of similar ones in conformal or integrable
field theories (or even finite group theory). Finally from
\compsi\
\eqn\compsuite{ V^{\ga\gb}=\sum_\gc V^{\ga \gc} \delta^{\gc \gb}=
\sum_{\gc} V^{\ga\gc} \psi^\gc_i\eta^{ij} \psi^\gb_j=
\sum_{ij} \mu_i \psi^\ga_i \eta^{ij} \psi^\gb_j}
\bigskip

\noindent{\bf 6.} The matrix $V=[\Gamma,U]$ fulfills
the master equation \reformat. Since ${\bf I} = \sum_\ga E_\ga$
it follows that
\eqn\follo{ \sum_\ga {\partial V \over \partial u^\ga} 
= \sum_\ga u^\ga {\partial V \over \partial \ga}=0}
Now 
\eqn\adjga{ \Gamma_{\ga\gb}= {V^{\ga\gb} \over u^\gb - u^\ga}=
-({\rm ad}_U^{-1} V)_{\ga\gb} }
where we recall that $\Gamma$ is symmetric with vanishing diagonal. This allows
to rewrite \reformat\ as
\eqn\rewmas{ {\partial V \over \partial u^\ga} =[{\rm ad}_{E_k}
{\rm ad}_U^{-1} V,V]}
For the $\IP_2$ quantum ring the index $\ga$ runs from $1$ to $3$.
One introduces a simplified notation by setting for $(\ga\gb\gc)$
a cyclic permutation of $(123)$
\eqn\setga{ V^{\ga\gb}= \Omega_\gc }
so that \follo\ becomes
\eqn\folbeco{ 
({\partial \over \partial u^1}+{\partial \over \partial u^2}+
{\partial \over \partial u^3}) \Omega_\ga =
(u^1{\partial \over \partial u^1}+u^2{\partial \over \partial u^2}+
u^3{\partial \over \partial u^3})\Omega_\ga =0}
Thus the $\Omega$'s, which are a priori functions of three (complex)
numbers, are invariant under a common translation or dilation of their
arguments, so they are only  functions of a single variable
\eqn\singvarf{ s={u^3 -u^1 \over u^2 -u^1}}
which is the value of $u^3$ when the coordinate system is chosen so that
$u^1=$, $u^2=1$. As a result the master equation \reformat\ reduces to
\eqn\master{\encadremath{ {d \Omega_1 \over ds} = -{\Omega_2 \Omega_3 \over s}
\quad {d \Omega_2 \over ds} = -{\Omega_3 \Omega_1 \over 1- s}
\quad  {d \Omega_3 \over ds} = {\Omega_1 \Omega_2 \over s(1-s)} }}
This is an evolution on a symplectic leaf
\eqn\simpleaf{ \Omega_1^2+\Omega_2^2+\Omega_3^2 = {\rm cst.} = -R^2 }
isomorphic to a sphere. Indeed with a Poisson structure on the (dual of the)
Lie algebra $so(3)$
\eqn\lieso{ \{ \Omega_\ga , \Omega_\gb \} =\Omega_\gc \quad (\ga\gb\gc) \ 
\quad\   
\hbox{cyclic permutation of }\ (123)}
the equations \master\ read
\eqn\liered{ {d \Omega_\ga \over ds}= \{ H, \Omega_\ga \} \qquad H={1 \over 2}
\bigg( {\Omega_1^2 \over s-1}+{\Omega_2^2 \over s} \bigg) }
To find the constant in \simpleaf\ we compute the characteristic
polynomial of the matrix $V$ with eigenvalues
$\{ \mu_i \}=\{0,\pm1\}$. The trace and determinant vanish, while
\eqn\taceV{ {\rm tr} V \wedge V =\sum_{i<j} \mu_i \mu_j = -1 =\sum_\ga
\Omega_\ga^2 = -R^2}
Following Dubrovin we keep the general notation $R$ in
\simpleaf\ which in our case can be taken equal to $1$.

\bigskip

\noindent{\bf 7.} The final step is to replace the system \master\ by a single 
non linear differential equation for $\Omega_3$ as follows.
We take the $s$--derivative of the third equation
$$ {d^2 \Omega_3 \over ds^2}={2s-1 \over s(1-s)} 
{d \Omega_3 \over ds} -{1 \over s(1-s)} \Omega_3(  
{\Omega_1^2 \over 1-s}+{\Omega_2^2 \over s} )$$
while $\Omega_1^2$ and $\Omega_2^2$ can be eliminated through
$$\Omega_1^2+\Omega_2^2= -R^2 -\Omega_3^2 \qquad (\Omega_1 \Omega_2)^2=
(s(1-s){d \Omega_3 \over ds})^2 $$
With
\eqn\defOm{ \Omega_3(s)= i \phi(s) }
this yields
\eqn\plphi{ \eqalign{
\bigg[ {d^2 \phi \over ds^2} +{1-2s \over s(1-s)} &{d \phi \over ds} +
{\phi(\phi^2 -R^2) \over 2s^2(1-s)^2} \bigg]^2= \cr
&={(1-2s)^2 \over 4s^4(1-s)^4} \phi^2 \big[ (\phi^2 -R^2)^2+4s^2(1-s)^2
({d \phi \over ds})^2 \big] \cr}}
To recognize that this is (a particular case of) a
Painlev\'e VI equation, one has to change both the argument and the function as follows.
First one trades $s$ for $z$ defined through\foot{Here we seem
to disagree with Dubrovin \DUB.}
\eqn\chvarsz{ 1-2s={z^{1/2} +z^{-1/2} \over 2} }
so that
\eqn\pltwo{\eqalign{\bigg[{d^2 \phi \over dz^2}+{3z-1 \over 2z(z-1)}
&{d \phi \over dz}+{2 \phi (\phi^2-R^2) \over z(z-1)^2} \bigg]^2= \cr
&= \bigg({(z+1)\phi \over z(z-1)}\bigg)^2 \bigg( \big({d \phi \over dz}
\big)^2 +{(\phi^2-R^2)^2 \over z(z-1)^2} \bigg) \cr}}
The final change amounts to replace the function $\phi(z)$ by $v(z)$
according to the following
\bigskip

\noindent{\bf Proposition A}
For $R \neq 0,-1/2$, the equation \pltwo, which summarizes the associativity 
and homogeneity conditions for the quantum ring of $\IP_2$ is in one
to one correspondence with the Painlev\'e equation
\eqn\pain{ \encadremath{
\eqalign{ {d^2 v \over dz^2}&={1 \over 2}({1 \over v}+{1 \over v-1}+
{1 \over v-z})\big( {dv \over dz} \big)^2-({1 \over z}+{1 \over z-1}+
{1 \over v-z})\big( {dv \over dz} \big) \cr
&+ {v(v-1)(v-z) \over z^2(z-1)^2} \bigg[ {(2R+1)^2 \over 8} (1-{z \over v^2})
+{z(z-1) \over 2(v-z)^2 } \bigg] }}}
through the field redefinition
\eqn\fielred{\encadremath{
2\phi+1 =2{z \over v} {dv \over dz} +{2R+1 \over z-1}({z \over v}-v) } }
and conversely 
\eqn\convpain{\eqalign{
v&={A \over B} \cr
A&=(z+1){d\phi \over dz} -{4R \over z-1}\phi \cr
&+{z(z-1)^2 \over z+1}\phi \bigg[{d^2\phi \over dz^2}+{3z-1 \over 2z(z-1)}
{d\phi \over dz}+{2 \phi (\phi^2-R^2)\over z(z-1)^2} \bigg] \cr
B&=2 {d\phi \over dz} -{1 \over z} \bigg[ \phi^2+2
\phi R {z+1 \over z-1} +R^2 \bigg] \cr}}
The proof, due to the authors of ref.\FOK, will not be reproduced here.
It follows from a very painful but straightforward calculation.

Some final remarks are in order.

\noindent{(i)} Our interest here is for $R=1$. For the exceptional
cases $R=0,-1/2$ see Dubrovin \DUB.

\noindent{(ii)} The most general Painlev\'e VI equation depends
on four parameters $a$, $b$, $c$ and $d$, and reads
\eqn\truepain{\eqalign{
{d^2v \over dz^2} &= {1 \over 2}({1 \over v}+{1 \over v-1}+
{1 \over v-z})\big( {dv \over dz} \big)^2-({1 \over z}+{1 \over z-1}+
{1 \over v-z})\big( {dv \over dz} \big) \cr
&+ {v(v-1)(v-z) \over z^2(z-1)^2} \bigg[ 
a+b {z \over v^2}+c{z-1 \over (v-1)^2}
+d{z(z-1) \over (v-z)^2 } \bigg] \cr}}
The special case occuring here corresponds to
\eqn\specabcd{ a=-b={(2R+1)^2 \over 8} \to {9 \over 8} \quad c=0 
\quad d={1 \over 2} }

\noindent{(iii)} It would seem futile to reconstruct from an appropriate 
solution of the Painlev\'e equation the (genus zero) free energy $F$.
The whole point of the exercise is to prepare the way for a future
generalization in the hope of including the badly missing 
contributions of higher genera.

\noindent{(iv)} Courageous readers are invited to describe quantum 
deformations of other cohomology rings in the language of this appendix.
This should be rather straightforward for higher dimensional projective 
spaces.

\listrefs
\end